\documentclass[twocolumn,aps,prb,superscriptaddress,usenames,dvipsnames]{revtex4-1}
\usepackage[latin9]{inputenc}
\usepackage{color}
\usepackage{amsmath}
\usepackage{amssymb}
\usepackage{graphicx}
\PassOptionsToPackage{normalem}{ulem}
\usepackage{ulem}

\makeatletter
\usepackage{bm}
\usepackage{psfrag}
\usepackage{xfrac}

\usepackage{xcolor}
\usepackage{siunitx}
\usepackage{placeins}
\usepackage{hyperref}

\newcommand{\teal}[1]{{\textcolor{black}{#1}}}

\newcommand{\carlos}[1]{{\textcolor{black}{#1}}}

\newcommand{\siggi}[1]{{\textcolor{black}{#1}}}

\newcommand{\dmz}[1]{{\textcolor{black}{#1}}}

\newcommand{\denis}[1]{{\textcolor{black}{#1}}}

\newcommand{\denisnew}[1]{{\textcolor{black}{#1}}}

\usepackage{setspace}

\makeatother

\begin{document}
\title{Quantum oscillations in 2D electron gases with spin-orbit and Zeeman interactions}
\author{Denis R. Candido}
\affiliation{Department of Physics and Astronomy, University of Iowa, Iowa City, Iowa 52242, USA}

\author{Sigurdur I. Erlingsson}
\author{Hamed Gramizadeh}
\affiliation{Department of Engineering, Reykjavik University, Menntavegi 1, IS-102 Reykjavik, Iceland}
\author{Jo\~ao Vitor I. Costa}
\affiliation{Instituto de F\'isica de S\~ao Carlos, Universidade de S\~ao Paulo, 13560-970 S\~ao Carlos, SP, Brazil}
\author{Pirmin~J.~Weigele}
\affiliation{Department of Physics, University of Basel, CH-4056, Basel, Switzerland}
\author{Dominik~M.~Zumb\"uhl}
\affiliation{Department of Physics, University of Basel, CH-4056, Basel, Switzerland}
\author{J. Carlos Egues}
\affiliation{Instituto de F\'isica de S\~ao Carlos, Universidade de S\~ao Paulo, 13560-970
S\~ao Carlos, SP, Brazil}
\affiliation{Department of Physics, University of Basel, CH-4056, Basel, Switzerland}

\begin{abstract}

\dmz{Shubnikov-de Haas (SdH) oscillations are the fingerprint of the Landau and Zeeman splitting level structure on the resistivity in presence of a moderate magnetic field before full quantization is manifest in the integer quantum Hall effect. These oscillations have served as a paradigmatic experimental probe and tool for extracting key semiconductor parameters such as carrier density, effective mass $m^*$, Zeeman splitting with g-factor $g^*$, quantum scattering time and \denis{Rashba $\alpha$ and Dresselhaus $\beta$} spin-orbit (SO) coupling parameters.} 
\denis{Analytical descriptions of the SdH oscillations are available for some special cases, but the generic case with all three terms simultaneously present has not been solved analytically so far, seriously hampering the analysis and interpretation of experimental data.} \carlos{Here, we bridge this gap by providing an analytical formulation for the SdH oscillations of 2D electron gases (2DEGs) with simultaneous and arbitrary Rashba, Dresselhaus, and Zeeman interactions. We use a Poisson summation formula for the density of states of the 2DEG, which affords a complete yet simple description of the oscillatory behavior of its magnetoresistivity. Our} analytical and numerical calculations allow us to extract the beating frequencies, \carlos{quantum lifetimes,} and also \carlos{to} understand the role of higher harmonics in the SdH oscillations. \dmz{More importantly, we derive a simple condition for the vanishing of SO induced SdH beatings for all harmonics in 2DEGs: $\alpha/\beta= [(1-\tilde \Delta)/(1+\tilde \Delta)]^{1/2}$, where $\tilde \Delta\propto g^* m^*$ is a material parameter given by the ratio of the Zeeman and Landau level splitting. This  condition is notably different from \carlos{that of} the persistent spin helix at $\alpha/\beta=1$ for materials with large $g^*$ such as InAs or InSb.} We also predict beatings in the higher harmonics of the SdH oscillations and elucidate the inequivalence of the SdH response of Rashba-dominated ($\alpha>\beta$) vs Dresselhaus-dominated ($\alpha<\beta$) 2DEGs in semiconductors with substantial $g^*$. We find excellent agreement with recent available experimental data of Dettwiler {\it et al.} Phys. Rev. X \textbf{7}, 031010 (2017), and Beukman {\it et al.}, Phys. Rev. B \textbf{96}, 241401 (2017). The new formalism builds the foundation for a new generation of quantum transport experiments and spin-orbit materials with unprecedented physical insight and material parameter extraction.

\end{abstract}
\date{\today}

\maketitle

\section{Introduction}
\dmz{The spin-orbit (SO) interaction couples the orbital and spin degrees of freedom, not only \carlos{forms} the basis for a range of spin related effects such as the spin Hall effect~\cite{d1971possibility,dyakonov1971current,PhysRevLett.83.1834,doi:10.1126/science.1114655} and the persistent spin helix~\cite{PhysRevLett.90.146801,PhysRevLett.97.236601,PhysRevLett.117.226401},} but also underlies the physical mechanisms of new phases of matter, \carlos{e.g.,} topological insulators, quantum spin Hall materials ~\cite{PhysRevLett.95.226801,PhysRevLett.96.106802,doi:10.1126/science.1133734}, and Majorana~\cite{Kitaev_2001,PhysRevB.79.161408,candido18:256804}, Dirac and Weyl fermions~\cite{doi:10.1119/1.3549729}. Accordingly, advancing techniques and methods to measure and extract SO couplings from experimental data are crucial for the development of these fields. 
\begin{figure}[ht!]
\centering \includegraphics[width=0.475\textwidth]{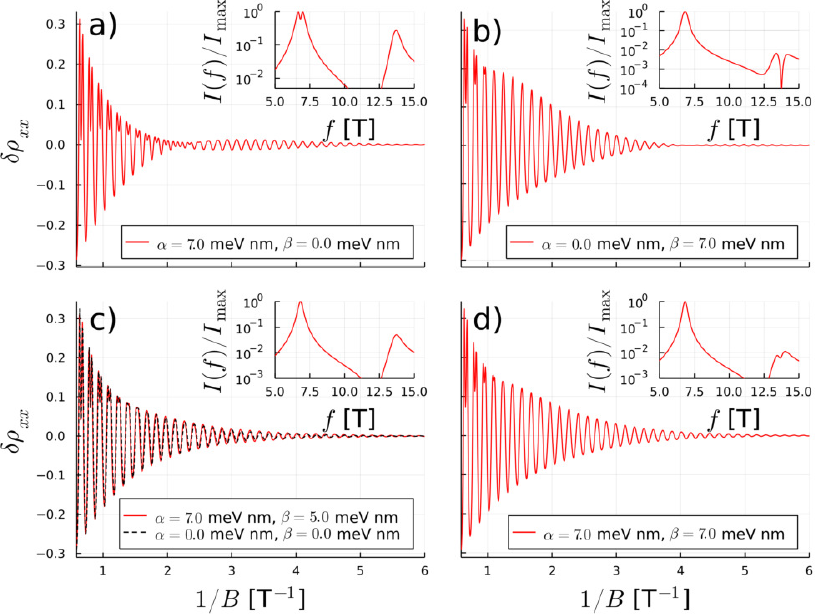}
\caption{Magnetoresistivity for a) pure Rashba $\alpha=7.0$\,meV\,nm and b) pure Dresselhaus $\beta=7.0$\,meV\,nm \dmz{with ${m^*=0.019 m_o}$ and $n_{2D}=3.3 \times 10^{11}$\,cm$^{-2}$ from Ref.~\onlinecite{gilbertsson08:165335}. The curves in a) and b) are not the same due to the large $g$--factor $g^{*}=-34$. The insets display the normalized FFT including the 2nd harmonic. 
The presence of beating nodes in $\delta \rho_{xx}$ are clearly visible in a) the fundamental and b) the 2nd harmonic, see Fig.\ \ref{fig:drxx_D}.
The condition for the {\em absence} of beatings (single peak for each harmonic) is $\alpha=7.0$\,meV\,nm and $\beta=5.0$\,meV\,nm, shown in c), but not $\alpha=\beta=7.0$\,meV\,nm, the persistant spin helix case, shown in d), clearly exhibiting a beating (here a splitting of 2nd harmonic peak). }}
\label{fig1} 
\end{figure}

Shubnikov-de Haas (SdH) oscillations\cite{sdh-original,sdh-original2} are among the best techniques to probe simultaneously spin- and charge-related quantities associated to electrons in semiconductors, including effective masses, gyromagnetic ratios, \dmz{quantum scattering times}, densities and SO couplings. \teal{Most recently, they have \carlos{been} crucial to the study and understanding of new materials, as for example, 2D-materials, transition metal dichalcogenides, van der Waals heterostructures~\cite{PhysRevLett.123.097403,Kormanyos_2015,cui2015multi,xu2016universal,cui2017low,rhodes2019disorder,PhysRevResearch.5.013113,slizovskiy2023kagom}, and also materials hosting new phases of matter e.g., topological insulators~\cite{Ferreira2022}, unconventional superconductivity~\cite{cao2018unconventional} and correlated insulator behavior~\cite{cao2018correlated}. \carlos{It} has also \siggi{been used to establish} the presence of nodal-lines~\cite{hu2016evidence}, Berry's phase~\cite{doi:10.1126/science.1242247,doi:10.1126/sciadv.aax6550}, and different topology of Fermi surfaces~\cite{alexandradinata2018revealing}. SdH oscillations} are  magneto-oscillations in the resistivity and originate from the sequential crossings of the discrete Landau Levels (LLs) through the Fermi energy. Without SO coupling and in the low-field regime, the period of the SdH oscillations can be related to the density of the electron gas\cite{ihn10:book}. In the presence of SO interaction, on the other hand, the energy spectrum changes dramatically thus leading to additional frequencies in the  magnetoresistivity and hence beatings, Figs.~\ref{fig1}(a). This was first theoretically described semiclassically by Das \textit{et. al.,} \cite{das89:1411}. In the so-called Onsager's picture, different sub-bands possess different Sommerfeld quantized orbits (playing the role of the LLs), which cross the Fermi energy with different frequencies in $B^{-1}$. The spin-split bands give rise to two distinct oscillating frequencies in the magnetotransport. The standard experiment relies on Fourier analyzing the measured SdH oscillations. An experimental method introduced in \denis{Refs.~\onlinecite{PhysRevLett.78.1335,engels97:1958R,10.1063/1.367192}} has often been used to estimate the strength of the Rashba coupling via the splitting of the Fourier frequency peaks.  However, these methods have been criticized for not accounting for the Zeeman splitting (through the g-factor $g^*$) nor for the additional Dresselhaus SO coupling~\citep{gilbertsson08:165335}.

There have been some attempts to analyze the SdH oscillations taking into account both $\alpha$, $\beta$ \carlos{\textit{and}} $g^{*}$. However, these mostly involved qualitative comparison with the energy spectrum of pure Rashba and pure Dresselhaus\cite{akabori06:413,akabori08:205320}. {In Ref.~\onlinecite{yang06:045303}, \carlos{fully} numerical calculations of  magneto-oscillations were performed but for relatively high magnetic fields and low electron densities, far away from the regime of recent experimental works~\cite{beukmann17:241401}.  Moreover, it was realized that in the absence of the Zeeman interaction, important features are absent. 
More specifically, without accounting for the spin mixing generated by the magnetic field (via the Zeeman interaction), predictions become imprecise \cite{winkler2000anomalous}, and even fail to describe phenomena such as magnetic inter-subband scattering \cite{raikh94:5531} and magnetic breakdown \cite{averkiev05:543}.} In general, full quantum mechanical numerics are generally done in order to check agreement with experiments, which are neither very practical nor elucidate much of the physics happening in those systems \cite{beukmann17:241401,cimpoiasu19:075704}. Finally, all the previous works have neglected the influence of higher harmonics, recently seen experimentally~\cite{psh-prx}.
\vskip 2cm
Here, we present a detailed investigation of SdH oscillations in the presence of SO couplings of both Rashba $\alpha$ and Dresselhaus $\beta$ types {\em and} Zeeman interaction \dmz{with} g-factor $g^*$. Our main result is the derivation, for the first time in the literature, of a simple analytical expression for the SdH oscillations in the presence of simultaneous arbitrary couplings $\alpha$ and $\beta$ in addition to $g^*$. {\carlos{We note that earlier analytical descriptions of SdH magnetoresistivity oscillations considered particular cases, namely, when either only one of the parameters $\alpha$, $\beta$ or $g^*$ was nonzero or any two of these parameters were nonzero, with the exceptions ($\alpha\neq\beta\neq 0$, $g^*=0$) and ($\alpha=\beta$, $g^*\neq 0$).} 

\carlos{Interestingly,} our analytical formula generalizes previous results\cite{averkiev05:543} for $g^*=0$ and predicts a new condition for the vanishing of the SdH magneto-oscillation beatings \carlos{in all harmonics [e.g., Figs.~\ref{fig1}(c)]} in Rashba-Dresselhaus coupled 2DEGs with substantial Zeeman splittings, \denis{namely,}
 \begin{equation}
\frac{\alpha}{\beta}=\sqrt{\frac{1-\tilde{\Delta}}{1+\tilde{\Delta}}}, 
\label{new-condition}
\end{equation}
\noindent
where $\tilde \Delta$ is \dmz{a material parameter given by} the ratio between the Zeeman splitting and the Landau-level spacing. As we discuss later on, Eq.~(\ref{new-condition}) \textit{is not} associated with a conserved quantity in our system; this contrasts with the persistent-spin-helix condition $\alpha=\beta$, which predicts spin conservation along particular axes~\cite{PhysRevLett.90.146801,PhysRevLett.97.236601,PhysRevLett.117.226401}. \carlos{We stress that this case with  $\alpha=\beta$ and generic $g^*\neq 0$ leads to beating in the frequency spectrum of our system, Figs.~\ref{fig1}(d), as opposed to our new condition in Eq.~\ref{new-condition}.} As we discuss below, our numerical and analytical approaches show excellent agreement with available data from Refs.~\onlinecite{beukmann17:241401} and \onlinecite{psh-prx}.

Our approach combines a semiclassical formulation for the resistivity of 2DEGs with a trace formula for the density of states (DOS) in a quantizing magnetic field.  The trace formula expresses the DOS using the usual Poisson summation formula\cite{brack97:book}. This formulation brings out the oscillatory part of the DOS, thus allowing us to clearly identity the higher harmonics of the SdH oscillations. It enables us to conveniently separate the frequency scales into ``fast'' and ``slow'' oscillations thus allowing for a clearer interpretation of the underlying physical phenomena, e.g., the slow beating SdH oscillations due to the SO coupling.

Our main results for the oscillatory part of magnetoresistivity $\delta \rho_{xx}(1/B)$ and its frequency spectra $I(f)$ [panel insets] are show in Fig.~\ref{fig1}. For pure Rashba [$\alpha \neq 0$, $\beta=0$, Fig.~\ref{fig1}a)] and pure 
Dresselhaus [$\alpha = 0$, $\beta \neq 0$, Fig.~\ref{fig1}b)], but non-zero Zeeman term ($g^*\neq0$), the frequency spectra, as 
usual, show two main peaks, which correspond to the first two Fourier components of $\delta \rho_{xx}(1/B)$. These two cases, 
however, exhibit a marked contrast: while the pure Rashba shows a peak splitting at the fundamental frequency, the pure Dresselhaus exhibits 
a peak splitting in the second harmonic. As we explain in detail in Sec.\ \ref{sec:SdH_beatingsPure}, this contrasting behavior arises from the interplay 
between the Zeeman and SO interactions, which makes the SdH magneto-responses with \textit{nonzero} g-factors $g^*$ inequivalent for Rashba-dominated 
($\alpha > \beta$) vs.\  Dresselhaus-dominated ($\alpha<\beta$) 2DEGs. \carlos{For $g^*=0$, the pure Rashba and pure Dresselhaus cases give identical results.}}

Figure 1(c) illustrates our prediction 
in Eq.~(\ref{new-condition}) thus showing no peak splitting in the frequency spectra -- at any harmonic --  when this condition is satisfied. 
To emphasize this condition emulates a situation with no SO coupling (i.e., no beating), we plot in Fig.~1(c) the $\alpha=\beta=0$ (with $g^*\neq 0$) 
case [dashed curve in 1(c)], which shows complete overlap with the case satisfying Eq.~(\ref{new-condition}).  In contrast and for completeness, Fig.~\ref{fig1}d) shows the $\alpha=\beta\neq 0$ case with $g^*\neq 0$, which exhibits peak splitting in the \textit{second} harmonic.

\carlos{We have applied our analytical description to low-density GaAs-based 
quantum wells for which there are experimental data\cite{psh-prx} 
showing several harmonics in 
the SdH magneto-oscillations. Figure 2 shows the excellent agreement obtained, 
thus illustrating that our  semiclassical formulas can satisfactorily capture 
the higher harmonics of the SdH oscillations. Moreover, we have applied our analytical approach to low-density InSb-based 2DEGs\cite{gilbertsson08:165335,akabori08:205320} where, unlike GaAs-based 2DEGs, a strong SO coupling manifests itself as beatings in the measured SdH oscillations, and find good agreement.
We have also implemented a 
detailed numerical calculation for high-density InAs-based 2DEGs for 
which an analytical description is not adequate. 
Here again we find very good agreement with available data\cite{beukmann17:241401} and are able to extract
SO coupling parameters.}

Next (Sec~\ref{sec:hamiltonian}), we present a description of the Hamiltonian of our system. In Sec.~III we discuss how to obtain the ``$F$--function'', the central quantity in our formulation, from the Landau-quantized energy spectrum of our system and its connection with the density of states (DOS). The formalism for obtaining the Shubnikov-de Haas oscillations in terms of the Poisson summation formula and the F-function is described in Sec.~IV. Finally, in Sec.~V we present and analyze different particular cases of SdH oscillations and, more important, derive the new condition in Eq.~(\ref{new-condition}) for the complete absence of beatings (all harmonics) in the SdH oscillations, for 2DEGs with non-zero Rashba, Dresselhaus, and Zeeman couplings. \carlos{The appendices present relevant details of our theoretical formulation.} 

\section{2DEG Hamiltonian}
\label{sec:hamiltonian}
%
%
Our starting point is the Hamiltonian for a 2DEG \carlos{confined in} a quantum well \carlos{($xy$ plane)} grown along the [001] crystallographic direction\carlos{, taken as $z$ axis}. In the presence of a perpendicular external magnetic field $\textbf{B}=(0,0,B)$ and both Rashba\citep{bychkov84:6039} and Dresselhaus\citealp{dresselhaus55:580} spin orbit interactions, the Hamiltonian reads
\begin{eqnarray}
{\cal H} & = & \frac{1}{2m_{}^{*}}\left(\Pi_{x}^{2}+\Pi_{y}^{2}\right)+\frac{1}{2}{g^{*}\mu_{B}B}{}\sigma_{z}\nonumber \\
 & + & \frac{\alpha}{\hbar}\left(\Pi_{x}\sigma_{y}-\Pi_{y}\sigma_{x}\right)+\frac{\beta}{\hbar}\left(\Pi_{x}\sigma_{x}-\Pi_{y}\sigma_{y}\right), \label{Hinit}
\end{eqnarray}
where $g^*$ is the g-factor, $m^*$ is effective mass, $\bm{\Pi}=\textbf{p}-q\textbf{A}$ is the canonical momentum, $q$ is the electric charge, $\mu_B$ is the Bohr magneton, \carlos{$\hbar$ the reduced Planck's constant and $\sigma_x$, $\sigma_y$, $\sigma_z$ denote the usual Pauli matrices.} \carlos{The parameters $\alpha$ and $\beta$ denote the linear-in-$k$ Rashba and Dresselhaus SO} couplings, respectively. \carlos{The $\beta$ coupling includes a density dependent correction arising from the cubic Dresselhaus term. As we discuss later on [Sec.~\ref{sec:cubic}], our numerical results will account for the full cubic Dresselhaus term.}

\carlos{Let us} introduce the annihilation and creation operators associated to the Landau level $\left|n\right\rangle$ 
\begin{eqnarray}
a & = & \frac{\ell_{c}}{\sqrt{2}\hbar}\left(\Pi_{x}-i \zeta \Pi_{y}\right), \label{a}\\
a^{\dagger} & = & \frac{\ell_{c}}{\sqrt{2}\hbar}\left(\Pi_{x}+i \zeta \Pi_{y}\right),\label{adagger}
\end{eqnarray}
obeying $\left[a,a^{\dagger}\right]=1$, ${a\left|n\right\rangle =\sqrt{n}\left|n-1\right\rangle}$, ${a^{\dagger}\left|n\right\rangle =\sqrt{n+1}\left|n+1\right\rangle}$, $\zeta=-{\rm sign}\left(qB\right)$, \carlos{with the magnetic length and  the center of the Landau orbit denoted by $\ell_{c}=\sqrt{\frac{\hbar}{\left|qB\right|}}$ and $y_0=\frac{ek_x}{|qB|}$, respectively}. In this work, we have $q=-e$, where $e>0$ is the absolute value of the elementary electronic charge, and we choose $B>0$, yielding $\zeta=1$. Using Eqs.~(\ref{a}) and (\ref{adagger}), our Hamiltonian [Eq.~(\ref{Hinit})] becomes
\begin{eqnarray}
{\cal H} & = & \hbar\omega_{c}(a^{\dagger}a+1/2)+\frac{\Delta}{2}\sigma_{z}-\frac{i\alpha}{\sqrt{2}\ell_{c}}(a^{\dagger}\sigma_{-}-a\sigma_{+})\nonumber \\
 &+  & \frac{\beta}{\sqrt{2}\ell_{c}}(a^{\dagger}\sigma_{+}+a\sigma_{-}),\label{eq:LL-RD}
\end{eqnarray}
where the cyclotron frequency is $\omega_{c}=eB/m^{*}$,  ${\Delta=g^{*}\mu_B B}$, which inherits its sign from  $g^{*}$, and $\sigma_{\pm}=\sigma_{x}\pm i\sigma_{y}$, with  $\sigma_x$ and $\sigma_y$ denoting Pauli matrices.  We now perform the canonical transformation  $\tilde{{\cal H}} = {\cal U} {\cal H} {\cal U}^\dagger$ with ${\cal U}=e^{-i\frac{\pi}{4}\left( \frac{\sigma_z}{2}+a^\dagger a \right)}$, which yields 
\begin{eqnarray}
{\cal U} \sigma_{\pm}{\cal U}^\dagger & =&  \sigma_{\pm}e^{\mp i\frac{\pi}{4}},\\
{\cal U} \sigma_{z}{\cal U}^{\dagger} & =& \sigma_{z},\\
{\cal U} a^{\dagger}{\cal U}^{\dagger}  & =&  e^{i\frac{\pi}{4}}a^\dagger,\qquad
\end{eqnarray}
and finally
\begin{eqnarray}
\frac{{\cal \tilde{H}}}{\hbar\omega_{c}} & = & (a^{\dagger}a+1/2)+\frac{\tilde{\Delta}}{2}\sigma_{z}+\alpha_{B}(a^{\dagger}\sigma_{-}+a\sigma_{+})\nonumber \\
 &  & +\beta_{B}(a^{\dagger}\sigma_{+}+a\sigma_{-}),\label{eq:LL-RD_real}
\end{eqnarray}
where we have introduced the real valued, dimensionless quantities $\alpha_{B}=\frac{\alpha}{\sqrt{2}\hbar\omega_{c}\ell_{c}}$,
$\beta_{B}=\frac{\beta}{\sqrt{2}\hbar\omega_{c}\ell_{c}}$ and $\tilde{\Delta}=\frac{\Delta}{\hbar\omega_{c}}=\frac{g^* m^*}{2m_0}$.

Analytical solutions for the above Hamiltonian [Eq.~(\ref{eq:LL-RD_real})] can be found for the cases with either pure Rashba or pure Dresselhaus\citealp{bychkov84:6039,winkler03:1}. 
The specific cases of $\alpha=\pm \beta$ turn out to be of great physical interest, where persistent spin helix (PSH)\cite{PhysRevLett.90.146801,PhysRevLett.97.236601,psh-prx} and persistent Skyrmion lattice (PSL) \cite{PhysRevLett.117.226401} were predicted. Interestingly, the case with $\alpha=\pm \beta$ maps to the Rabi model in quantum optics and was recently solved exactly \citep{braak11:100401}. The exact solution relies on obtaining zeros of a transcendental function. Moreover, previous studies of the Rabi model have important implications for our system. 
For instance, we have shown that the Rabi parity symmetry \citep{casanova2010deep,braak11:100401}
remains valid in our problem for arbitrary $\alpha$ and $\beta$ (See Appendix \ref{sec:Orthogonal-subspaces-}). This enables us to separate the Hilbert space in two subspaces with different parities, which can be individually analyzed and compared. As for general couplings $\alpha$ and $\beta$, similar systems have been studied before
in the framework of Landau levels, using either variational (Hartree-Fock)
methods \citep{schliemann03:085302}, second order perturbation \citep{zarea05:085342,wang05:085344}
or obtaining the spectrum in terms of solutions of transcendental
equations \citep{zhang06:L477}. A perturbation scheme based on 4th order Schrieffer-Wolff transformation has also been used to find approximate analytical solutions \citep{erlingsson10:155456}. However, we are unaware of any exact analytical solution for general
Rashba, Dresselhaus and Zeeman coupling.

\section{$F$-function and its connection with the energy spectrum and DOS}

For our 2DEG in the presence of perpendicular magnetic field, the low magnetic field regime corresponds to having a very large number of Landau levels below the Fermi energy $\varepsilon_{F}$ (\carlos{taken as} constant and equal to \carlos{its} zero-field value), i.e.,\ many occupied states. The system is thus assumed to be far away from the integer quantum Hall regime where few Landau levels are occupied and the effects of electron-electron
interaction become important \cite{ihn10:book}. Let us denote the eigenenergies of our \textit{\carlos{dimensionless}} Hamiltonian Eq.~(\ref{eq:LL-RD_real}) by $\varepsilon_{n,s}$, where $n\in\mathbb{N}_0$ represents the LL number and $s = \pm$  represents a pseudo-spin associated to the presence of two spin-split bands (due to \carlos{the Zeemann and SO interactions}). With this notation, the density of states (DOS) reads
\begin{equation}
\mathcal{D}(\varepsilon,B)=\frac{\carlos{\tilde D}}{A}\sum_{n,s}\delta(\varepsilon-\hbar\omega_{c}\varepsilon_{n,s}),\label{eq:DOS}
\end{equation}
\carlos{where $\tilde D=A/2\pi\ell_{c}^{2}$ is the LL degeneracy and $A$ the 2DEG area. This LL degeneracy is the same for all 2DEGs studied here in the presence or absence of Zeemann and SO interactions.}

As we show in the next section, the magnetotransport properties of the system can be determined by the Landau
levels sequentially crossing $\varepsilon_{F}$. The rate at which these crossings
happen will determine a periodic behavior of the magnetotransport properties
of the system as the magnetic field is varied. In order to describe
this periodicity, we introduce the $F$-function \cite{ihn10:book} \carlos{(see Appendix A for details)}, which is defined by the relation 
\begin{equation}
\varepsilon_{n,s}(B)=\varepsilon\leftrightarrow n=F_{s}(\varepsilon,B).\label{eq:defFfun}
\end{equation}
The $F$-function gives the Landau level index $n$ of the state that
has energy $\varepsilon$ and pseudo-spin $s$ at magnetic field $B$.\cite{brack97:book,tarasenko02:1769}
It is important to notice that the equation for $n$ [Eq.~(\ref{eq:defFfun})]  can also return non-integer values
for $n$. In such cases the $F$-function provides a measure of how
close a Landau level $n$ is to the energy $\varepsilon$, for a given pseudo-spin $s$ and magnetic field $B$.

Since one can relate transport phenomena with the density
of states, we rewrite  the DOS of our system in a way that
highlights its oscillatory behavior dependence on both $\alpha$ and
$\beta$. First we introduce the $F_{s}$ function into Eq.\ (\ref{eq:DOS})
\begin{align}
\mathcal{D}(\varepsilon,B) 
 & \approx\frac{m^{*}}{2\pi\hbar^{2}}\sum_{n,s}\delta(n-F_{s}(\varepsilon/\hbar\omega_{c},B)),
 \label{eq12-dos-main-text}
\end{align}
which neglects terms $\mathcal{O}[(\alpha m^*\ell_c)^{2}/\hbar]+\mathcal{O}[(\beta m^*\ell_c)^2/\hbar)]$.  This holds for typical values of $\alpha$, $\beta$,  $m^*$ and small magnetic fields $B\lesssim 1$T. Using the Poisson summation formula $\sum_{n=0}^{\infty}\delta(n-F_{s})=1+2\sum_{l=1}^{\infty}\cos(2\pi l F_{s})$ and defining the relevant quantities
\begin{equation} 
\mathcal{F}_{\pm}=\frac{1}{2}(F_{+}\pm F_{-}), 
\label{fast-slow}
\end{equation}
we obtain
\begin{eqnarray}
\frac{\mathcal{D}(\varepsilon,B)-\denis{2}\mathcal{D}_{0}}{\denis{2}\mathcal{D}_{0}} & \approx & 2\sum_{l=1}^{\infty}\cos(2\pi l \mathcal{F}_{+})\cos(2\pi l \mathcal{F}_{-}),\label{eq8}
\end{eqnarray}
where $\mathcal{D}_{0}=\frac{m^{*}}{2\pi\hbar^{2}}$ is the (constant)
density of states per spin for the 2DEG at zero magnetic field \carlos{(see Appendix A for details)}. As we are going to see later, $\mathcal{F}_{+}$ contains the fast oscillations with respect to $1/B$, which is is proportional to the electron density $n_{2D}$. On the other hand, $\mathcal{F}_{-}$ contains the slow oscillations that are determined by the spin-orbit coupling terms, $\alpha$ and $\beta$. Moreover, the fast oscillations arising from the terms with $l>1$ correspond to the higher harmonics, and have be seen in experiments \cite{psh-prx}.

\section{SdH oscillations in the magnetoresisitivity}
\label{sec:SdH} 
As already mentioned, the oscillations in the magnetoresistivity as a function of the magnetic field are called SdH oscillations\citep{ihn10:book}. They appear as a consequence of the sequential depopulation of the LLs when the magnetic field is increased.  For low magnetic fields where multiple LL are occupied, i.e., far from the integer quantum Hall regime\cite{ihn10:book}, a semi-classical description of the magneto-oscillations can be used.

In experiments, the measurement of the SdH oscillations is accessed via the longitudinal differential resistivity. In general, the resistivity tensor is defined as the inverse matrix of the conductivity tensor,
\begin{equation}
    \boldsymbol{\rho}=\left(\begin{array}{cc}
\sigma_{xx} & \sigma_{xy}\\
\sigma_{xy} & \sigma_{xx}
\end{array}\right)^{-1}.
\end{equation}
\carlos{The relevant magnetoresistivity component for us is
\begin{equation}
    \rho_{xx} = \frac{\sigma_{xx}}{\sigma_{xx}^2+\sigma_{xy}^2},
\end{equation}
where}
\begin{equation}
\sigma_{xx(xy)}(B,T)=\int d\varepsilon \left(-\frac{d f_0(\varepsilon)}{d\varepsilon}\right) \sigma_{xx(xy)}(B,\varepsilon,T=0),
\end{equation}
where $f_0 (\varepsilon)$ is the Fermi-Dirac distribution. Using a semi-classical approach, we account for the 
magnetic field dependence of the conductivity via the electron scattering time $\tau(\varepsilon,B)$, 
which is proportional to the DOS ${\cal D}(\varepsilon,B)$ via Fermi's golden rule.
Accordingly, up to linear order on the deviation of the DOS, we obtain
\begin{equation}
    \tau(\varepsilon,B)\approx\tau_{0}\left(\varepsilon\right)\left[1-\frac{{\cal D}(\varepsilon,B)-{\cal D}_0(\varepsilon)}{{\cal D}_0(\varepsilon) }\right],
\end{equation}
with ${\cal D}_0(\varepsilon)={\cal D}(\varepsilon,B=0)$ and $\tau_0(\varepsilon)=\tau (\varepsilon,B=0)$. Using the Drude semi-classical equations for the frequency-independent current\cite{ihn10:book}, the normalized longitudinal resistivity reads
\begin{align}
\delta\rho_{xx}(B) & =\frac{\rho_{xx}(B)-\rho_{xx}(B=0)}{\rho_{xx}(B=0)}\\
 & =\int d\varepsilon\left(-\frac{df_{0}(\varepsilon)}{d\varepsilon}\right)\frac{{\cal D}(\varepsilon,B)-{\cal D}_0(\varepsilon)}{{\cal D}_0(\varepsilon) }.
\end{align}

For the DOS in the presence of Landau level broadening due to scattering processes, the relation in Eq.~(\ref{eq:DOS}) is replaced by
\begin{equation}
\mathcal{D}(\varepsilon,B)=\frac{\carlos{\tilde D}}{A}\sum_{n,s}L_\Gamma(\varepsilon-\hbar\omega_{c}\varepsilon_{n,s}),
\label{eq:DOSdefinition}
\end{equation}
where $L_\Gamma(x)$ describes the broadening function, e.g., Lorentzian or Gaussian, and $\Gamma$ is parameter defining the broadening of the levels \carlos{(see Appendix A for details)}.
After applying the Poisson summation formula, we obtain a result that resembles Eq.~(\ref{eq8}), apart from the appearance of the
the cosine Fourier transform of $L_\Gamma(x)$, denoted with $\tilde{L}_\Gamma(x)$,
\begin{equation}
\frac{{\cal D}(\varepsilon,B)-{\cal D}_0(\varepsilon)}{{\cal D}_0(\varepsilon) }\approx 2\sum_{l=1}^{\infty}
\tilde{L}_\Gamma \left ( l
\frac{\Gamma}{\hbar \omega_c}\right )
\cos(2\pi l\mathcal{F}_{-})\cos(2\pi l\mathcal{F}_{+}). \label{sdh}
\end{equation}
The so-called Dingle factor $\tilde{L}_\Gamma(x)$ \cite{ihn10:book} sets the limit of validity of the semi-classical approximation, i.e., that the oscillatory part of the resistivity should be much
smaller than the constant term. It also gives the regime
where it is valid to consider only the lowest harmonic.
Higher harmonics have been observed in magnetoresistivity measurements \cite{psh-prx} in GaAs-based 2DEGs. The $\mathcal{F}_{-}$ function can be related to the envelope of the SdH oscillations. 
\carlos{The general form of the} temperature-dependent normalized resistivitity reads 
\begin{align}
\delta\rho_{xx}(B,T)&=2\sum_{l=1}^{\infty} \int d\varepsilon 
{\tilde{L}_\Gamma \left ( l
\frac{\Gamma}{\hbar \omega_c}\right )} \left(-\frac{df_{0}(\varepsilon,T)}{d\varepsilon}\right)\\ \nonumber
&\times\cos(2\pi l\mathcal{F}_{-})\cos(2\pi l\mathcal{F}_{+}).\label{normdiffmagEnInt}
\end{align}
\carlos{Even though we only consider the zero-temperature limit in the present work, for completeness, below we present the temperature-dependence of $\delta\rho_{xx}(B,T)$ valid in the relevant parameter range considered in this work and for all the systems studied here. As show in Appendix~\ref{app-temp}, we find}
\begin{eqnarray}
\delta \rho_{xx}(B,T)&\approx&2\sum_{l=1}^{\infty} {\tilde{L}_\Gamma \left ( l
\frac{\Gamma}{\hbar \omega_c}\right )} 
{\cal A}_{l}(T)
\nonumber \\
& & \times \left.\cos(2\pi l\mathcal{F}_{-})\cos(2\pi l\mathcal{F}_{+})\right|_{\varepsilon=\varepsilon_F},\label{normdiffmag}
\end{eqnarray}
\carlos{where the temperature-dependent coefficient
\begin{equation}
{\cal A}_{l}(T)=\frac{2\pi^{2}lk_{B}T/\hbar\omega_{c}}{\sinh\left(2\pi^{2}lk_{B}T/\hbar\omega_{c}\right)}  
\label{T-coef}
\end{equation}
 accounts} for the temperature dependence of the SdH oscillations.  In the limit of vanishing $\alpha$ and $\beta$ this reduces to the result in Ref. \onlinecite{laikhtman94:1994332}, and in the case of {\em both} vanishing $\beta$ and broadening ($\Gamma=0$)  gives the result in Ref. \onlinecite{winkler03:1} [Eq.\ (9.28)]. Here we assume that $\varepsilon$ is close to the zero magnetic field Fermi energy $\varepsilon_F=\hbar^2 k_F^2/2m$.

A widely used method to extract spin-orbit couplings and electronic densities is to analyze
the oscillations by calculating the quantity
\begin{equation}
I(f)=  \left|\int_{B_{2}^{-1}}^{B_{1}^{-1}}d\left(\frac{1}{B}\right)
 \frac{\rho_{xx}(B)-\rho_{xx}(B_1)}{\rho_{xx}(B_1)} 
e^{i2\pi f/B }\right|^{2},
\label{eq:power}
\end{equation}
which defines the power spectrum of the normalized magneto-resistivity with a trivial background value $\rho_{xx}(B=B_1)$ removed.  Note that $B_1$ should be small enough such that the semiclassical regime of a constant $\rho_{xx}(B \rightarrow 0)$ is reached. 

\begin{figure}[tbh]
\includegraphics[clip=true,width=1\columnwidth]{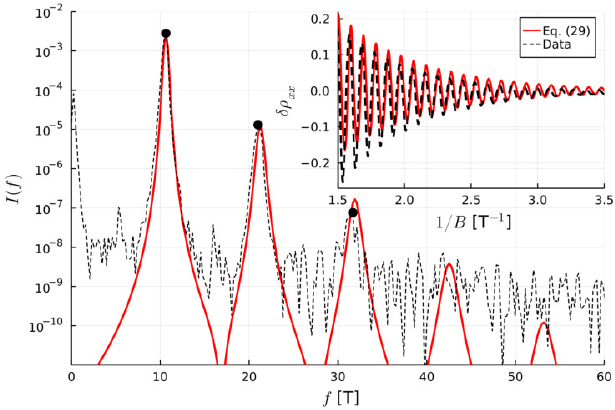} 
\caption{The power spectrum $I(f)$ for $\delta \rho_{xx}$ measurements on a GaAs 2DEG in Ref.\ \onlinecite{psh-prx} obtained using Eq.\ (\ref{eq:power}).  The calculated results used Eq.\ (\ref{sdhzeeman}) with one fitting parameter: $ \tau_q$.  The inset \carlos{shows the magneto-resistivity data and the corresponding calculated $\delta \rho_{xx}$}.}
\label{figFFT} 
\end{figure}

In Fig.\ \ref{figFFT} the power spectrum is shown for data from Fig.\ S11a in Ref.\ \onlinecite{psh-prx}, where magnetoresistivity SdH oscillations were measured in a GaAs 2DEG over a magnetic field interval \siggi{$[0.20, 1.5]$\,T}.  The power spectrum shows a SdH peak at $f\approx 10.5$\,T (the fundamental frequency), and higher harmonics are clearly visible at $21.0$\,T and $31.5$\,T, corresponding to the first and second harmonic, respectively. The experimental data was fitted with Eq.\ (\ref{sdhzeeman}) with one fit parameter: $\tau_q$. \carlos{The resulting fit matches very well the harmonics of the SdH signal. To acccount for the small background shift in the experimental data as seen in the inset a more elaborate modeling of the data would be required.}
The fitting was done using six harmonics, and resulted in  $\tau_q=0.8$\,ps, using standard GaAs parameters $m=0.067{m_0}$ and $g^*=-0.44$. \carlos{Note that we have used Eq.(\ref{sdhzeeman}), which does not include SO coupling, for our fitting procedure here. This is justifiable because GaAs-based 2DEGs have relatively small SO couplings, not accessible via SdH measurements. Weak anti-localization measurements can access the SO parameter in these systems\cite{beukmann17:241401}. However, GaAs-based 2DEGs have relatively high mobilities thus making it possible to see many harmonics.}

\section{Results and discussions}
\label{sec:Analysis}
In this section we present the energy spectrum, $F$--function and magnetoresistivity SdH oscillations for different parameter regimes of our Hamiltonian,  Eq.~(\ref{eq:LL-RD_real}). Additionally, we discuss in detail the interpretation of the SdH oscillations within the trace formula description (e.g., contribution of higher harmonics) and show how to extract relevant spin-orbit couplings from it. The results are presented in order of simplicity, i.e., from the simplest to the more complex case.

\subsection{Landau levels with \carlos{only} Zeeman interaction}
\label{zeeman}
\begin{figure}[]
\includegraphics[clip=true,width=1\columnwidth]{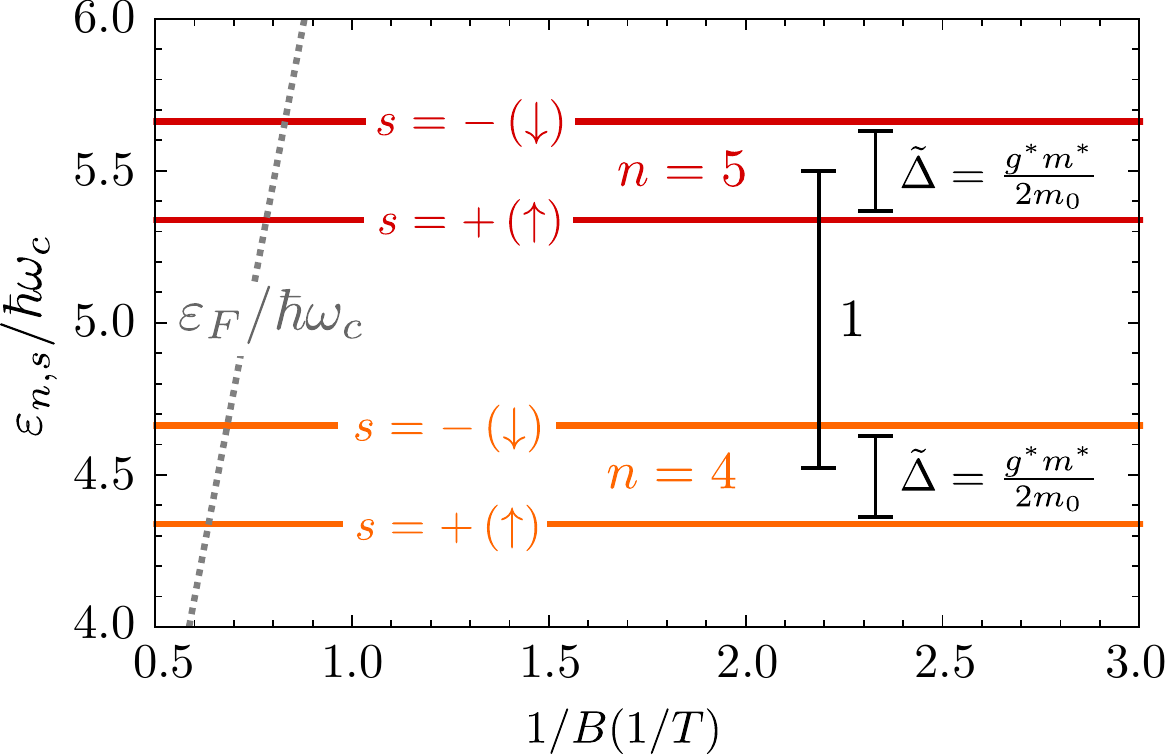} 
\caption{\carlos{Landau levels $n=4$, 5  [Eq.\ (\ref{eq:EnsZ})] as a function of $1/B$ for a 2DEG with only Zeeman interaction and no SO couplings ($\alpha=\beta=0$). The dotted line  shows $\varepsilon_F/\hbar \omega_c$}. \carlos{Here, we use} $m^*=0.019m_o$, $g^*=-34$ and $n_{2D}=3.3 \times 10^{-3}$\,nm$^{-2}$  \carlos{for InSb-based wells}\denis{\cite{gilbertsson08:165335,akabori08:205320}}. }
\label{fig:EnsZ} 
\end{figure}

In the presence of Zeeman and no Rashba and Dresselhaus SO couplings, i.e., $\alpha=\beta=0$, the eigenenergies of our Hamiltonian [Eq.~(\ref{eq:LL-RD_real})] are given by 
\begin{equation}
    \frac{\varepsilon_{n,s}}{\hbar \omega_c}=n+\frac{1}{2} + \frac{\tilde{\Delta}}{2}s. \label{eq:EnsZ}
\end{equation}
with $n \in \mathbb{N}_0$ and $s=1$ ($s=-1$) representing the pure spin state $\left| \uparrow \right \rangle$ ($\left| \downarrow \right \rangle$).  
In Fig.~\ref{fig:EnsZ} we plot the four energy levels corresponding to $n=4,5$ and $s=\pm1$, along with $\varepsilon_F/\hbar \omega_c$, using the following InSb QW parameters from Refs.\ \onlinecite{gilbertsson08:165335,akabori08:205320}:  $m^*=0.019m_o$, $g^*=-34$ and electron density $n_{2D}=3.3 \times 10^{-3}$\,nm$^{-2}$. For these parameters, the ordering of the energies obeys $\varepsilon_{n+1,-1}>\varepsilon_{n+1,1}>\varepsilon_{n,-1}>\varepsilon_{n,1}$. Figure~\ref{fig:EnsZ} shows how successive levels cross the Fermi energy as a function of the magnetic field. This, in turn, will reflect on the oscillations of the resistivity once for $\varepsilon_F\approx \varepsilon_{n,s}$, an increase on the conductivity will happen due to the resonance condition between the corresponding LL and the Fermi energy.

From the energy expressions above [Eq.\ (\ref{eq:EnsZ})], we can obtain the $F$--functions through Eq.~(\ref{eq:defFfun}), \carlos{namely, 
\begin{equation}
{F_s\left(\varepsilon\right) = \frac{\varepsilon}{\hbar \omega_c}-\frac{\tilde{\Delta}}{2}s-\frac{1}{2}}, \text{ with }
\frac{dF_s(\varepsilon)}{d\varepsilon}=\frac{1}{\hbar \omega_c}, 
\label{FF-z-der}
\end{equation}
yielding} the fast and slow components [Eq.~(\ref{fast-slow})]

\begin{equation}
    \mathcal{F}_+ \left(\varepsilon,B\right) =\frac{\varepsilon}{\hbar \omega_c} -\frac{1}{2}, \qquad \mathcal{F}_-\left(\varepsilon,B\right)  =-\frac{\tilde{\Delta}}{2}. \label{FpmZ}
\end{equation}
\carlos{At $\varepsilon=\varepsilon_F$ these can be expressed (to a very 
good approximation)} as 
$\mathcal{F}_+=\frac{h n_{2D}}{2e}\frac{1}{B}-\frac{1}{2}$ and 
$\mathcal{F}_{-}=-\frac{g}{4}\frac{m^{*}}{m_{0}}$, where we assume that 
$n_{2D}=\frac{k_{F}^{2}}{2\pi}$ \carlos{is the 2DEG electron density at $B=0$}. 

\begin{figure}[ht]
\includegraphics[clip=true,width=1\columnwidth]{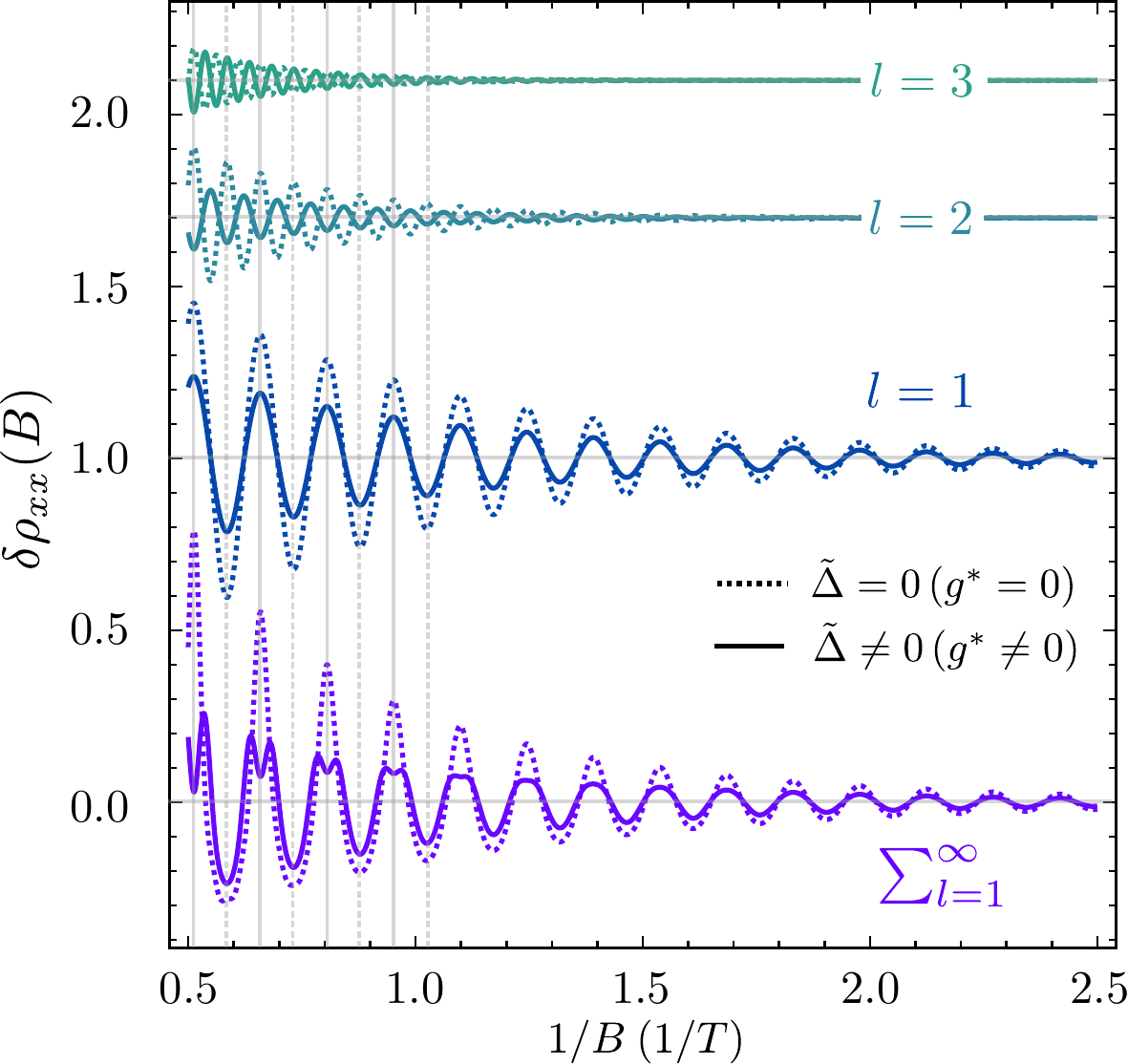} 
\caption{Magnetoresistivity deviation $\delta\rho_{xx}(B)$
as a function of $1/B$ for a 2DEG with only Zeeman coupling \carlos{and no SO couplings}. The lowest curve corresponds to $\delta\rho_{xx}(B)$ and the curves labelled by $l$ are the individual frequency components in Eq. (\ref{sdhzeeman}). The solid (dashed) line corresponds to $g^{*}=-34$ ($g^{*}=0$), ${m^*=0.019 m_o}$ and $n_{2D}=3.3 \times 10^{-3}$\,nm$^{-2}$.  \carlos{These parameters are for InSb-based wells} \denis{\cite{gilbertsson08:165335,akabori08:205320}}.}
\label{fig3} 
\end{figure}

The corresponding resistivity can now be determined through Eq.~(\ref{sdh}) and reads
\begin{widetext}
\begin{equation}
\delta\rho_{xx}(B)=2\sum_{l=1}^{\infty}e^{-l\pi\frac{\hbar/\tau_{q}}{\hbar\omega_{c}}} \frac{2\pi^{2}lk_{B}T/\hbar\omega_{c}}{\sinh\left(2\pi^{2}lk_{B}T/\hbar\omega_{c}\right)}\cos\left[2\pi l  \left(\frac{f^{\rm SdH }}{B}-\frac{1}{2}\right)\right]\cos\left(\pi l{g^*}\frac{m^*}{m_0}\right), \label{sdhzeeman}
\end{equation}
\end{widetext}
where $f^{\rm SdH}=\frac{h n_{2D}}{2e}$ and we have assumed a Lorentzian form for the $L_\Gamma$ broadening.
For small magnetic fields, both effective mass and $g$-factor nominal values do not depend on the magnetic field \cite{lei20:033213}.  As a result, the $1/B$-dependence of the resistivity in a 2DEG with only Zeeman coupling, displays oscillations with frequencies multiple of $f^{\rm{SdH}}$, and absence of beating. This can be seen from Fig.~\ref{fig3}, where we plot $\delta\rho_{xx}(B)$ vs $1/B$ for the harmonics $l=1,2,3$ and clearly see oscillations with the respective frequencies $f^{\rm{SdH}}$,$2f^{\rm{SdH}}$, and $3f^{\rm{SdH}}$. The solid (dotted) curves correspond to $g^*=-34$ and $m^*=0.019 m_o$ ($g^*=0$ and $m^*=0.019 m_o$) \cite{gilbertsson08:165335,akabori08:205320}. Note that the higher harmonics have smaller resistivity amplitudes. This occurs due to the Dingle factor $\propto e^{-l/B}$, which suppresses the higher harmonic components. 

We should stress that the effects of the Zeeman coupling within the plot of $\delta \rho_{xx}(B)$ are not immediately obvious. For instance, it can be seen that for $g^*=0$ and $g^*\neq0$, the corresponding $\delta \rho_{xx}^{l=1}(B)$ (blue curves depicting the first harmonic) only differ from themselves by the amplitude of the oscillation. For $\tilde{\Delta}=-0.323$, $\cos(2\pi\tilde{\Delta}/2)$ is smaller than one, thus yielding a reduction of the total amplitude for $g^*\neq0$ as compared to $g=0$. As a consequence, the presence of Zeeman is not readily evident from the oscillations of $\delta \rho_{xx}^{l=1}(B)$. Conversely, the Zeeman is only manifested within the resistivity when one considers many harmonics, as we discuss below. 

The definition of DOS in Eq.\ (\ref{eq:DOSdefinition}) 
gives broadened Landau levels separated by $\hbar \omega_c$ , which are in turn spin split by the Zeeman term $\tilde{\Delta}$ [See Eq.~(\ref{eq:EnsZ}) and Fig.~\ref{fig:EnsZ}]. This spin splitting can only be seen in the resistivity [Eq.\ (\ref{sdhzeeman})] when the contributions from the first {\em and} second harmonics, $\cos (2\pi f^{\rm{SdH}}/B-\pi)\cos(2 \pi \tilde{\Delta}/2)$ and $\cos (4\pi f^{\rm{SdH}}/B-2\pi)\cos(4 \pi \tilde{\Delta}/2)$, respectively, have opposite signs. For the parameters of Fig.~\ref{fig:EnsZ} $\tilde{\Delta}=-0.323$ the Zeeman term significantly affects the maximum of the resistivity. This can be seen in Fig.~\ref{fig3}, where the resistivities associated to harmonics $l=1$ and $l=2$ (blue and cyan solid curves, respectively), interfere in a destructive way, producing the double-peak feature in the total resistivity (purple solid lines), \carlos{characteristic of the incipient spin splitting in such data}. We emphasize, however, that this feature can be absent depending on the broadening of the energy levels (due to the overlap of the spin-split levels). This is the reason why the double-peak feature is not seen on the other maximum peaks.

Although the $g^*$-factor term does not depend explicitly on magnetic field, it can manifest itself in the magneto-oscillations. 
\carlos{More specifically, Zeeman-only  effects can have a pronounced effect on the magneto-oscillations, controlling the amplitude and sign of how subsequent harmonics are added, either constructively or destructively, before being damped by the quantum life time.}
Furthermore, it is important to say that the Zeeman can give rise to interesting features and affect drastically the understanding of the magneto-oscillations. For instance, if one could engineer a material \footnote{InAs has the value of $\tilde{\Delta} \approx 0.24$, but one would have a symmetric structure to minimize the spin-orbit contribution} such that $\tilde{\Delta}=\frac{g^*}{2}\frac{m^*}{m_o}=0.5 + m$ with $m\in \mathbb{Z}$, then the main weight of the resistivity would be due to the second harmonic with SdH frequently $2 f^{\rm SdH} $ as $\cos(l\pi \tilde{\Delta})=0$ for $l=1$.

\subsection{Landau Levels with Zeeman and Rashba interactions}
\label{zeemanR}
We now analyze the case where we have the presence of both Zeeman and  Rashba terms, i.e., $\tilde{\Delta}\neq0$, $\alpha\neq0$ and no Dresselhaus coupling $\beta=0$ in Eq.~(\ref{eq:LL-RD_real}). In the spin basis $\left\{\left| \uparrow \right\rangle, \thinspace \left| \downarrow \right\rangle \right\}$, the corresponding Hamiltonian assumes the following matrix form
\begin{equation}
\frac{{\cal \tilde{H}}}{\hbar\omega_{c}}=\left(\begin{array}{cc}
a^{\dagger}a+\frac{1}{2}+\frac{\tilde{\Delta}}{2} & 2 \alpha_{B}a\\
2 \alpha_{B}a^{\dagger} & a^{\dagger}a+\frac{1}{2}-\frac{\tilde{\Delta}}{2}
\end{array}\right). \label{Hrashba}
\end{equation}
Interestingly, the operator ${\cal N}_+ = a^\dagger a + \sigma_{z}/2$ commutes with the Hamiltonian above, i.e., ${[{\cal \tilde{H}},{\cal N}_+]=0}$, and hence  ${\cal \tilde{H}}$ and ${\cal N}_+$ share the same eigenstates. Hence we have ${{\cal N}_+\left|n,\uparrow\right\rangle =\left(n+1/2\right)\left|n,\uparrow\right\rangle}$ and 
${{\cal N}_+\left|n+1,\downarrow\right\rangle =\left(n+1/2\right)\left|n+1,\downarrow\right\rangle }$ , i.e., for $n\in\mathbb{N}$, $\left|n,\uparrow\right\rangle$ and $\left|n+1,\downarrow \right\rangle$ are degenerate with respect to the operator ${\cal N}_+$, except for $\left|0,\downarrow\right\rangle$ with corresponding energy $\frac{\varepsilon_{0,\downarrow}}{\hbar\omega_c}=\frac{1}{2} (1- \tilde{\Delta})$. As a consequence, a linear combination of $\left|n,\uparrow\right\rangle$ and $\left|n+1,\downarrow\right\rangle$ is also an eigenstate of our Hamiltonian Eq.~(\ref{Hrashba}). This motivates us to rewrite the total Hamiltonian as a direct sum of $2\times2$ block Hamiltonians in the basis $\left\{ \left|n,\uparrow\right\rangle,\thinspace \left|n+1,\downarrow\right\rangle \right\}$(${\cal \tilde{H}}_{\left|n,\uparrow\right\rangle ;\left|n+1,\downarrow\right\rangle }$), in addition to the non-degenerate decoupled Hamiltonian (${\cal \tilde{H}}_{\left|0,\downarrow\right\rangle }$), namely
\begin{equation}
   {\cal \tilde{H}}= {\cal \tilde{H}}_{\left|0,\downarrow\right\rangle}\oplus \bigoplus_{n=0}^{\infty}{\cal \tilde{H}}_{\left|n,\uparrow\right\rangle ;\left|n+1,\downarrow\right\rangle }, \label{Hrt}
\end{equation}
with ${\cal \tilde{H}}_{\left|0,\downarrow\right\rangle}=\varepsilon_{0,\downarrow}$ and
\begin{equation}
{\cal \tilde{H}}_{\left|n,\uparrow\right\rangle ;\left|n+1,\downarrow\right\rangle }=\hbar \omega_c\left(\begin{array}{cc}
n+\frac{1}{2}+\frac{\tilde{\Delta}}{2} & 2 \alpha_{B}\sqrt{n+1}\\
2 \alpha_{B}\sqrt{n+1} & n+1+\frac{1}{2}-\frac{\tilde{\Delta}}{2}
\end{array}\right). \label{pureR}
\end{equation}
The diagonalization of the Hamiltonian Eq.~(\ref{pureR}) yields energies
\begin{align}
    \frac{\varepsilon_{n,s}}{\hbar\omega_{c}}=&\left(n+\frac{1}{2}+\frac{s}{2}\right)\label{eq:EnsR}\\
    &-\frac{s}{2}\frac{1-\tilde{\Delta}}{|1-\tilde{\Delta}|}\sqrt{\left(1-\tilde{\Delta}\right)^{2}+16 \alpha_{B}^{2}\left(n+\frac{1}{2}+\frac{s}{2}\right)},\nonumber 
\end{align}
with $s=\pm$ and $n\in\mathbb{N}_0$, which already incorporates the energy of the decoupled state $\left|0,\downarrow \right \rangle$, $\varepsilon_{0,-}\equiv\varepsilon_{0,\downarrow}$ $(\varepsilon_{0,+}\equiv\varepsilon_{0,\downarrow})$ {if ${1-\tilde{\Delta}>0}$} ${(1-\tilde{\Delta}<0)}$.
\begin{figure}[ht]
\centering \includegraphics[clip=true,width=1\columnwidth]{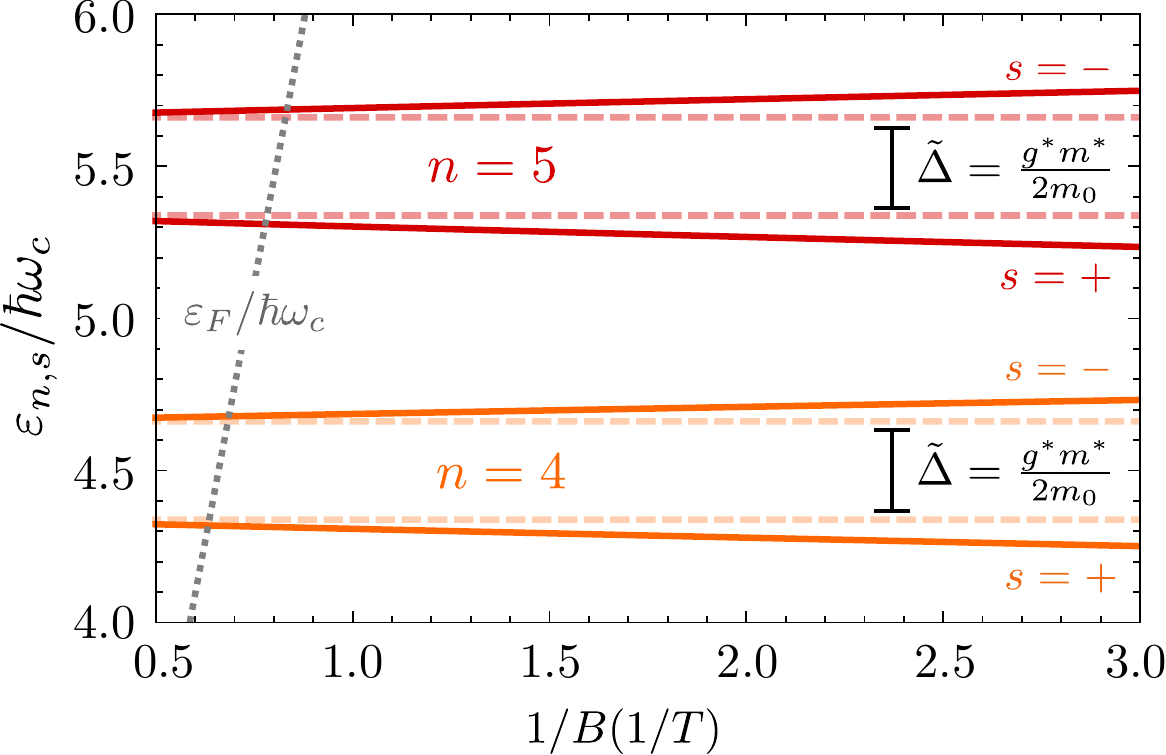}
\caption{\carlos{Landau levels $n=4$, 5  [Eq.~(\ref{eq:EnsR})] as a function of $1/B$ for a 2DEG with non-zero Zeeman and Rashba interactions but no Dresselhaus coupling ($\beta=0$).
The dotted line denotes $\varepsilon_F/\hbar \omega_c$}. The parameters here are $\alpha=10$\,meV\,nm, $m^*=0.019m_o$, $g^*=-34$ and $n_{2D}=3.3 \times 10^{-3}$\,nm$^{-2}$ \carlos{for InSb-based wells}\denis{\cite{gilbertsson08:165335,akabori08:205320}}. The dashed lines show the corresponding levels for $\alpha=0$. }
\label{fig:EnsR} 
\end{figure}
These LLs are plotted in Fig.~\ref{fig:EnsR} as a function of $1/B$ for parameters $\alpha=10$\,meV\,nm, $m^*=0.019m_o$ and $g^*=-34$ \denis{\cite{gilbertsson08:165335,akabori08:205320}}. 
Due to the spin-orbit coupling, the energy levels $\frac{\varepsilon_{l,s}}{\hbar\omega_{c}}$ are no longer equidistant, and their separation changes as function of $1/B$. On this scale, the energy dispersion appears linear in $1/B$.  In fact, for $\tilde{\Delta}<0$ ($\tilde{\Delta}>0$) the spin-splitting is enhanced (suppresses) relative to the case with $\alpha=0$ (See Fig.\ \ref{fig:EnsZ}).  This can be seen through the expansion of the term $(1-\tilde{\Delta})^2$ within the square root of Eq. (\ref{eq:EnsR}), yielding $-2\tilde{\Delta}$, which enhances the Zeeman splitting in the presence of Rashba SO coupling.\cite{akabori08:205320}
\begin{widetext}
 \begin{center}
\begin{table}
\begin{centering}
\caption{\carlos{Definitions of the Zeeman and SO-related quantities used is this work. }}
\begin{tabular}{ccccc}
 &  &  &  & \tabularnewline
\hline 
\hline 
Zeeman $\left(g^{*}\right)$ & $\omega_{c}=\frac{eB}{m^{*}}$ & $\ell_{c}=\sqrt{\frac{\hbar}{eB}}$ & $\Delta={g^{*}\mu_B B}$ & $\tilde{\Delta}=\frac{\Delta}{\hbar\omega_{c}}=\frac{g^* m^*}{2m_0}$\tabularnewline
Rashba $\left(\alpha\right)$ & $\varepsilon_{R}=\frac{\alpha^{2}m^{*}}{2\hbar^{2}}=\frac{\hbar^{2}k_{R}^{2}}{2m^{*}}$ & $\alpha_{B}=\frac{\alpha}{\sqrt{2}\hbar\omega_{c}\ell_{c}}$ & $\frac{\varepsilon_{R}}{\hbar\omega_{c}}=\alpha_{B}^{2}$ & \tabularnewline
Dresselhaus $\left(\beta\right)$ & $\varepsilon_{D}=\frac{\beta^{2}m^{*}}{2\hbar^{2}}=\frac{\hbar^{2}k_{D}^{2}}{2m^{*}}$ & $\beta_{B}=\frac{\beta}{\sqrt{2}\hbar\omega_{c}\ell_{c}}$ & $\frac{\varepsilon_{D}}{\hbar\omega_{c}}=\beta_{B}^{2}$ & \tabularnewline
\carlos{SO parameters} & $\gamma=\alpha_{B}+\beta_{B}$ & $\delta=\alpha_{B}-\beta_{B}$ & $\Omega=\frac{2\varepsilon_{R}/\hbar\omega_{c}}{1-\tilde{\Delta}}+\frac{2\varepsilon_{D}/\hbar\omega_{c}}{1+\tilde{\Delta}}$ & $\Lambda=\frac{2\varepsilon_{R}/\hbar\omega_{c}}{1-\tilde{\Delta}}-\frac{2\varepsilon_{D}/\hbar\omega_{c}}{1+\tilde{\Delta}}$\tabularnewline
\hline 
\hline 
 &  &  &  & \tabularnewline
\end{tabular}
\par\end{centering}
\end{table}
\par\end{center}
\end{widetext}
Accordingly, for this case we obtain 
\begin{align}
{\cal F}_{+} (\varepsilon,B) & =\frac{\varepsilon}{\hbar\omega_{c}}-\frac{1}{2}+2 \alpha_{B}^{2},\label{frp}\\
{\cal F}_{-} (\varepsilon,B) & =-\frac{1}{2}+\frac{1}{2}\frac{1-\tilde{\Delta}}{|1-\tilde{\Delta}|}\sqrt{(1 -\tilde{\Delta})^{2}+16 \alpha_{B}^{2}\left(\alpha_{B}^{2}+\frac{\varepsilon}{\hbar\omega_{c}}\right)}.\label{frm}
\end{align}
Differently from the results in the previous section, here both $\mathcal{F}_{\pm}$ functions depend on the magnetic field. As a consequence, we will have more complex oscillations in $\rho_{xx}(B)$ as compared to the case without Rashba coupling [Fig.~\ref{fig3}].

\begin{figure}[ht]
\centering \includegraphics[clip=true,width=1\columnwidth]{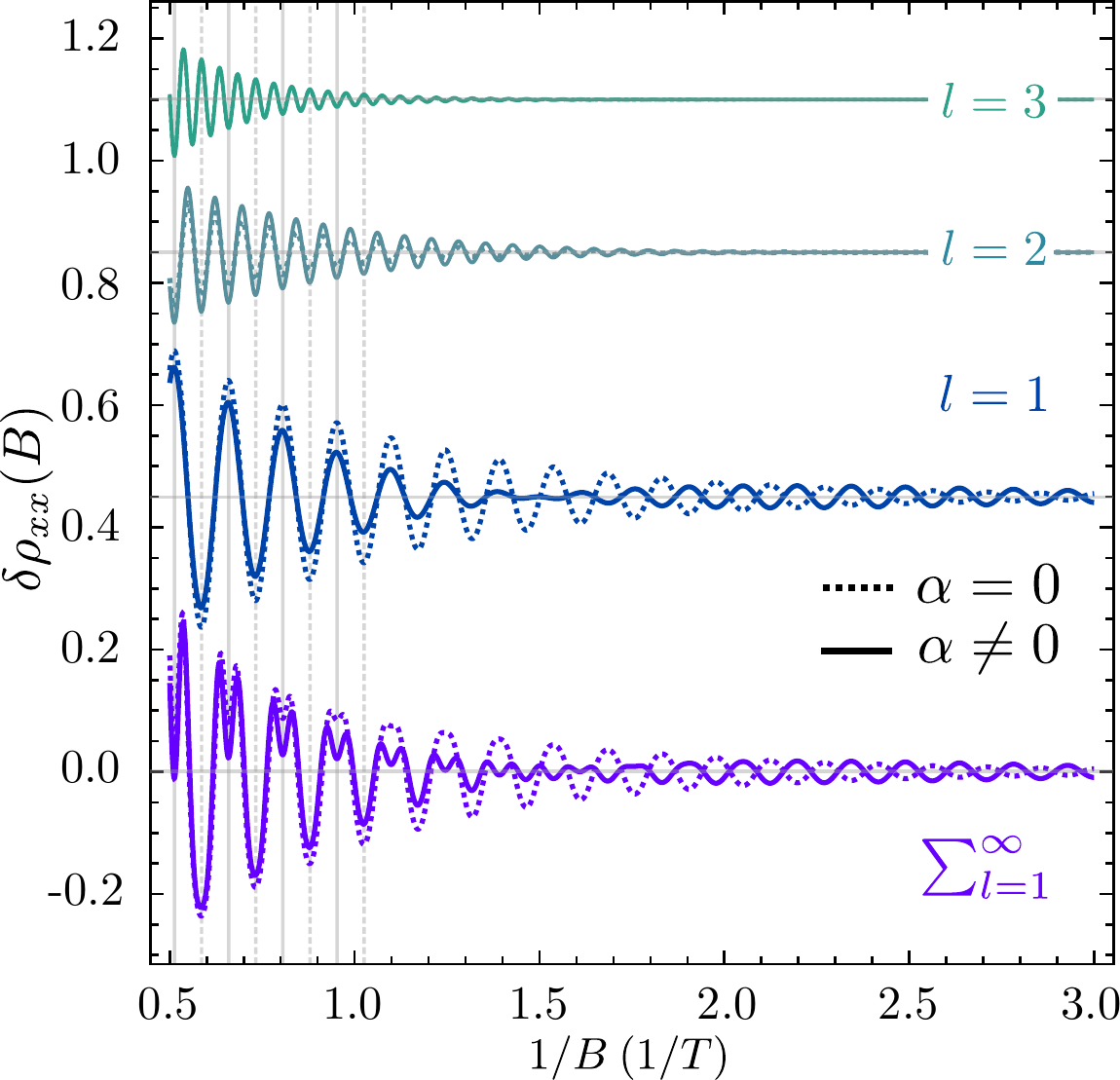}
\caption{Magnetoresistivity deviation $\delta\rho_{xx}(B)$
as a function of $1/B$ for a 2DEG with Zeeman  and Rashba \carlos{interactions and no Dresselhaus coupling ($\beta=0$)}. The lowest curve corresponds to $\delta\rho_{xx}(B)$ and the curves labelled by $l$ are the individual frequency components \carlos{(i.e., harmonics)} in Eq. (\ref{sdhzeeman}). \carlos{The solid lines are obtained with  $g^{*}=-34$, $\alpha=10$\,meV\,nm, ${m^*=0.019 m_o}$ and $n_{2D}=3.3 \times 10^{-3}$\,nm$^{-2}$ \cite{gilbertsson08:165335,akabori08:205320}; the dotted lines show the corresponding $\alpha=0$ case.}}
\label{fig:drxx_R} 
\end{figure}

In Fig.~\ref{fig:drxx_R}, we plot the total differential magneto-resistivity $\delta \rho_{xx}(B)$, and the independent contributions from harmonics $l=1,2$ and $l=3$. Here we use $\alpha=10$\,meV\,nm, $m^*=0.019m_o$, $g^*=-34$ and $n_{2D}=3.3 \times 10^{-3}$\,nm$^{-2}$ \cite{gilbertsson08:165335,akabori08:205320}. Similarly to the case with $\alpha=0$ (\carlos{dashed line in Fig.~\ref{fig:drxx_R}}), here we also see oscillations for the $l=1,2,3$ harmonics with respective frequencies $f^{\rm{SdH}},2f^{\rm{SdH}}$ and $3f^{\rm{SdH}}$. However, for $l=1$ we observe beating, which can be expected as both ${\cal F}_{-}(\varepsilon,B)$ and ${\cal F}_{+}(\varepsilon,B)$ frequencies now depend on $1/B$. More specifically, this beating appears here because in the magnetic range considered we have $2\pi l \mathcal{F}_-(B)= \frac{\pi}{2}$, which leads to a node in $\delta \rho_{xx}$ as $\delta \rho_{xx}\propto \cos\left[2\pi l \mathcal{F}_-(B)\right]$.  Note that this only occurs for $l=1$, since for higher harmonics this condition is not satisfied. Due to the larger amplitude of the harmonic $l=1$, this beating is also seen in the total magneto-resistivity. 

\subsection{Landau Levels with Zeeman and Dresselhaus interaction}
\label{zeemanD}
In the case of Zeeman with pure Dresselhaus, i.e., $\tilde{\Delta}\neq0$, $\alpha=0$ and $\beta\neq0$, the Hamiltonian Eq.~(\ref{eq:LL-RD_real}) in the spin basis is given by 
\begin{equation}
\frac{{\cal \tilde{H}}}{\hbar\omega_{c}}=\left(\begin{array}{cc}
a^{\dagger}a+\frac{1}{2}-\frac{\tilde{\Delta}}{2} & 2\beta_{B}a^{\dagger}\\
2\beta_{B}a & a^{\dagger}a+\frac{1}{2}+\frac{\tilde{\Delta}}{2}
\end{array}\right).
\end{equation}
Differently from the case of pure Rashba, here the operator ${\cal N}_- = a^\dagger a - \sigma_{z}/2$ commutes with the Hamiltonian above. For this case we have ${{\cal N}_-\left|n,\downarrow\right\rangle =\left(n+1/2\right)\left|n,\downarrow\right\rangle}$ and 
${{\cal N}_-\left|n+1,\uparrow\right\rangle =\left(n+1/2\right)\left|n+1,\uparrow\right\rangle }$, i.e., for $n\in\mathbb{N}$, $\left|n,\downarrow\right\rangle$ and $\left|n+1,\uparrow \right\rangle$ are degenerate with respect to the operator ${\cal N}_-$, except for the state $\left|0,\uparrow\right\rangle$ with corresponding energy $\frac{\varepsilon_{0,\uparrow}}{\hbar\omega_c}=\frac{1}{2} (1+ \tilde{\Delta})$. As a consequence, a linear combination of $\left|n,\downarrow\right\rangle$ and $\left|n+1,\uparrow\right\rangle$ is also an eigenstate of our Hamiltonian. Therefore, differently from the previous case here the Hamiltonian reads,   
\begin{equation}
   {\cal \tilde{H}}={\cal \tilde{H}}_{\left|0,\uparrow\right\rangle} \oplus \bigoplus_{n=0}^{\infty}{\cal \tilde{H}}_{\left|n,\downarrow\right\rangle ;\left|n+1,\uparrow\right\rangle }, \label{Hdt}
\end{equation}
with ${\cal \tilde{H}}_{\left|0,\uparrow\right\rangle}=\varepsilon_{0,\uparrow}$ and
\begin{figure}[ht]
\centering \includegraphics[width=0.5\textwidth]{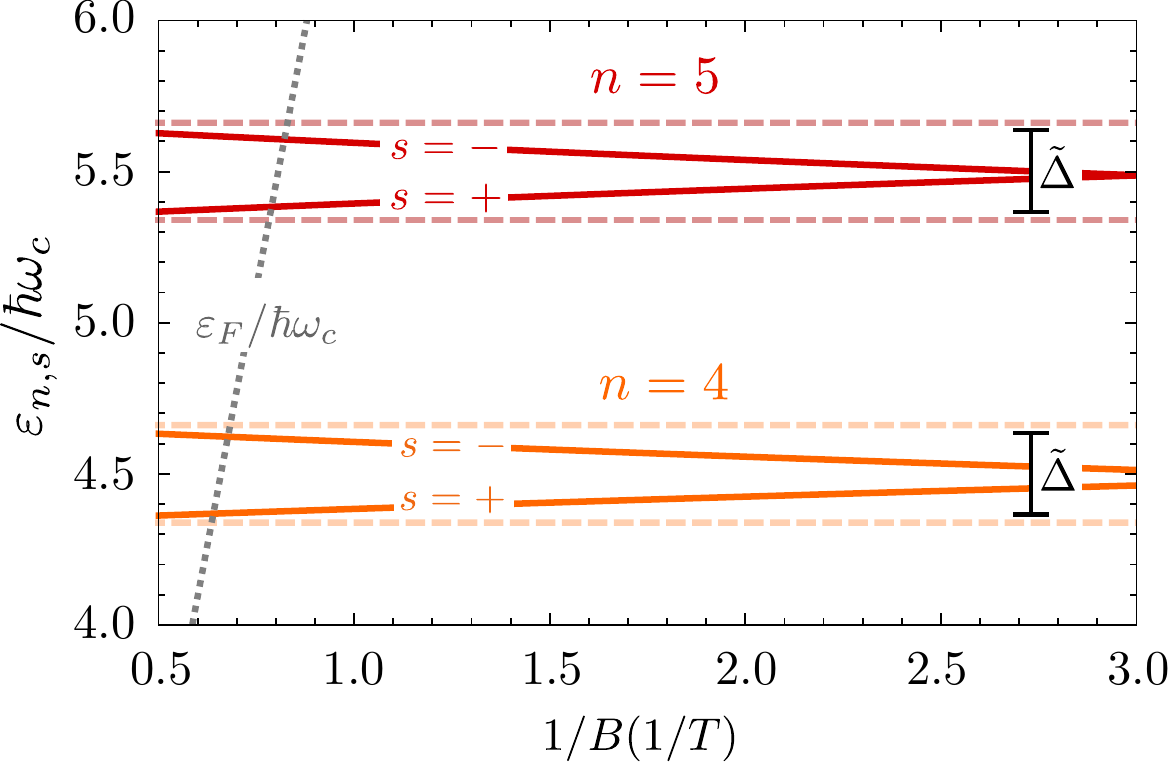}
\caption{\carlos{Landau levels $n=4$, 5 [Eq.~(\ref{eq:EnsD})] as a function of $1/B$ for a 2DEG with Zeeman and Dresselhaus interations but no Rashba coupling ($\alpha=0$). The dotted line denotes $\varepsilon_F/\hbar \omega_c$. }The parameters \carlos{here} are $\beta=10$\,meV\,nm, $m^*=0.019m_o$, $g^*=-34$ and $n_{2D}=3.3 \times 10^{-3}$\,nm$^{-2}$  \cite{gilbertsson08:165335,akabori08:205320} dashed lines show the corresponding levels for $\beta=0$. }
\label{fig:EnsD} 
\end{figure}
\begin{equation}
{\cal \tilde{H}}_{\left|n,\downarrow\right\rangle ;\left|n+1,\uparrow\right\rangle }=\hbar \omega_c\left(\begin{array}{cc}
n+\frac{1}{2}-\frac{\tilde{\Delta}}{2} & 2 \beta_{B}\sqrt{n+1}\\
2 \beta_{B}\sqrt{n+1} & n+1+\frac{1}{2}+\frac{\tilde{\Delta}}{2}
\end{array}\right).\label{pureD}
\end{equation}
The diagonalization of the Hamiltonian Eq.\ (\ref{pureD}) yields
energies
\begin{align}
    \frac{\varepsilon_{n,s}}{\hbar\omega_{c}}=&\left(n+\frac{1}{2}-\frac{s}{2}\right)  \label{eq:EnsD}\\
    &+\frac{s}{2}\frac{1+\tilde{\Delta}}{|1+\tilde{\Delta}|}\sqrt{\left(1+\tilde{\Delta}\right)^{2}+16  \beta_{B}^{2}\left(n+\frac{1}{2}-\frac{s}{2}\right)}, \nonumber 
\end{align}
with $s=\pm$ and $n\in\mathbb{N}_0$, which already incorporates the energy of the decoupled state $\left|0,\uparrow \right \rangle$, $\varepsilon_{0,+}\equiv\varepsilon_{0,\uparrow}$ ($\varepsilon_{0,-}\equiv\varepsilon_{0,\uparrow}$) if $1+\tilde{\Delta}>0$ ($1+\tilde{\Delta}<0$). Here, it is important to notice the opposite sign of $s$ with respect to Eq.~(\ref{pureR}). This happens because the pure Dresselhaus Hamiltonian Eq.~(\ref{pureD}) has opposite basis ordering of the spin states as compared to the pure Rashba Hamiltonian Eq.~(\ref{pureR}). Accordingly, the ${\cal F}_{\pm}$ functions change slightly and read
\begin{align}
{\cal F}_{+} (\varepsilon,B) & =\frac{\varepsilon}{\hbar\omega_{c}}-\frac{1}{2}+2 \beta_{B}^{2},\label{frpD}\\
{\cal F}_{-} (\varepsilon,B) & =\frac{1}{2}-\frac{1}{2}\frac{1+\tilde{\Delta}}{|1+\tilde{\Delta}|}\sqrt{\left(1 +\tilde{\Delta}\right)^{2}+16 \beta_{B}^{2}\left(\beta_{B}^{2}+\frac{\varepsilon}{\hbar\omega_{c}}\right)}\label{frmD}.
\end{align}
Due to the: i) similarity of the Dresselhaus expression Eqs.~(\ref{eq:EnsD}), (\ref{frpD}) and (\ref{frmD}) to the ones arising from the pure Rashba case, Eqs.~(\ref{eq:EnsR}), (\ref{frp}) and (\ref{frm}); ii)  cosine dependence of the ${\cal F}_{\pm}$ functions within the resistivity Eq.~(\ref{sdh}); all the results and equations in the last section also holds here by making $\alpha_B \rightarrow \beta_B$, $\tilde{\Delta} \rightarrow -\tilde{\Delta}$ and $s\rightarrow -s$.  This can also be seen on the level of  the Hamiltonian in Eq.~(\ref{Hinit}) where applying the  unitary transformation $W=e^{i\frac{\pi}{2}\sigma_x}e^{i\frac{\pi}{4}\sigma_z}$ results in
\begin{eqnarray}
W \frac{{\cal \tilde{H}}}{\hbar\omega_{c}} W^\dagger& = & (a^{\dagger}a+1/2)+\frac{(-\tilde{\Delta})}{2}\sigma_{z}+\beta_{B}(a^{\dagger}\sigma_{-}+a\sigma_{+}),\nonumber \\
 &  & +\alpha_{B}(a^{\dagger}\sigma_{+}+a\sigma_{-}),\label{eq:LL-RD_rotated}
\end{eqnarray}
which is the expected result. This mapping from $(\alpha,\tilde{\Delta})$ to $(\beta,-\tilde{\Delta})$ has visible consequences on the energy levels. In Fig.~\ref{fig:EnsD} we plot the corresponding LLs [Eq.~(\ref{eq:EnsD})] as a function of $1/B$ for parameters $\beta=10$\,meV\,nm, $m^*=0.019m_o$ and $g^*=-34$  \cite{gilbertsson08:165335,akabori08:205320}. Due to the spin-orbit coupling, the energy levels $\frac{\varepsilon_{l,s}}{\hbar\omega_{c}}$ are no longer equidistant, and their separation changes as function of $1/B$. However, differently from the pure Rashba case, now the Dresselhaus competes with the Zeeman coupling, even leading to LL-dependent  crossings. This can be seen through the expansion of $(1+\tilde{\Delta})^2$ within the square root [Eq.\ (\ref{eq:EnsD})], which will give rise to $2\tilde{\Delta}<0$, thus suppressing the spin splitting in the presence of Dresselhaus SO coupling.

\begin{figure}[hb]
\centering \includegraphics[clip=true,width=1\columnwidth]{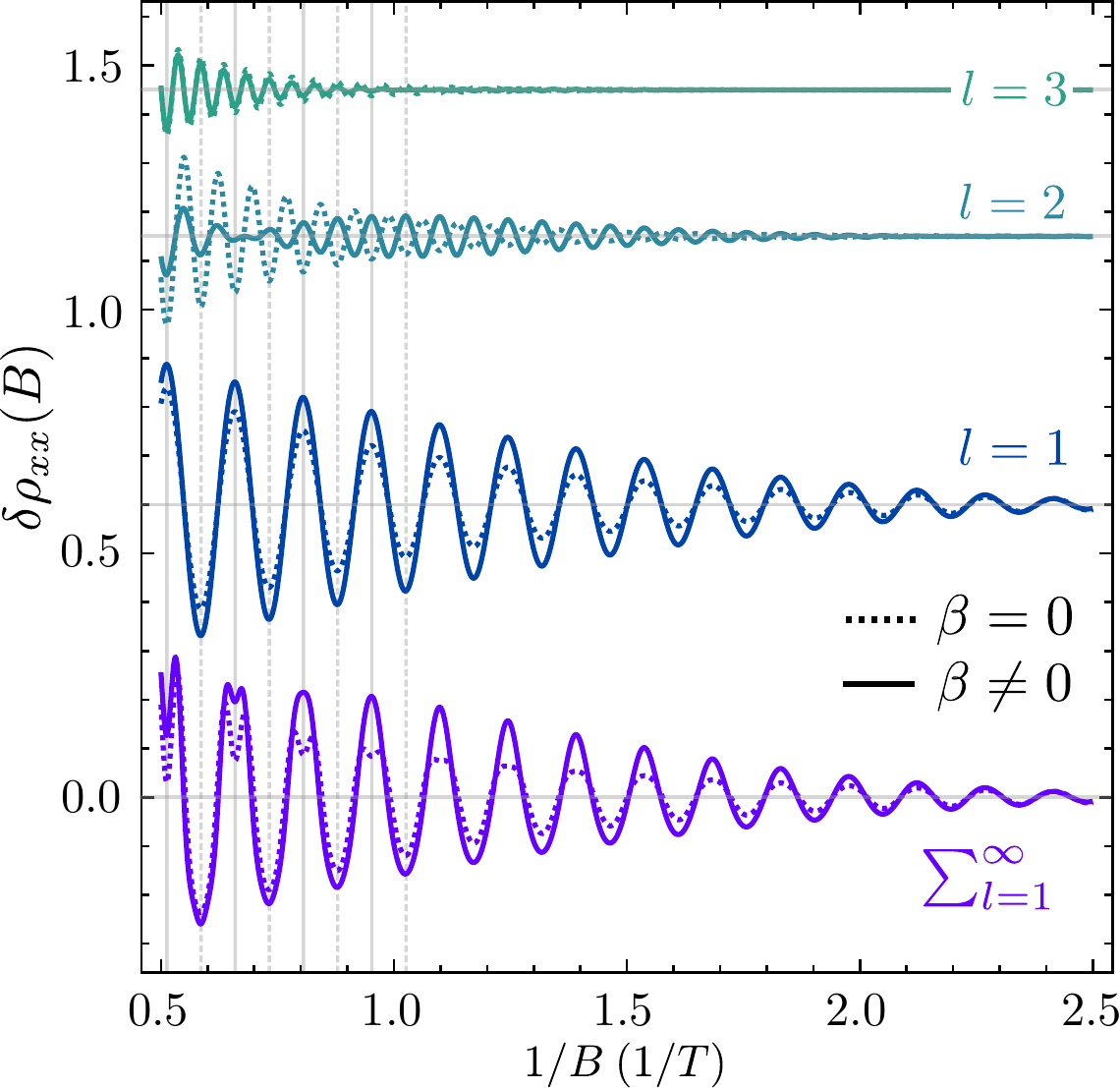}
\caption{Magnetoresistivity deviation $\delta\rho_{xx}(B)$ as a function of $1/B$ for a 2DEG with Zeeman and Dresselhaus \carlos{interactions and no Rashba coupling ($\alpha=0$). The lowest curve corresponds to $\delta\rho_{xx}(B)$ and the curves labelled by $l$ are the individual frequency components in Eq.~(\ref{sdhzeeman}). The solid lines are calculated for $g^{*}=-34$, $\beta=10$\,meV\,nm, ${m^*=0.019 m_o}$ and $n_{2D}=3.3 \times 10^{-3}$\,nm$^{-2}$  \cite{gilbertsson08:165335,akabori08:205320}; the dotted line shows the corresponding $\beta=0$ case.}}
\label{fig:drxx_D}
\end{figure}

In Fig.~\ref{fig:drxx_D} we plot the total differential magneto-resistivity $\delta \rho_{xx}(B)$, and the individual contributions from the harmonics ${l=1,2}$ and $l=3$. We use $\beta=10$\,meV\,nm, $m^*=0.019m_o$, $g^*=-34$ and $n_{2D}=3.3 \times 10^{-3}$\,nm$^{-2}$  \cite{gilbertsson08:165335,akabori08:205320}. First, similarly to the previous cases, here we can also clearly see oscillations with  frequencies $f^{\rm{SdH}},2f^{\rm{SdH}}$, $3f^{\rm{SdH}}$. Differently from the previous case with $\alpha=10$\,meV\,nm and $\beta=0$, now we see no beating for the $l=1$ harmonic but find beating for $l=2$. This happens as $2\pi  l \mathcal{F}_-(B)=\frac{\pi}{2}$ -- the condition to observe beating --  is only satisfied for $l=2$. Even though the beating appears within the second harmonic, it is not manifested in the total differential magneto-resistivity $\delta\rho_{xx}(B)$ for our choice of parameters. This is due to the smaller oscillation amplitude of $l=2$ with respect to $l=1$. 

\subsection{Beatings in the SdH oscillations with nonzero Zeeman and in the presence of either Rashba or Dresselhaus: \carlos{a unified description}}
\label{sec:SdH_beatingsPure}
In this section we will discuss more thoroughly the conditions for the appearance of beatings. The two functions $\mathcal{F}_+$ and $\mathcal{F}_-$, Eq.~(\ref{fast-slow}), determine the fast and slow component, respectively, of the SdH oscillations.  To highlight this point and its connection to the power spectrum in Eq.\ (\ref{eq:power}), we start by rewriting Eqs.\ (\ref{frp})- (\ref{frm}), and  Eqs.\ (\ref{frmD})- (\ref{frmD}) in a unified way
\begin{align}
{\cal F}_{+} (\varepsilon,B) & =f_{{R(D)}}^\mathrm{SdH}\frac{1}{B}
-\frac{1}{2},\label{eq:frpSdH}\\
{\cal F}_{-} (\varepsilon,B) & ={\mp}\frac{1}{2}{\pm}{ \frac{1}{2}\frac{1\mp\tilde{\Delta}}{|1\mp\tilde{\Delta}|}}\sqrt{(1 {\mp}\tilde{\Delta})^{2}+4 \left ( f_{ R(D)} \frac{1}{B} \right )^2
},\label{frm2}
\end{align}
where we have introduced the magneto-oscillation frequencies 
\begin{align}
f_{{R(D)}}^\mathrm{SdH}&=\frac{h}{2e}
\left (
n_{2D}+\frac{k_{R(D)}^2}{\pi}
\right ),  \label{eq:fSdHR} \\
f_{R(D)} &=\frac{h}{2e} \sqrt{\frac{2k_{R(D)}^2}{\pi}}\sqrt {n_{2D}+\frac{k_{R(D)}^2}{2\pi}
}, \label{eq:fR}
\end{align}
where the $R$ $(D)$ index refers to either pure Rasbha (Dresselhaus) case, with
$k_R=\frac{m\alpha}{\hbar^2}$  ($k_D=\frac{m\beta}{\hbar^2}$). Here, the upper (lower) sign refers to the Rashba (Dresselhaus) case. 
\begin{figure}[hb]
\begin{center}
\includegraphics[clip=true,width=1\columnwidth]{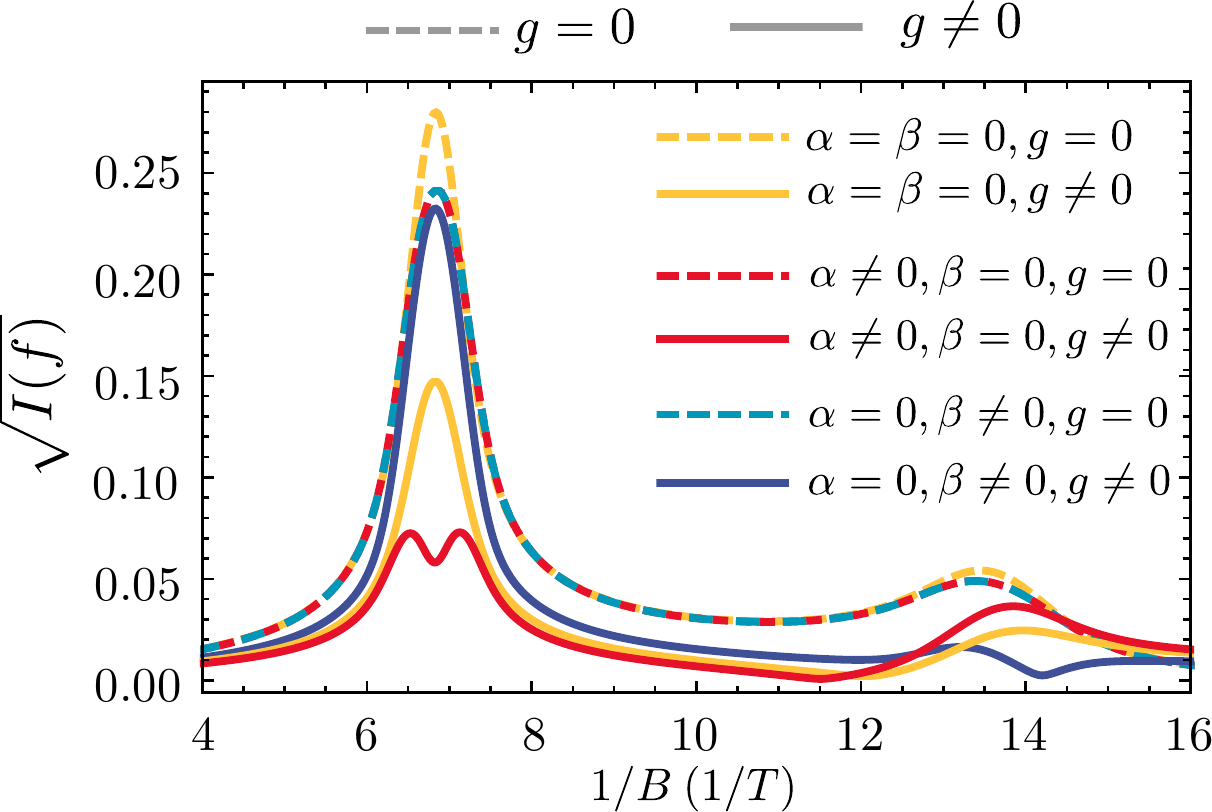}
\caption{Frequency response $\sqrt{I(f)}$ for $\alpha=10.0$\,meV\,nm and $\beta=0.0$ [red curve], and $\alpha=0.0$ and $\beta=10.0$\,meV\,nm [blue curve]. Other parameters are $m^*=0.019 m_o$, $g^*=-34$ and $n_{2D}=3.3 \times 10^{-3}$\,nm$^{-2}$ \cite{gilbertsson08:165335,akabori08:205320}. The solid black shows corresponds to no spin-orbit coupling ($\alpha=\beta=0$) and black dashed corresponds to $\alpha=\beta=g=0$.}
\label{fig:If_pure} 
\end{center}
\end{figure}
In the case where $n_{2D} \gg {k_{R(D)}^2}/{2\pi}$, and ${f_{R(D)}}/{B} \gg 1$, the beating frequency takes the standard form $f_{R(D)} =\frac{h}{2e} \sqrt{{2k_{R(D)}^2 n_{2D}}/{\pi}}$, in which case $\tilde{\Delta}$ becomes irrelevant for the magnitude of the beating frequency\cite{engels97:1958R}.

{The frequency $f_{R(D)}^{\rm{SdH}}$  [Eq.~(\ref{eq:fSdHR})] is the main SdH frequency of the magneto-resistance oscillations, usually extracted from experiments to infer the 2D electronic density $n_{2D}$. On the other hand, the frequency $f_{R(D)}$ [Eq.~(\ref{eq:fR})] is the one allowing for possible beatings in the magneto-oscillation. As previously discussed in the last two sections, the presence of beating happens when $2\pi l \mathcal{F}_-(B)=\frac{\pi}{2}$ is satisfied, which} depends on the value of both $f_{R(D)}$ and $\tilde{\Delta}$. 

The presence or absence of beatings can also be visualized through  the power spectrum defined by Eq.~(\ref{eq:power}).  From interference of waves, we know that the presence of beatings correspond to sum of cosines waves with slightly different frequencies. Accordingly, the power spectrum for this case would show two peaks located at slightly different frequencies. In Fig. \ref{fig:If_pure}   we plot $\sqrt{I(f)}$  for $m^*=0.019 m_o$ and $n_{2D}=3.3 \times 10^{-3}$\,nm$^{-2}$, using different spin-orbit parameters and $g$-factor values.  
For all different sets of parameters, we always have the presence of two main peaks located at both $1/B\approx 6.8$ T$^{-1}$ and $1/B\approx 13.6$ T$^{-1}$. These correspond to the main SdH frequencies for the first and second harmonics, $f_{R(D)}^{\rm{SdH}}$ and $2f_{R(D)}^{\rm{SdH}}$, respectively. In the absence of both Rashba, Dresselhaus and $g$-factor (dashed yellow curve), we observe no beating in the $\delta\rho_{xx}$ (Fig.~\ref{fig3}). 

 On the other hand, for the case of pure Rashba $\alpha=10$~meV\,nm with $g=-34$ (solid red curve), the presence of the beating in Fig.~\ref{fig:drxx_R} is made clear by the splitting of the peak of the power spectrum around $f=f_{R}^{\rm{SdH}}$ in Fig.~\ref{fig:If_pure}. Interestingly, for $\alpha=10$~meV\,nm with $g=0$ (dashed red curve), the splitting of the peak is not seen anymore, thus highlighting the important role of the Zeeman on the visualization of beatings. For the pure Dresselhaus case with $\beta=10$~meV\,nm and $g=-34$ (solid blue line), we do not see a peak splitting at the $f=f_{D}^{\rm{SdH}}$  but rather at $f=2f_{D}^{\rm{SdH}}$, which is consistent with the presence of the beating seen on the second harmonic in Fig.~\ref{fig:drxx_D}. Similarly to the pure Rashba case, for $\beta=10$~meV\,nm with $g=0$  (dashed blue line), the splitting of the peak is not seen anymore, corroborating again the role of the Zeeman term on the presence of beatings.

\begin{figure}[ht]
\centering \includegraphics[width=1\columnwidth]{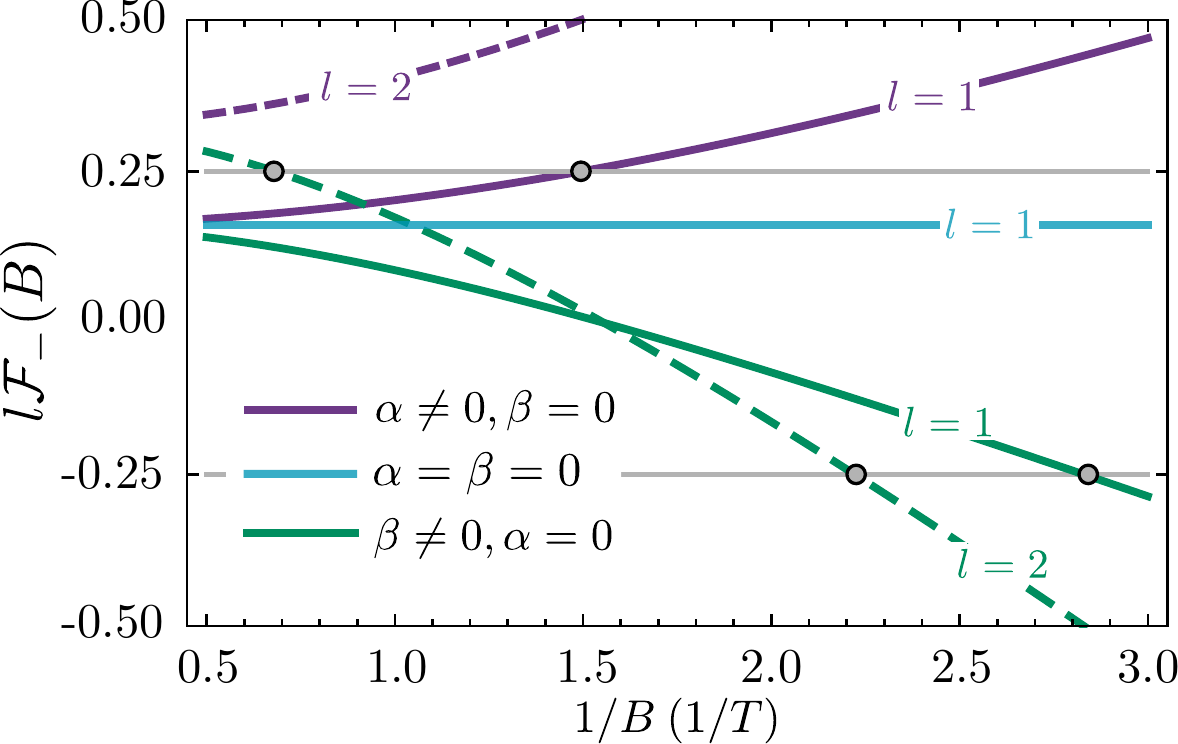}
\caption{Plot of $l \mathcal{F}_-(B)$ vs $1/B$ for $l=1$ and $l=2$ using $\alpha=10$\,meVnm {with $ \beta=0$ (purple lines), $\alpha=\beta=0$ (cyan lines), and $\beta=10$\,meVnm with $ \alpha=0$ (green lines). The solid gray lines indicate $ \pm1/4$ and}  the gray circles indicate where beating nodes occur. For all curves, we use
 $m^*=0.019${$m_o$}, $g^*=-34$ and $n_{2D}=3.3 \times 10^{-3}$\,nm$^{-2}$,  \carlos{parameters for InSb-based 2DEGs\cite{gilbertsson08:165335,akabori08:205320}}.}
\label{fig:FpmExplain} 
\end{figure}

{The apparent ``asymmetry'' in having peak-splitting for Rashba spin-orbit coupling but not for Dresselhaus (even when they have same SO strength) can be understood from the  behavior of the $l \mathcal{F}_-$-function { vs $1/B$}, shown in Fig.\ \ref{fig:FpmExplain}.  As already discussed previously in Secs.~\ref{zeemanR} and \ref{zeemanD}, the condition for beating happens when $\cos ( 2 l \pi \mathcal{F}_-)=0$ or equivalently, $l \mathcal{F}_-=\pm1/4$ ($\pm1/4$ plotted as gray lines).} In the case of Rashba {(green lines)}  one has $(1-\tilde{\Delta})>1$, and the condition for a beating node, $\cos ( 2 l \pi \mathcal{F}_-)=0$, is reached in the interval of $1/B$ {for $l=1$ (solid purple)} (gray circles).  In the Dresselhaus case, $(1+\tilde{\Delta})<1$, such that  $l\mathcal{F}_-$  for $l=1$ only crosses $-1/4$ for large values of $1/B$, where the amplitude of the SdH has already been suppressed. Conversely, $l\mathcal{F}_-$ crosses 1/4 for $l=2$ at smaller values of $1/B$, thus guaranteeing the presence of a beating within the magnetic field range, as shown in Fig.~\ref{fig:drxx_D}. 

\subsection{Landau Levels with \carlos{simultaneous} Zeeman, Rashba and Dresselhaus interactions: Analytical results}
\label{sec:RDanalytical}
As mentioned earlier, to the best of our knowledge, there are no general \emph{exact} analytical results \carlos{for the energies and SdH oscillations corresponding to the case with simultaneous and arbitrary} Zeeman, Rashba and Dresselhaus couplings. Therefore, in this section we will outline how to derive an effective \emph{approximate} solution that can be used to shed light on magnetotransport results for materials, e.g.\ GaAs or InAs, in which all the three couplings are \carlos{present}. For convenience, we define the sum and difference of the spin-orbit couplings
\begin{eqnarray}
\gamma&=&\alpha_B+\beta_B, \\
\delta&=&\alpha_B-\beta_B,
\end{eqnarray}
[see definitions of $\alpha_B$ and $\beta_B$ following Eq. (\ref{eq:LL-RD_real})]
which allows us to rewrite Eq.~(\ref{eq:LL-RD_real}) as
\begin{eqnarray}
\frac{{\cal \tilde{H}}}{\hbar\omega_{c}} & = & a^{\dagger}a+\frac{1}{2}+\frac{\tilde{\Delta}}{2}\sigma_{z}+\frac{\gamma+\delta}{2} \left (a^{\dagger}\sigma_{-}+a\sigma_{+} \right )\nonumber \\
 &  & +\frac{\gamma-\delta}{2}\left (a^{\dagger}\sigma_{+}+a\sigma_{-} \right). \label{eq:LL-ga_de}
\end{eqnarray}
Note that both the pure Rashba and pure Dresselhaus cases are recovered from the equation above for  $\gamma = \delta$ and $\gamma = -\delta$, respectively. Next, we define the Hamiltonian for $\gamma = \delta$ and $\gamma = -\delta$
\begin{eqnarray}
\frac{{\mathcal{\tilde{H}}}_\mathrm{\pm}}{\hbar \omega_c}&=& a^\dagger a+\frac{1}{2}  + \frac{\tilde{\Delta}}{2} \sigma_z \pm \delta(a^\dagger \sigma_\mp + a \sigma_\pm) ,
\label{eq:Hpm}
\end{eqnarray}
which describes the pure Rashba ($+$) and pure Dresselhaus ($-$) cases in the presence of the Zeeman coupling.
As we already discussed in the previous sections, by defining the operator $\mathcal{N}_\pm=a^\dagger a \pm \frac{1}{2}\sigma_z$, we obtain $[{{\cal \tilde{H}}}_{\pm},\mathcal{N}_\pm]=0$\denisnew{,} so the eigenstates of ${{\cal \tilde{H}}}_{\pm}$ are also eigenstates of $\mathcal{N}_\pm$.  The eigenstates of $\mathcal {N}_{+}$ ${({\mathcal N}_-)}$ are then constructed from the pair $\{|n,\uparrow\rangle,|n+1,\downarrow \rangle  \}$  ($\{|n,\downarrow\rangle,|n+1,{\uparrow }\rangle  \}$). The above statement is true except for the decoupled eigenstates $|0,\uparrow \rangle$ ($|0,\downarrow \rangle$) with corresponding eigenenergy $\hbar \omega_c (1-\tilde{\Delta})/2$ $[\hbar \omega_c (1+\tilde{\Delta})/2]$.   The diagonalization of each two state subspace results in  

\begin{eqnarray}
\frac{\varepsilon_{n,s}}{\hbar \omega_c}&=& \left(n+\frac{1}{2} +\frac{\delta}{|\delta|}\frac{s}{2}\right) -\frac{\delta}{|\delta|}\frac{s}{2}\left(1-\frac{\delta}{|\delta|}\tilde{\Delta}\right)  \nonumber \\
&\times& \sqrt{1+\frac{16\delta^2}{(1-\frac{\delta}{|\delta|}\tilde{\Delta})^2}{\left(n+\frac{1}{2}+\frac{\delta}{|\delta|}\frac{s}{2}\right)}}, 
\label{eq:eRD}
\end{eqnarray}
with $s=+$ $(-)$ and $n\in\mathbb{N}_0$.
Note that this form is valid for both pure Rashba \carlos{($\delta=\gamma>0$)} and Dresselhaus ($\delta=-\gamma<0$), Eqs.~(\ref{eq:EnsR}) and (\ref{eq:EnsD}), respectively, thus also including the corresponding decoupled state with the lowest eigenvalues of ${{\cal N}}_\pm$. \denis{Note that to recover the pure Zeeman case with no Rashba and Dresselhaus, we should take $\delta\rightarrow0$ with $\delta/|\delta|\rightarrow1$.}

When both Rashba and Dresselhaus are present, we can use second order perturbation theory with respect to $\delta,\gamma\ll1$ (See Appendix~\ref{perturbation}), to obtain the approximate eigenvalues of the Hamiltonian in Eq.~(\ref{eq:LL-ga_de}), namely 
\begin{eqnarray}
\frac{\varepsilon_{n,s}}{\hbar \omega_c}&=&
n+ \sfrac{1}{2}+s\frac{\tilde{\Delta}}{2} -2s \Lambda (l +\sfrac{1}{2})
-\Omega 
\label{eq:pertEn}
\end{eqnarray}
where the quantities $\Lambda$ and $\Omega$ are defined as
\begin{eqnarray}
\Lambda&=&\frac{(\gamma^2+\delta^2)\tilde{\Delta}+2\gamma \delta}{(1-\tilde{\Delta}^2)}=\frac{{2}\frac{\varepsilon_R}{\hbar \omega_c}}{(1-\tilde{\Delta})}-\frac{{2}\frac{\varepsilon_D}{\hbar \omega_c}}{(1+\tilde{\Delta})},
\label{eq:La}\\
\Omega&=&\frac{(\gamma^2+\delta^2)+2\gamma \delta\tilde{\Delta}}{(1-\tilde{\Delta}^2)}
=\frac{{2}\frac{\varepsilon_R}{\hbar \omega_c}}{(1-\tilde{\Delta})}+\carlos{\frac{{2}\frac{\varepsilon_D}{\hbar \omega_c}}{(1+\tilde{\Delta})},}
\label{eq:Om}
\end{eqnarray}
where we have introduced ${\varepsilon_{R}}/{\hbar\omega_{c}}=\alpha_{B}^2$ and $\varepsilon_{D}/\hbar\omega_{c}=\beta_{B}^2$.

Our goal now is to rewrite Eq.\ (\ref{eq:pertEn}) in a form that recovers the already obtained exact results for pure Rashba and pure Dresselhaus cases.  First, we write $\Lambda=
\frac{\Lambda}{|\Lambda|}|\Lambda|$ since $\Lambda$ changes sign depending on the relative strengths of $\alpha$ and $\beta$,  similarly to the sign of $\delta$ that enters into Eq.\ (\ref{eq:eRD}). By adding and subtracting a term $\frac{s}{2}\, \frac{\Lambda}{|\Lambda|}$ in Eq.\ (\ref{eq:pertEn}) and after some straightforward algebra we obtain
\begin{eqnarray}
\frac{\varepsilon_{n,s}}{\hbar \omega_c}&=&
 \left ( n+ \frac{1}{2}+\frac{\Lambda}{|\Lambda|}\frac{s}{2} \right ) -\frac{\Lambda}{|\Lambda|}\frac{s}{2} {\left(1-\frac{\Lambda}{|\Lambda|}\tilde{\Delta}\right)} \nonumber \\
 &\times& \left\{1+ 
\frac{{4}}{1-\frac{\Lambda}{|\Lambda|}\tilde{\Delta}}{ \left[|\Lambda|\left(n+\frac{1}{2}\right)+\, \Omega\frac{\Lambda}{|\Lambda|}\frac{s}{2} \right ]}{} \right \}. \nonumber \\
\label{eq:pertEnSgn}
\end{eqnarray}
In the case of pure Rashba we have $\Lambda=\Omega=\frac{\delta^2}{1-\tilde{\Delta}}>0$ while for pure Dresselhaus $\Lambda=-\Omega=-\frac{\delta^2}{1+\tilde{\Delta}}<0$; these neatly reduce to the exact results when using second order Taylor expansion of Eq.\ (\ref{eq:eRD}). \carlos{Note that Eq.~(\ref{eq:pertEnSgn}) also reproduces the exact result for when $\alpha=\beta$ and $g^*=0$ \denis{\cite{tarasenko02:552}, represented here by $\Lambda\rightarrow0$ with $\Lambda/|\Lambda|\rightarrow1$, $\tilde{\Delta}=0$, and $\Omega=2\varepsilon_{D/R}/\hbar\omega_c$} .}  \carlos{The \denis{mathematical expression of Eqs.~(\ref{eq:EnsR}) and (\ref{eq:EnsD})} motivate} us to rewrite Eq.\ (\ref{eq:pertEnSgn}) as
\begin{widetext}
\begin{align}
 & \frac{\varepsilon_{n,s}}{\hbar\omega_{c}}=\left(n+\frac{1}{2}+\frac{\Lambda}{|\Lambda|}\frac{s}{2}\right)-\frac{\Lambda}{|\Lambda|}\frac{s}{2}\frac{1-\frac{\Lambda}{|\Lambda|}\tilde{\Delta}}{|1-\frac{\Lambda}{|\Lambda|}\tilde{\Delta}|}\times \sqrt{\left(1-\frac{\Lambda}{|\Lambda|}\tilde{\Delta}\right)^{2}+8\left(1-\frac{\Lambda}{|\Lambda|}\tilde{\Delta}\right)\left[|\Lambda|\left(n+\frac{1}{2}\right)+\,\Omega\frac{\Lambda}{|\Lambda|}\frac{s}{2}\right]}\label{eq:pertEnSqrt} ,
\end{align}
\end{widetext}
where we have used $1+\frac{x}{2}\approx \sqrt{1+x}$ \footnote{Here we emphasize that depending on the values of $\Lambda$ and $\Omega$, the argument of square root of Eq.~(\ref{eq:pertEnSqrt}) becomes negative, thus yielding imaginary energies. This happens when $x\gg1$, which violates the assumption of writing Eq.~(\ref{eq:pertEnSqrt})}.  It is important to note that although $|\Lambda|\ll 1$, $\Lambda$ enters the square root multiplied by $n$, the Landau level index. This  means that for high enough $n$, the product $|\Lambda| n$ is not necessarily a small quantity. Accordingly, although the equation above becomes exact for either pure Rashba or Dresselhaus case, for $\alpha,\beta\neq0$ Eq.\ (\ref{eq:pertEnSqrt}) \carlos{is} only valid when $|\Lambda| n\lesssim 1$, \carlos{besides $\alpha_B,\beta_B,\delta,\gamma\ll1$ already assumed in Appendix~\ref{perturbation} to obtain Eq.~(\ref{eq:pertEn})}.

\carlos{We reiterate that Eq.~(\ref{eq:pertEnSqrt}) satisfies the exact results for (i) the \denis{Zeeman-only case [Eq.~(\ref{eq:EnsZ})], (ii) the pure} Rashba plus nonzero $g^*$ \denis{[Eq.~(\ref{eq:EnsR})]} and (iii) the pure Dresselhaus plus nonzero $g^*$  \denis{[Eq.~(\ref{eq:EnsD})]}. The case $\alpha=\beta$ with $g^*=0$, for which there is also an exact solution\denis{\cite{tarasenko02:552}}, is satisfied to leading order \denis{using $\sqrt{1+x}\approx 1+x/2$ for with $x=8\Omega (s/2)/(1-\tilde{\Delta})\ll1$}. That is, as mentioned in the previous paragraph, the approximate solution  given by Eq.~(\ref{eq:pertEnSgn}) reproduces the exact solution for $\alpha=\beta$ with $g^*=0$\denis{\cite{tarasenko02:552}}.}

As in the case of pure Zeeman, Rashba or Dresselhaus, we can now calculate the $F$-function from Eq.\ (\ref{eq:pertEnSqrt}). The corresponding results are presented in Appendix~\ref{appox-ffunction}, and by neglecting \carlos{SO} contributions higher or equal than \denis{second order in the
spin-orbit parameters $\Lambda$ and $\Omega$ (or fourth order in $\gamma$ and
$\delta$)}, we obtain 
\begin{widetext}
\begin{align}
{\cal F}_{+}  &=\frac{\varepsilon}{\hbar\omega_{c}}-\frac{1}{2}+\Omega-\Lambda\tilde{\Delta},
\label{Fp-app-RD}\\
{\cal F}_{-} & =-\frac{1}{2}\frac{\Lambda}{|\Lambda|}+\frac{1}{2}\frac{\Lambda}{|\Lambda|}\frac{1-\frac{\Lambda}{\left|\Lambda\right|}\tilde{\Delta}}{\left|1-\frac{\Lambda}{\left|\Lambda\right|}\tilde{\Delta}\right|}
\sqrt{\left(1-\frac{\Lambda}{\left|\Lambda\right|}\tilde{\Delta}\right)^{2}+8\left|\Lambda\right|\left(1-\frac{\Lambda}{\left|\Lambda\right|}\tilde{\Delta}\right)\left[\frac{\varepsilon}{\hbar\omega_{c}}+\frac{1}{2}\left|\Lambda\right|\left(1-\frac{\Lambda}{\left|\Lambda\right|}\tilde{\Delta}\right)\right]}.
\label{Fm-app-RD}
\end{align}
\end{widetext}
It is easy to see that these equations recover all the previous results: pure Zeeman [Eq.~(\ref{FpmZ})], Zeeman with pure Rashba [Eqs.~(\ref{frp}) and (\ref{frm})], and Zeeman with pure Dresselhaus [Eqs.~(\ref{frpD}) and (\ref{frmD})]. 
\carlos{Additionally, in the case of $\Lambda \approx 0$, ${\cal F}_-\denis{\approx -\tilde{\Delta}/2}$ \denis{, which reduces to the pure Zeeman case. Accordingly, here ${\cal F}_-$} becomes independent of $B$ (for $B\lesssim 1$~T), and therefore, we expect the absence of beatings in the magneto-resistivity, previously seen for both pure Rashba and pure Dresselhaus cases. }

\subsection{Generalized SdH magneto-resistivity for
arbitrary $\alpha$, $\beta$ and $g^*$ : new prediction for the
absence of beatings.}

Using the Eqs.\ (\ref{Fp-app-RD}) and (\ref{Fm-app-RD}) in 
Eq.\ (\ref{normdiffmag}), we can derive the magnetoresistivity $\delta \rho_{xx}(B)$ for the case with arbitrary Rashba and Dresselhaus couplings and simulta-
neous nonzero Zeeman field, 
\begin{widetext}
\begin{align}
\delta\rho_{xx}(B) =&  2\sum_{l=1}^{\infty}e^{-l\pi\frac{\hbar/\tau_{q}}{\hbar\omega_{c}}}\frac{2\pi^{2}lk_{B}T/\hbar\omega_{c}}{\sinh(2\pi^{2}lk_{B}T/\hbar\omega_{c})}\nonumber\\
 & \times\cos\left[2\pi l\left(\frac{\varepsilon_F}{\hbar\omega_{c}}+\frac{2\varepsilon_{R}}{\hbar\omega_{c}}+\frac{2\varepsilon_{D}}{\hbar\omega_{c}}\right)\right]\cos\left\{ \pi l\sqrt{\left(1-\frac{\Lambda}{\left|\Lambda\right|}\tilde{\Delta}\right)^{2}+16\lambda_{B}^{2}\left(\lambda_{B}^{2}+\frac{\varepsilon_F}{\hbar\omega_{c}}\right)}\right\} ,
 \label{ana-rd}
\end{align}
\end{widetext}
with $\lambda_{B}^{2}=\frac{\left|\Lambda\right|}{2}\left(1-\frac{\Lambda}{\left|\Lambda\right|}\tilde{\Delta}\right)$.
From Eq. (\ref{ana-rd}), we can derive the condition for the absence of beatings for any $l$ by finding the condition for the second cosine being independent of $1/B$ ; this implies \carlos{$|\Lambda| = 0$, which leads to}
\begin{equation}
    \frac{\alpha}{\beta} =\sqrt{\frac{1-\tilde{\Delta}}{1+\tilde{\Delta}}},
    \label{no-beating}
\end{equation}
thus \carlos{yielding} Eq.\ (\ref{new-condition}) presented in the introduction.
For  $\tilde{\Delta}\ll1$, the above condition is reduced to $\alpha \approx \beta$, corresponding to the situation where the total SO $k$-dependent effective field becomes unidirectional~\cite{PhysRevLett.90.146801,PhysRevLett.97.236601,PhysRevLett.117.226401}. 

Note that the above condition does not correspond to any fundamental symmetry, since there is no new conserved quantity in our Hamiltonian with both non-zero Zeeman ($g^* \neq 0$) and  Rashba-Dresselhaus couplings. We reiterate that Eq.~(\ref{no-beating}) is entirely  distinct from the persistent-spin-helix condition $\alpha=\beta$. As shown in Fig.~1(d), the case $\alpha=\beta$ and $g^*\neq 0$ does not show peak splitting in the first harmonic but ehxibits beating (or peak splitting) in the second harmonic. Only when $g^*=0$ (no Zeeman) and $\alpha=\beta$ there are peak splittings absent altogether \cite{tarasenko02:552,averkiev05:543}.

\subsection{Beatings for both $\alpha$ and $\beta$ non-zero}
\label{sec:SdH_beatings}
In the previous sections, we studied the effect of the Zeeman interaction on the frequency splitting of the power spectrum peaks, which represents the beatings in the SdH oscillations. Here we study the interplay of {\em both} the Dresselhaus and Rashba \carlos{interactions} on the beatings of the SdH oscillations.

Similarly to what we did leading up to Eq.\ (\ref{eq:fR}), we can obtain the effective beating frequency from the $\mathcal{F}_-$--function in Eq.\ (\ref{Fm-app-RD}) which results in 
\begin{align}
f_{R+D}&=\frac{h}{2e}\sqrt{ \left | \frac{2 k_{R+D}^2}{\pi} 
\left (n_\mathrm{2D}+\frac{k_{R+D}^2}{2\pi} \right ) \right | },
\label{eq:kRandD}
\end{align}
where the effective SO momentum is 
\begin{equation}
    k_{R+D}=\frac{m^*}{\hbar^2} \sqrt{\left (
    1-\frac{\Lambda}{|\Lambda|} \tilde{\Delta}
    \right )
    \left (
    \frac{\alpha^2}{1-\tilde{\Delta}}-\frac{\beta^2}{1+\tilde{\Delta}}
    \right )}.
\end{equation}
\begin{figure}[ht]
\centering \includegraphics[width=0.5\textwidth]{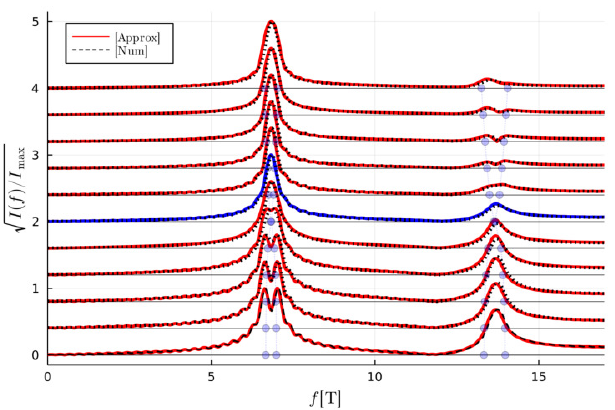}
\caption{Normalized power spectrum $\sqrt{I(f)}$ for a {\em fixed} $\alpha=7.0$\,meV\,nm, and for $\beta=0.0$ to $10$\,meV\,nm, from bottom to top. [red curves], along with {\em full} numerical results [black dashed].  The curve corresponding to $k_{RD}=0$ with $\beta=5.0$\,meV\,nm, is shown [blue curve]. The gray circles indicate the frequency splitting in Eq.\ (\ref{eq:kRandD}). Other parameters are $m^*=0.019m_o$, $g^*=-34$, $n_{2D}=3.3 \times 10^{-3}$\,nm$^{-2}$, and $\hbar \tau_q^{-1}=1.75$\,meV, \carlos{for InSb-based 2DEGs\cite{gilbertsson08:165335,akabori08:205320}}}
\label{fig:If1} 
\end{figure}
\begin{figure}[ht]
\centering \includegraphics[width=0.5\textwidth]{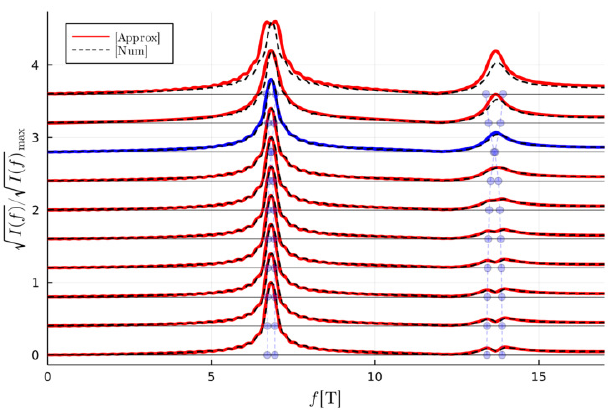}
\caption{Normalized power spectrum $\sqrt{I(f)}$ for a fixed $\beta=5.0$\,meV\,nm, and for $\alpha=0.0$ to $9.0$\,meV\,nm, from bottom to top. [red curves] along with {\em full} numerical results [black dashed].  The curve corresponding to $k_{RD}=0$ with $\alpha=7.0$\,meV\,nm [blue curve]. 
The gray circles indicate the frequency splitting in Eq.\ (\ref{eq:kRandD}).
Other parameters are $m^*=0.019 m_o$, $g^*=-34$, $n_{2D}=3.3 \times 10^{-3}$\,nm$^{-2}$, and $\hbar \tau_q^{-1}=1.75$\,meV, \carlos{for InSb-based 2DEGs\cite{gilbertsson08:165335,akabori08:205320}}}
\label{fig:If2} 
\end{figure}
We start with the pure Rashba case plus Zeeman, $\alpha=7.0$~meV\,nm and $g^*=-34$. The corresponding power spectrum yields the bottom curve in Fig.~\ref{fig:If1}, similar to the one plotted in Fig.~\ref{fig:If_pure}. This curve shows two main peaks representing the first two harmonics, and the presence of a split main peak. We assume a Lorentzian broadening $\hbar \tau_q^{-1}=1.75$\,meV. When the Dresselhaus coupling $\beta$ increases, we see the splitting of the main peak reduces until it vanishes for $\beta=5.0$~meV\,nm (the frequency splitting from Eq.\ (\ref{eq:kRandD}) is indicated by the gray circles). The absence of beating is indeed expected \carlos{as predicted by the condition $\beta=\alpha\sqrt{\frac{1+\tilde{\Delta}}{1-\tilde{\Delta}}}=5.0$~meV in Eq.~(\ref{no-beating})}. For larger $\beta$, we see that the splitting of the main peak remains neglible.  However, in the second harmonic a clear splitting opens up.  The condition for having no peak splitting at any harmonics is indeed the condition in Eq.~(\ref{no-beating}), where the effects of the SO couplings basically disappear [there are still small SO terms $\varepsilon_R$ , $\varepsilon_D$ in Eq. (\ref{ana-rd})].
The power spectrum using full numerical calculations are also shown [black dashed], and for this parameter regime the analytical and numerical results agree well.

A similar analysis can be done for the case of pure Dresselhaus with Zeeman, $\beta=5.0$~meV\,nm and $g^*=-34$,  see Fig.~\ref{fig:If2}. Here the splitting is not observed in the main peak, but rather in the second harmonic.  As $\alpha$ is increased from 0.0 to 9.0\,meV\,nm, the splitting in the second harmonic decreases, and vanishes at $\alpha=7.0$\,meV\,nm, which again corresponds to the condition in Eq.\ (\ref{no-beating}).
Despite the good accuracy of the approximate analytical solution for $\alpha \lesssim 8.0$~meV\,nm, it starts to deviate from the exact one ({\em full} numerics) for higher values of $\beta$. This happens because for these values, the combined effect Rashba and Dresselhaus is more pronounced, producing an anti-crossing between different energy levels (see blue curves Fig.~\ref{fig:EnFfun}, discussed further below). 
While the approximate energies obtained here are always monotonic with respect to $1/B$, around the anti-crossing the numerical ones are not. Accordingly, our $F$-function calculation will not be able to fully describe the SdH oscillations and frequencies around the anti-crossing regions, specifically the approximate solution misses a central peak that starts developing, which will be discussed in the next section. In terms of the $F$-function, the occurance of level anticrossings corresponds to $|{\cal F}_-|\approx1/2$.
Since the power spectrum is obtained by integrating $\delta \rho_{xx}$ over a range of $1/B$, there is no simple condition determining the validity of the approximate solution.  However, looking at the $\lambda_B$ term in Eq.\ (\ref{Fm-app-RD}) the condition
\begin{equation}
    8 \pi n_\mathrm{2D} \left (
\frac{k_R^2}{1-\tilde{\Delta}}-\frac{k_D^2}{1+\tilde{\Delta}}
    \right ) l_c^4 \lesssim 1 ,
    \label{eq:FminusCondition}
\end{equation}
yields a useful estimate for the $1/B$ values where the Dingle factor has not suppressed $\delta \rho_{xx}$. 
Equation (\ref{eq:FminusCondition}) generalizes a similar condition derived in Ref.\ \onlinecite{erlingsson10:155456}.
It is also interesting to note that the analytical result is more accurate for higher harmonics, as the Dingle-factor helps diminishing the amplitude of the anti-crossing at higher-fields (see Fig.~\ref{fig:EnFfun}).

\section{Landau Levels with Zeeman, Rashba and Dresselhaus interactions: Numerical results}
\label{sec:RDnumerical}
In the previous section, we have derived an approximate analytical result for the magnetoresistance oscillations in the presence of both Rashba, Dresselhaus \textit{and} Zeeman interactions. \carlos{The assumptions and approximations underlying the derivation involved the relatively small SO coupling and the low number of occupied Landau levels. These are satisfied in the low electron density InSb-based 2DEGs of Refs. \onlinecite{gilbertsson09:235333,akabori08:205320}. For higher electron density systems (but still with just a singly-occupied subband at $B=0$), such as the InAs/GaSb wells in  Ref.~\onlinecite{beukmann17:241401}, a numerical approach is needed. Below we outline the numerical procedure. The numerical approach also allows us to account for the full form of cubic Dresselhaus term, see Sec.~\ref{sec:cubic}.}

For the case of either pure Rashba or Dresselhaus with Zeeman, the absence of anti-crossing in the LL spectrum allow us to obtain \emph{exact} analytical results for the problem. As we explain below, this does {\em not} hold in the presence of both Rashba and Dresselhaus with the Hamiltonian (in the spin basis) Eq.~(\ref{eq:LL-RD_real}) 
\begin{equation}
\frac{{\cal \tilde{H}}}{\hbar\omega_{c}}=\left(\begin{array}{cc}
a^{\dagger}a+\frac{1}{2}+\frac{\tilde{\Delta}}{2} & 2 \alpha_{B}a+2 \beta_{B}a^{\dagger}\\
2\alpha_{B}a^{\dagger}+2\beta_{B}a & a^{\dagger}a+\frac{1}{2}-\frac{\tilde{\Delta}}{2}
\end{array}\right).
\label{eq:HRD_2x2}
\end{equation}
Therefore, here we calculate the magnetotransport numerically via the diagonalization of the Hamiltonian above. The $F$-function method used for the analytical cases can be extended to allow for numerical methods for calculating the energy spectrum, see App.\ \ref{app:numerical}. 

As opposed to both the pure Rashba and pure Dresselhaus cases, ${\cal N}_{\pm}$ do not commute with the Hamiltonian above, and therefore, the diagonal basis cannot be described by any linear combination of the previous degenerate eigenstates of ${\cal N}_\pm$. However, there is still a  unitary operator, ${\cal P}=\exp \left \{i\pi  \left ({\cal N}_\pm-\frac{1}{2} \right )  \right \}$ that commutes with this Hamiltonian, called the \emph{parity} operator ~\citep{casanova2010deep,braak11:100401}, which is discussed in detail in App.~\ref{app:chi}. The corresponding unitary transformation gives $\mathcal{P} a \mathcal{P}^\dagger=-a$, $\mathcal{P} a^\dagger \mathcal{P}^\dagger=-a^\dagger$ and $\mathcal{P} \sigma_\pm \mathcal{P}^\dagger=-\sigma_\pm$, which clearly makes the Hamiltonian Eq.~(\ref{eq:LL-RD_real}) invariant due to presence of only $a^\dagger a$, $a^\dagger \sigma_\pm$ and $a \sigma_\pm$ terms. The eigenvalues of $\mathcal{P}$, $\pm 1$, help analyze the energy spectrum behavior.

To understand the influence on the spectrum of both Rashba and Dresselhaus contributions, we first recall that in the absence of the latter, the Rashba term is responsible for coupling $\left|n,\uparrow  \right \rangle$ to $\left|n+1,\downarrow  \right \rangle$, for \denis{$n\in\mathbb{N}_0$} , thus yielding decoupled $2\times2$ block diagonal Rashba Hamiltonians (shown by the red boxes in the Hamiltonian below). When we account for the Dresselhaus contribution, we obtain a coupling between states $\left|n,\downarrow  \right \rangle$ and $\left|n+1,\uparrow  \right \rangle$ for $n\in\mathbb{N}_0$, which belongs to different Rashba blocks. More specifically, the Dresselhaus term produces a coupling between blocks $\left\{ \left|n,\uparrow  \right \rangle,\left|n+1,\downarrow  \right \rangle \right\}$ and $\left\{ \left|n+\Delta n,\uparrow  \right \rangle,\left|n+1+\Delta n,\downarrow  \right \rangle \right\}$ with $\Delta n =2$, which is indicated by the blue box in the Hamiltonian below (See App.\ \ref{app:chi}). As a consequence, we have two decoupled orthogonal basis set given by $\left\{ \left| 0^{+} \right \rangle \right\} =\left\{ \left|n,\uparrow\right\rangle, \left|n+1,\downarrow\right\rangle, \thinspace \ldots \right\}$  and $\left\{ \left| 0^{-} \right \rangle \right\} =\left\{ \left|n,\downarrow\right\rangle ,\left|n+1,\uparrow\right\rangle, \thinspace \ldots \right\}$ with $n\in\mathbb{N}_0$. Interestingly, these decoupled basis have different eigenvalues with respect to the parity operator, i.e., ${\cal P} \left| 0^\pm \right \rangle = \pm 1\left| 0^\pm \right \rangle $ and therefore, represent different parity subspace.

\begin{figure*}[htb] 
 \centering
  \includegraphics[width=1\textwidth]{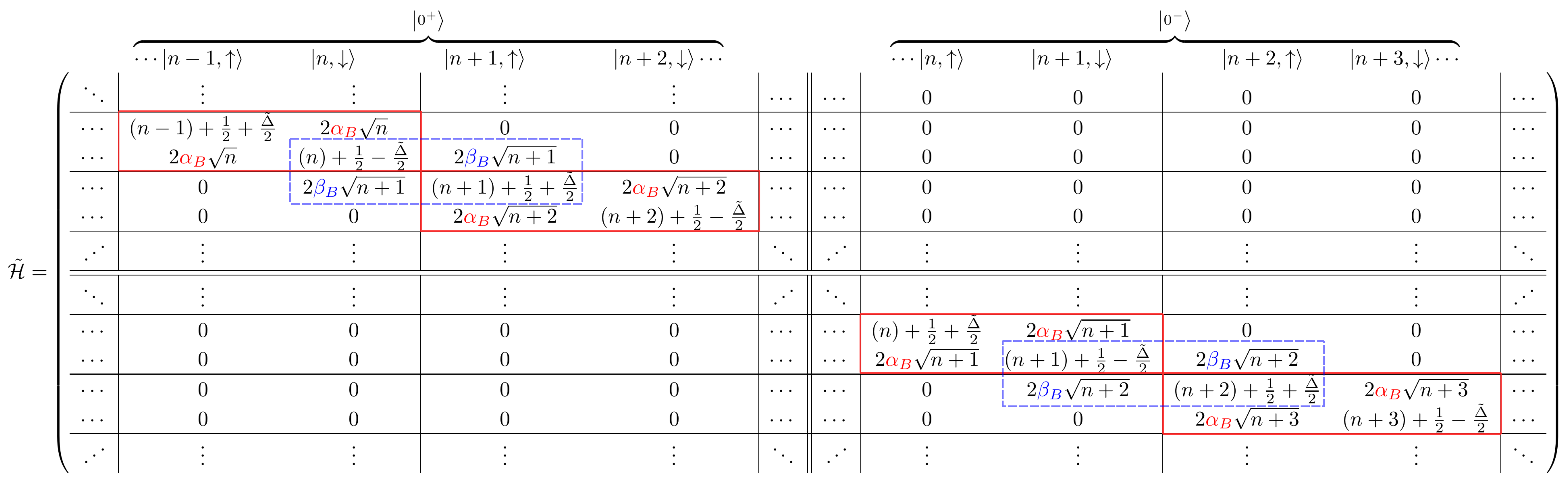}
  \caption{Graphical illustration of the parity subspaces in the {\em matrix} representation of the Hamiltonian Eq.\ (\ref{eq:HRD_2x2}). \denis{Here we see the Rashba interaction couples $\left|n-1,\uparrow\right\rangle$ to $\left|n,\downarrow\right\rangle$ (red boxes), while the Dresselhaus interaction couples $\left|n,\downarrow\right\rangle$ to $\left|n+1,\uparrow\right\rangle$ (blue boxes).}}
\label{fig:Hamiltonian}
\end{figure*}

\begin{figure}[ht]
\centering \includegraphics[width=0.48\textwidth]{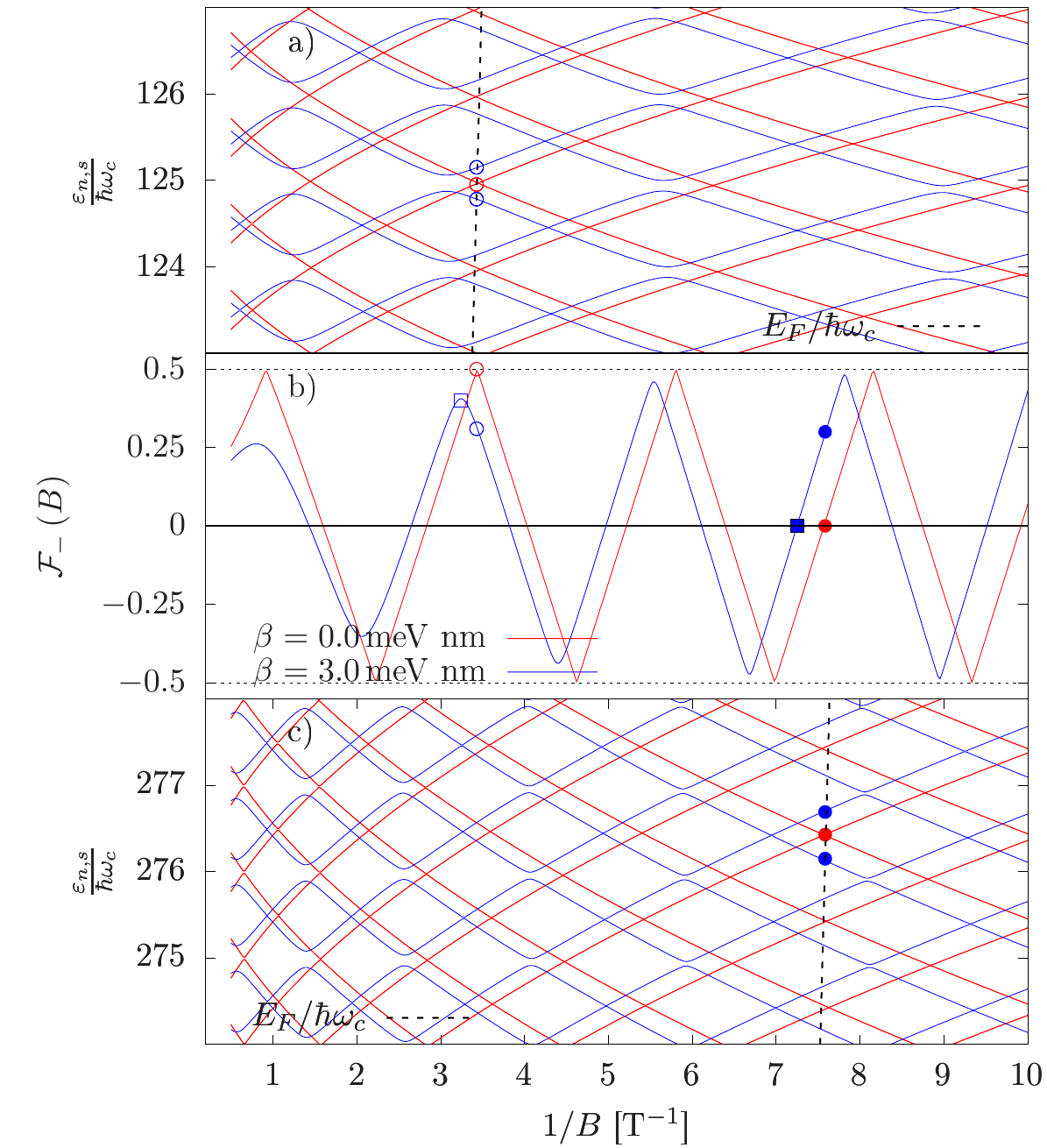}
\caption{The energy spectrum for two sets of $(\alpha,\beta)=(7.5, 0.0)$\,meV\,nm [red] and $(\alpha,\beta)=(7.5, 3.0)\,$meV\,nm [blue], along with $\varepsilon_F / \hbar \omega_c$ (black dashed), a) around $n=125$ and c) $n=255$.  b) $\mathcal{F}_c$ for the same pair of parameters.  Note sawtooth form for pure Rashba [red], and for $(\alpha,\beta)=(7.5, 3.0)\,$meV\,nm [blue] a rounding, and translation, of the cusps due to level anticrossing [solid circles]. Other parameters are $m^*=0.04$, $g^*=-12$ and $n_{2D}=17.6 \times 10^{-3}$\,nm$^{-2}$, \carlos{for InAs-based quantum wells\cite{beukmann17:241401}}.
}
\label{fig:EnFfun} 
\end{figure}

In terms of the spectrum, in the presence of only Rashba SO coupling, we observe multiple crossing between the Rashba eigenstates $\{|n,- \rangle, |n,+ \rangle \}$ for different $n \in \mathbb{N}_0$, with energy given by Eq.~(\ref{eq:EnsR}), obtained through the diagonalization of the Rashba blocks (red boxes within the  Hamiltonian matrix in Fig.\ \ref{fig:Hamiltonian}). This is shown by the red solid lines in Fig.~\ref{fig:EnFfun}(a) for $\alpha=7.5$\,meV\,nm. In the presence of Dresselhaus SO coupling, the states $|n,- \rangle$ and $|n+\Delta n,+ \rangle $ with $\Delta n\in\mathbb{N}_{\mathrm{odd}}$ belong to the same parity subspace and adding a Dresselhaus contribution will yield anti-crossing, which open up gaps in the spectrum (blue curves). Conversely, the decoupling between the different parity sets, i.e., $|n,- \rangle$ and $|n+\Delta n,+ \rangle $ with $\Delta n\in\mathbb{N}_{\mathrm{even}}$, implies multiple crossing between their corresponding energy states. 
These features are shown by the blue curve in Figs.~\ref{fig:EnFfun}(a) and (c), where we have used $\beta=3.0$\,meV\,nm. Other parameters are $m^*=0.04$, $g^*=-12$ and $n_{2D}=17.6 \times 10^{-3}$\,nm$^{-2}$. \carlos{These parameters are for InAs/GaSb-based (double) quantum wells\cite{beukmann17:241401} in the electron regime. This regime, as emphasized in Ref.~\onlinecite{beukmann17:241401}, corresponds to the configuration in which the GaSb well is depleted and the system is effectively a single InAs-based asymmetric quantum well with electrons only.}
Furthermore, we also observe that the effect of the Dresselhaus term is to simply shift the crossing point to a different magnetic field  and energy 
(the crossing-point energy remains constant to lowest order in $\beta$ but does in general shift for higher values of $\beta$). 

The contrasting behavior of crossings vs.\ anti-crossings has direct consequences on the $F$-function, which will be analyzed in the next paragraphs. First we consider the crossing between states  $|n,- \rangle$ and $|n+\Delta n,+ \rangle $, with {\em even} $\Delta n$ (corresponding to states belonging to different parity subspaces).  The  $F$-function are
 \begin{eqnarray}
  \varepsilon_{n,-}(B)=\frac{\varepsilon}{\hbar \omega_c} \leftrightarrow n=F_- \left (\frac{\varepsilon}{\hbar \omega_c},B;\alpha, \beta \right ),
  \label{eq:Fm_aRaD}
 \end{eqnarray}
 and 
 \begin{eqnarray}
  \varepsilon_{n+\Delta n,+}(B)=\frac{\varepsilon}{\hbar \omega_c} \leftrightarrow n+\Delta n=F_+ \left (\frac{\varepsilon}{\hbar \omega_c},B;\alpha, \beta \right),
   \label{eq:Fp_aRaD}
 \end{eqnarray}
 where we have explicitly added their dependence on $\alpha$ and $\beta$. This results in an $F$-function difference [see Eq.\ \ref{fast-slow}] at the crossing $\varepsilon=\varepsilon_c$ and $B=B_c$
 \begin{equation}
 \mathcal{F}_-\left (\frac{\varepsilon_c}{\hbar \omega_c},B_c;\alpha, \beta \right )=\frac{\Delta n}{2}\in \mathbb{Z}.
 \label{eq:Nintegar}
 \end{equation}
 Note that since the SdH oscillation is dependent on $\mathcal{F}_{\pm}$ in the form of $\cos(2 \pi \mathcal{F}_\pm)$, we can re-define $\mathcal{F}_{-}$ to lie within an unit interval, e.g.,  $\mathcal{F}_- \in [-1/2 , 1/2]$. Accordingly, integer values of $\mathcal{F}_-$ are equivalent to $\mathcal{F}_-=0$\, and therefore, the vanishing of ${\cal F}_-$ provides the field values where the crossing happens. The curves for ${\cal F}_-$ are plotted in Fig.~\ref{fig:EnFfun}(b) for the same parameters as in Fig.~\ref{fig:EnFfun}(a). It presents a sawtooth pattern because values of $|{\cal F}_-|>1/2$ are shifted back to the $[-1/2 , 1/2]$ interval. The role of the Dresselhaus coupling for these crossings  is evident in Fig.\ \ref{fig:EnFfun}(b), where the zeros of $\mathcal{F}_-$ remain zeros for any value of $\beta$, but are simply shifted to new values of magnetic field, open circle moves to open rectangle Fig.~\ref{fig:EnFfun}(b).

 Next, we look at the crossing between states belonging to the same parity subspace, i.e., $|n,- \rangle$ and $|n+\Delta n,+ \rangle $ for odd $\Delta n$. We recall that this crossing only exists for the pure Rashba case, shown in both Figs.~\ref{fig:EnFfun}(a) and (c). Here the relations in Eqs.\ (\ref{eq:Fm_aRaD}) and (\ref{eq:Fp_aRaD}) still hold, the only difference being the value of $\Delta n$, which results in 
 \begin{equation}
 \mathcal{F}_-\left (\frac{\varepsilon_c}{\hbar \omega_c},B_c;\alpha, \beta=0 \right )=\frac{\Delta n}{2}  \in \mathbb{Z}+\frac{1}{2}.
 \label{eq:Nhalfinteger}
 \end{equation}
 Adding a non-zero Dresselhaus contribution will couple these states and lead to an anti-crossing, shown in Figs.~\ref{fig:EnFfun}(a) and (c).   The anti-crossing result in non half-integer values of $\mathcal{F}_\pm$ in Eqs.\ (\ref{eq:Fm_aRaD}) and (\ref{eq:Fp_aRaD}) and will lead to a rounding of the sawtooth pattern as seen in Fig.\ (\ref{fig:EnFfun})(b) (blue curves). 

 The conditions in Eqs.\ (\ref{eq:Nintegar}) and (\ref{eq:Nhalfinteger}) lead to values of $\cos(2 \pi \mathcal{F}_-)=1$ [filled circle and rectangle in Fig.\ \ref{fig:EnFfun}b)] and $\cos(2 \pi \mathcal{F}_-)=-1$ [open cirlce circle in \ref{fig:EnFfun}b)], respectively, in the case of either pure Rashba or pure Dresselhaus.  However, when both Rashba and Dresselhaus are present only the former condition $\cos(2 \pi \mathcal{F}_-)=1$ holds (crossing of states with opposite parity) but the latter condition changes such that  $\cos(2 \pi \mathcal{F}_-)> -1$ due to anticrossings of states with same parity eigenvalue [open rectangle in Fig.\ \ref{fig:EnFfun}b)].  This, in turn, affects the shape of the magneto\denis{-}oscillations leading to an asymmetry in the maximum and minimum values of $\cos(2 \pi \mathcal{F}_-)$. 
 
\begin{figure}[ht]
\centering \includegraphics[width=0.48\textwidth]{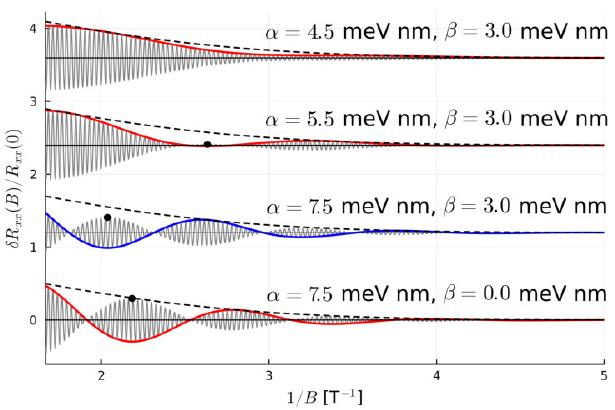}
\caption{Magnetooscillations for four different parameter values, including pure Rashba, then different combinations of $(\alpha,\beta)$.  The anti-crossings in the spectrum complicates the beating behavior, which eventually vanishes for around $(\alpha,\beta)=(4.5, 3.0)\,$meV\,nm. Other parameters are $m^*=0.04$, $g^*=-12$ and $n_{2D}=17.6 \times 10^{-3}$\,nm$^{-2}$.}
\label{fig:dRxxcomparison} 
\end{figure}

\begin{figure}[ht]
\centering \includegraphics[width=0.48\textwidth]{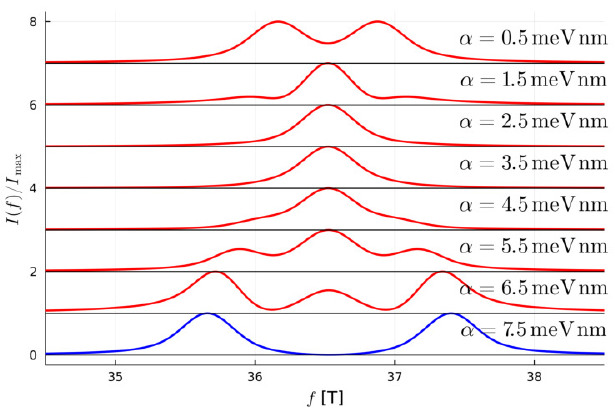}
\caption{Power spectrum for af {\em fixed} $\beta=3.0$\,meV\,nm for $\alpha=7.5$ down to 0.5\,meV\,nm.  Other parameters are $m^*=0.04$, $g^*=-12$ and $n_{2D}=17.6 \times 10^{-3}$\,nm$^{-2}$, \carlos{from Ref.~\onlinecite{beukmann17:241401}}.}
\label{fig:Ifcomparison} 
\end{figure}

In Fig. \ref{fig:dRxxcomparison} this asymmetry is visible in the magneto-osillations.  Here we assume Gaussian broadening with $B_q=0.50$\,T which forms an envelope (black dashed curve).  The lowest curve is the pure Rashba $(\alpha,\beta)=(7.5, 0.0)\,$meV\,nm and there {\em all} maximas intersect the envelope [black circles].  The curve for $(\alpha,\beta)=(7.5, 3.0)\,$meV\,nm shows that only {\em some} maxima intersect the envelope, the other maximas correspond to $\cos(2 \pi \mathcal{F}_-)>-1$ do not (black circle).  This is a direct consequence of the anti-crossing in the spectrum in Fig. \ref{fig:EnFfun}.  The curves for $(\alpha,\beta)=(5.5, 3.0)\,$meV\,nm and $(4.5, 3.0)\,$meV\,nm show how the anti-crossing becomes larger, eventually leading to an absence of beatings.
This can also be seen in the frequency spectrum shown in Fig. \ref{fig:Ifcomparison}, for the $f \approx f^\mathrm{SdH}$ peak.  The lowest curve (blue) corresponds to $(\alpha,\beta)=(7.5, 3.0)\,$meV\,nm where the spectrum shows well separated peaks.  However, as the strength of the Rashba coupling is decreased all the way down to $\alpha=0.5$\,meV\,nm for a fixed value of $\beta=3.0$\,meV\,nm a central peak develops and for $\alpha$ between 4.5 and 1.5\,meV\,nm, the two split peaks are barely visible.

\subsection{Extracting $\alpha$ and $\beta$ from SdH data}
\label{sec:cubic}
The magneto-oscillations can be thought of as a fingerprint of the sample parameters, including  Fermi energy $\varepsilon_F$, effective mass $m^*$, $g^*$, and $\alpha$ and $\beta$.
To better capture the influence of the spin-orbit couplings for higher electron density, the {\em full} form of the Dresselhaus interaction will be used. For non-zero magnetic fields, this corresponds to having Dresselhaus SO term in Eq.\ (\ref{eq:LL-RD_real}) replaced with 
\begin{eqnarray}
 \left  [\frac{1}{\sqrt{2} \hbar \omega_c \ell_c} \left (
 \beta_1
 -\gamma\frac{a^\dagger a}{2  \ell_c^2} \right )
 a^{\dagger}\sigma_{+}+\frac{\gamma}{2  \ell_c^2}a^3\sigma_{-} \right ] + \mathrm{h.c.} , 
 \label{eq:D3}
\end{eqnarray}
where $\beta_1=\gamma \langle k_z^2 \rangle $, $\gamma$ is material-dependent parameter describing the SO interaction due to bulk inversion asymmetry, and $\langle k_z^2 \rangle$ is the expectation value of the $z$-component
of the square of momentum operator (divided by $\hbar$), 
see App.\ \ref{app:D3} for details of {\em full} Dresselhaus coupling.    
Note that $\beta$ in Eq.\ (\ref{Hinit}) is assumed to include the first harmonic of the cubic Dresselhaus \cite{PhysRevLett.117.226401,dresselhaus55:580}, which makes it linearly dependent on the electron density.
For instance, if the potential confining the 2DEG is assumed to be an infinite well of width $d_\mathrm{QW}$ then $\langle k_z^2 \rangle=\pi^2/d_\mathrm{QW}^2$. 
To model the magnetoresistance data we start from Eq.\ (\ref{sdh}), which features $(i)$ a sum over higher harmonics , $(ii)$ rapid oscillations coming from $\mathcal{F}_+$, and $(iii)$ damping due to Landau level broadening $\tilde{L}_\Gamma$.  The analysis introduced in the previous section was based on the study of the properties of $\cos(2 \pi\mathcal{F}_-)$, which forms an envelope on top of the rapid oscillations.
Note that in the case having both Rashba and Dresselhaus  coupling the rapid oscillations are still dominated by the normal SdH oscillations, i.e.
\begin{eqnarray}
\mathcal{F_+}(B)&=&-\frac{1}{2}+\frac{\varepsilon_F}{\hbar \omega_c} \left( 1+\mathcal{O} \left ( \frac{\varepsilon_R}{\varepsilon_F} ,\frac{\varepsilon_D}{\varepsilon_F}  \right) \right )  \nonumber \\
&\approx&-\frac{1}{2}+\frac{f^\mathrm{SdH}}{B},
\label{order-mag}
\end{eqnarray}
so the SO coupling does not affect the rapid oscillations.  The resulting lowest harmonic form of the magneto-resistivity is 
\begin{equation}
\delta \rho_{xx}(B)
 = -2\tilde{L}_\Gamma(B) \cos(2\pi l\mathcal{F}_{-}(B))\cos \left (2 \pi \frac{f_\mathrm{SdH}}{B} \right ) ,
\label{eq:Ffit}
\end{equation}
which can be fitted to available data.

\begin{figure}[ht]
\centering \includegraphics[width=0.50\textwidth]{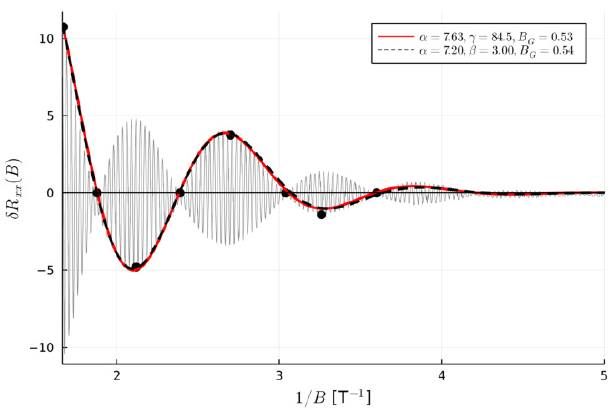}
\caption{The black dots are reference points for curve C1, solid black line.  The black dashed curve is the linear Dresselhaus result and solid red curve {\em full} Dresselhaus result. Parameter values from fitting are shown in the inset. Other \carlos{parameters\cite{beukmann17:241401}} are $m^*=0.019$, $g=-12$ and $n_\mathrm{2D}=0.0176$\,nm$^{-2}$.}
\label{fig:Rxx_c01_Gfit_D3} 
\end{figure}

\begin{figure}[ht]
\centering \includegraphics[width=0.5\textwidth]{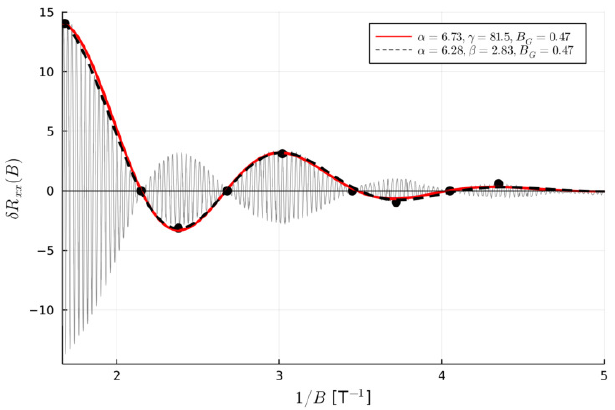}
\caption{Similar to Fig.\ (\ref{fig:Rxx_c01_Gfit_D3}), but for curve C5, solid black line. The black dashed curve represents the linear Dresselhaus result, while the solid red curve, the {\em full} cubic Dresselhaus term.
Extracted fitting parameters are shown in the inset. Other \carlos{parameters\cite{beukmann17:241401}} are $m^*=0.019$, $g=-12$ and $n_\mathrm{2D}=0.0176$\,nm$^{-2}$. }
\label{fig:Rxx_c05_Gfit_D3} 
\end{figure}

\begin{figure}[ht]
\centering \includegraphics[width=0.5\textwidth]{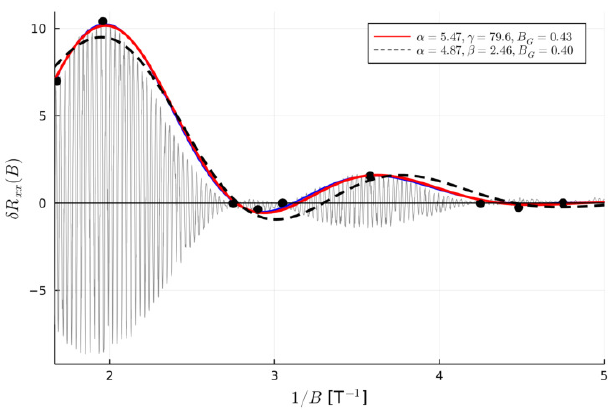}
\caption{Similar to Fig.\ (\ref{fig:Rxx_c01_Gfit_D3}), but for curve C10, solid black line. The black dashed curve represents the linear Dresselhaus result, which fails to fit the data.  However, the {\em full} cubic Dresselhaus term (solid red curve) results in a good fit.
Extracted fitting parameters are shown in the inset. Other \carlos{parameters\cite{beukmann17:241401}} are $m^*=0.019$, $g=-12$ and $n_\mathrm{2D}=0.0176$\,nm$^{-2}$.}
\label{fig:Rxx_c10_Gfit_D3} 
\end{figure}

Figures \ref{fig:Rxx_c01_Gfit_D3}-\ref{fig:Rxx_c10_Gfit_D3} show the experimental data from Ref.\ \onlinecite{beukmann17:241401} \carlos{for InAs/GaSb quantum wells in the electron regime)} along with our theoretical fits [Eq.\ (\ref{eq:Ffit})].  We focus on the experimental curves 1, 5 and 10, of Fig.\ S4 of Ref.\ \onlinecite{beukmann17:241401} that we label as C1, C5 and C10 in Fig.\ \ref{fig:Rxx_c01_Gfit_D3}-\ref{fig:Rxx_c10_Gfit_D3}.  The data was fitted to $\delta \rho_{xx}(B)$ in Eq.\ (\ref{eq:Ffit}), where $\mathcal{F}_-$ was calculated numerically. 
For the fitting we consider both the Dresselhaus coupling in Eq.\ (\ref{eq:LL-RD_real}) [black dashed lines], and also with the {\em full} Dresselhaus term in Eq.\ (\ref{eq:D3}) [solid red lines].  The black dots are reference points extracted from the data, which are used in the fitting of $\tilde{L}_\Gamma(B) \cos(2\pi \mathcal{F}_-)$. The best fittings were produced by assuming Gaussian broadening, namely.
\begin{equation}
 \tilde{L}_\Gamma(B) =\exp \left (
 -2 \pi^2 \frac{\Gamma ^2}{(\hbar \omega_c)^2}
 \right )=\exp \left (
 -\frac{B_q^2}{B^2}
 \right ),
 \label{eq:GaussBq}
\end{equation}
where $B_q=\sqrt{2}\pi \frac{m^*\Gamma}{\hbar e}$ and $\Gamma$ is a constant Landau level broadening.

For curve C1 in Fig.\ \ref{fig:Rxx_c01_Gfit_D3} the fitting with linear Dresselhaus  yields values $\alpha=7.2$\,meV\,nm and $\beta=3.0$\,meV\,nm. On the other hand, for fitting  to the {\em full} model we obtain $\alpha=7.6$\,meV\,nm, and $\gamma=85$\,meV\,nm$^3$.  
We see that both fits produce equally good curves fitting the experimental data points, with comparable values for the extracted Rashba SO coupling.  This indicates  that when the Rashba coupling dominates the cubic Dresselhaus term ($a^3$-term in Eq. (\ref{eq:D3})), fitting the data with the addition of the cubic term does not strongly affect the result.
The results for curve C5 in Fig.\ \ref{fig:Rxx_c05_Gfit_D3} behave similarly, i.e.\ we find fitted values of the Rashba coefficient, $\alpha=6.7$\,meV\,nm for the linear Dresselhaus with $\beta=2.8$\,meV\,nm, and $\alpha=6.3$\,meV\,nm for the {\em full} cubic Dresselhaus, with $\gamma=82$\,meV\,nm$^3$.

However, the story is different for the curve C10 shown in Fig.\ \ref{fig:Rxx_c10_Gfit_D3}.  Here the value of Rashba and Dresselhaus coupling are closer, and then the details of the linear vs.\ cubic Dresselhaus become relevant.  Indeed, the linear Dresselhaus model fitting yields  $\alpha=5.5$\,meV\,nm and $\beta=2.5$\,meV\,nm while the cubic fit gives $\alpha=4.9$\,meV\,nm.  More importantly the error in the linear fit is quite high, and the fit [black dashed curve] fails to describe the data points.  
However, the cubic model gives a good fit , with $\gamma=80$\,meV\,nm$^3$. This clearly shows the importance of the cubic contributions in samples with high density, where the Rashba and Dresselhaus contributions are of comparable magnitudes.

The fit results in Fig. \ref{fig:Rxx_c01_Gfit_D3}-\ref{fig:Rxx_c10_Gfit_D3} were done for $\langle k_z^2 \rangle=\pi^2/d_\mathrm{QW}^2$ where $d_\mathrm{QW}=12.5$\,nm  \cite{beukmann17:241401}.  To fully model the sample a self-consistent Poisson-Schr\"odinger calculation is required\cite{psh-prx,fu20:134416,PhysRevLett.117.226401}, which is beyond the scope of this work.  We can however use different values of $\langle k_z^2 \rangle$, which {\em indirectly} emulate self-consistent potential details, i.e.\ increasing the value of  $\langle k_z^2 \rangle$ suggests a stronger confinement in the InAs quantum well, and decreased value of $\langle k_z^2 \rangle$ would correspond to wavefunctions being less localized in the InAs quantum well.  

\begin{figure}[ht]
\centering \includegraphics[width=0.5\textwidth]{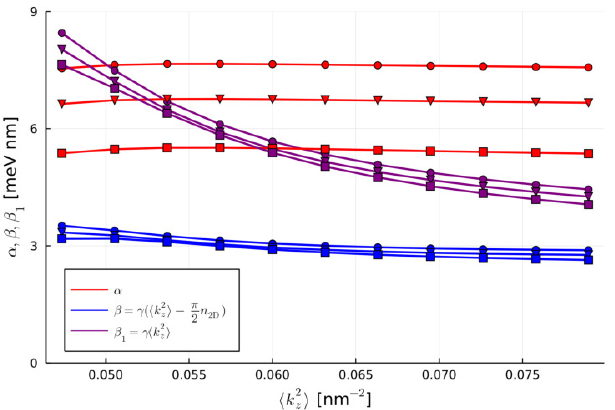}
\caption{The spin-orbit parameters that result from the fitting as a function of $\langle k_z^2 \rangle $. 
Other parameters are $m^*=0.019$, $g=-12$ and $n_\mathrm{2D}=0.0176$\,nm$^{-2}$. The three different symbols represent different curves: curve 1: circle, 5: triangle, and 10: square. 
%
}
\label{fig:figS4_aRaDga} 
\end{figure}

In Fig.\ \ref{fig:figS4_aRaDga} the values of $\alpha$, $\beta_1$, and $\beta$ are shown as a function of $\langle k_z^2 \rangle$ from 0.75$\frac{\pi^2}{d_\mathrm{QW}^2}$ to 1.25$\frac{\pi^2}{d_\mathrm{QW}^2}$. The data from the three curves 
are indicated by different forms: C1: circle, C5: triangle, and C10: square.  For each value of $\langle k_z^2 \rangle$, specific values of $\alpha$ $\beta_1$, and $\beta$ are obtained from the fit. 
The fit results for $\alpha$ and $\beta$ for each curve remain relatively insensitive to $\langle k_z^2 \rangle$-variations.  Note that as $\langle k_z^2 \rangle$ varies $\beta_1$ changes quite rapidly via the fitted value of $\gamma$.  This is to be expected since lower values $\langle k_z^2 \rangle$, correspond to the electron leaking out the InAs quantum well $\gamma$ into the GaSb, which has a higher bulk value of $\gamma$.  For higher values of $\langle k_z^2 \rangle$ the system becomes more strongly confined in the InAs quantum well and the value of $\gamma$ should tend to the value corresponding to bulk InAs.

The fact that  the values of $\alpha$ and $\beta$ change only slightly as function of $\langle k_z^2 \rangle$, as can be seen in Fig. \ref{fig:figS4_aRaDga}, has important consequences on the fitting proceedure.
For this reason a fitting with $\gamma$ and $\langle k_z^2 \rangle$ {\em both} being independent fitting parameters can not be performed, since if $\beta_1 = \gamma \langle k_z^2 \rangle$ is  the dominant contribution to the Dresselhaus couplings then there are multiple (infinite) solutions to the equation $\gamma \langle k_z^2 \rangle=\mathrm{const.\ } $ and fitting the data with $\gamma$ and $\langle k_z^2 \rangle$ independent will not converge \cite{beukmann17:241401}.

\section{Summary}

We investigated the SdH magneto-oscillations in the resistivity $\rho_{xx}$ of 2DEGs in the presence of spin-orbit (Rashba-Dresselhaus) and Zeeman couplings. We used a semiclassical approach for the resistivity combined with a Poisson summation formula for the Landau-quantized DOS. Our approach allows for an intuitive separation of the slow and fast quantum oscillations in terms of ``F-functions'', central quantities in our description, essentially being the inverse functions of the spin-resolved  Landau-level structure of the system. We study a variety of exact cases such as the pure Zeemann, pure Dresselhaus, and pure Rashba cases -- all of which provide analytical expressions for the magnetoresistivity. 

More importantly, from our unified and general formulation we also derive, for the first time, an analytical solution for the case with arbitrary Rashba and Dresselhaus couplings and simultaneous non-zero Zeeman coupling ($g^* \neq0$).Interestingly, this allows us to derive a  unique new condition for the vanishing of the SO-induced beatings in the SdH signals: $\alpha/\beta= [(1-\tilde \Delta)/(1+\tilde \Delta)]^{1/2}$, where $\tilde \Delta =g^*m^*/2m_0$ (i.e., ratio (Zeeman energy)/$\hbar\omega_c$). This new condition does not correspond to any conserved quantity in our Hamiltonian, unlike the persistent-spin-helix condition $\alpha=\beta$ which is associated with the conservation of spin along some particular axes. We emphasize that our new condition precludes beatings in all harmonics of the quantum oscillations. 

\carlos{We have applied our analytical formulation to describe low-density data for SdH oscillations showing many harmonics in GaAs-based 2DEGs (see SM in Ref. \onlinecite{psh-prx}) and found an excellent agreement, Fig.~2. We have also applied our theory to low-density InSb-based 2DEGs\cite{gilbertsson08:165335,akabori08:205320}.  In addition, we have also developed a detailed numerical calculation for high-density InAs-based 2DEGs, in which an analytical description is not satisfactory. We find excellent agreement with available data for high-density InAs-based 2DEGs \cite{beukmann17:241401,psh-prx}. We have also pointed out} an inequivalence between the  Rashba-dominated + Zeeman vs the Dresselhaus-dominate + Zeeman cases, with only the former showing beatings. This follows from a distinct interplay between the SO and Zeeman terms in these two regimes. 
We hope our detailed study and unified general formulation will stimulate furhter experimental investigations aiming at verifying our theoretical predictions.

\section{Acknowledgments}
The authors would like to thank Arjan Beukman and Leo Kouvenhowen for sharing experimental data from Ref.\ \onlinecite{beukmann17:241401} that was used for fitting.  \carlos{We also thank Thomas Schaepers and Makoto Kohda for useful discussions}. \carlos{The authors acknowledge funding from the Reykjavik University Research Fund, the S\~ao Paulo Research Foundation (FAPESP) Grants No. 2016/08468-0 and No. 2020/00841-9, Conselho Nacional de Pesquisas (CNPq), Grants No. 306122/2018-9 and 301595/2022-4, \denis{the Swiss Nanoscience Institute (SNI), the NCCR SPIN and grant no. 179024 of the Swiss NSF, the Georg H. Endress Foundation, the EU H2020 European Microkelvin Platform EMP (grant no. 824109) and FET TOPSQUAD (grant no. 862046).}}

\pagebreak

\appendix

\section{\carlos{Density of states and F-Functions}}

\carlos{Here we follow closely the discussion (and notation) in Sec.~3.2.2 of the book \textit{Semiclassical Physics} by Brack and Bhaduri \cite{brack97:book}.}

\carlos{For simplicity, we first consider the case  with a discrete spectrum $\varepsilon_n = f(n)$, $n=0,1,2,...$ in which each level has a degeneracy 
$d_n=\tilde D(n)$, with $\tilde D(n)$ being an arbitrary function of $n$. Later on we will account for a (pseudo) spin index. As an example, we note that for the usual 2DEG Landau levels (LLs) (in 
the absence of both Zeeman or SO interaction), $\varepsilon_n= \hbar\omega_c (n+1/2)$ and $d_n= AeB/h=\tilde D(n)$ (A: area 
of the 2DEG, $e>0$); in this 
case, $d_n=\tilde D(n) $ denotes the LL degeneracy and is independent of $n$. This same Landau degeneracy holds 
in the presence of Zeeman and SO interactions. For later convenience, we define $D(n)=\tilde D(n)/A$
to be the level degeneracy per unit area [e.g., for LLs $D(n)=n_{LL}(B)=eB/h$].  
As in Ref.~\onlinecite{brack97:book}, let $f(n)$ be an arbitrary monotonic function with 
a differentiable inverse $f^{-1}(x)=F(x)$, the relevant ``F-function" in our discussion. In this case, 
because $f^{-1}(f(x)) = x = f(f^{-1}(x))$ it follows that $n=F(\varepsilon_n)$. Next we define the DOS of our system and relate it the to the F-function, which ultimately allows us to calculate the oscillatory part of the DOS relevant for our semiclassical transport calculation.} 

\subsection{Density of states without LL broadening}

\carlos{Quite generally we can define the DOS of our system as,
\begin{equation}
g(\varepsilon) = \frac{1}{A}\sum_{n=0}^{\infty} \tilde D(n) \delta(\varepsilon-\varepsilon_n) \label{dos}.
\end{equation}
Note that the above DOS is defined per area and energy. In Ref. \cite{brack97:book} the DOS is defined just per energy. 
Motivated by the property $\delta[\denis{y}(x)]=\frac{1}{|\denis{y}^\prime(x_0)|}\delta(x-x_0)$ where $x_0$ denotes a root of $\denis{y}(x)$, i.e., $\denis{y}(x_0)=0$ and $\denis{y}^\prime(x)=d\denis{y}(x)/dx$, we define $h(\varepsilon)=n-F(\varepsilon)$, which obeys $h(\varepsilon_n)=0$ as $n=F(\varepsilon_n)$ by construction. We can then write
\begin{equation}
\delta[h(\varepsilon)]=\delta(n-F(\varepsilon)) = \frac{1}{|F^\prime(\varepsilon_n)|}\delta(\varepsilon - \varepsilon_n)=\frac{1}{|F^\prime(\varepsilon)|}\delta(\varepsilon - \varepsilon_n), \label{delta1}
\end{equation}
or
\begin{equation}
\delta(\varepsilon - \varepsilon_n)=|F^\prime(\varepsilon)|\delta(n-F(\varepsilon)). \label{delta2}
\end{equation}
Substituting (\ref{delta2}) into (\ref{dos}), we find
\begin{eqnarray}
g(\siggi{\varepsilon}) &=& D(\varepsilon) |F^\prime(\varepsilon)| \sum_{n=0}^{\infty} \delta(n-F(\varepsilon)), \label{dos1}
\end{eqnarray}
where $D(\varepsilon)\equiv D(F(\varepsilon))$. Noting that 
\begin{equation}
\sum_{n=0}^{\infty} \delta(\varepsilon - n) = \sum_{l=-\infty}^{\infty}e^{2\pi i l \varepsilon}, \qquad (\varepsilon>0) \label{delta3}
\end{equation}
we can straightforwardly cast (\ref{dos1}) in the form
\begin{equation}
g(\varepsilon) = D(\varepsilon) |F^\prime(\varepsilon)| \sum_{l=-\infty}^{\infty}e^{2\pi i l F(\varepsilon)}. \label{dos2}
\end{equation}
Now we introduce the (pseudo) spin index $s=\pm 1$ by adding a subscript 
$s$ to all quantities [except $D(\varepsilon)$ for it is not (pseudo) spin dependent]. 
This $s$ index accounts for the spin-dependent Zeeman and SO 
interactions in our 2DEG.  With this new index, the DOS in Eq.~(\ref{delta3}), viewed as per spin now,  becomes
\begin{equation}
g_s(\varepsilon) = D(\varepsilon) |F_s^\prime(\varepsilon)| \sum_{l=-\infty}^{\infty}e^{2\pi i l F_s(\varepsilon)}, \label{dos22}
\end{equation}
or
\begin{equation}
g_s(\varepsilon) = D(\varepsilon) |F_s^\prime(\varepsilon)| \left\{1+2\sum_{l=1}^{\infty}\cos[2\pi  l F_s(\varepsilon)]\right\}. \qquad (\varepsilon>0) \label{dos3}
\end{equation}
By summing over $s$, we obtain the total DOS, 
\begin{equation}
g(\varepsilon) = D(\varepsilon) \sum_s |F_s^\prime(\varepsilon)| \left\{1+2\sum_{l=1}^{\infty}\cos[2\pi  l F_s(\varepsilon)]\right\}. \label{tot-dos}
\end{equation}
For the systems investigated in our work, 
$F_s^\prime (\varepsilon) \simeq 1/\hbar\omega_c$. This is actually exact for the Zeeman-only case, see Eq.~(\ref{FF-z-der}), main text, but only approximate in the presence of SO interaction [see Eq.~(\ref{F-deriv})]. In this case and $D(\varepsilon)|F_s^\prime (\varepsilon)| =  \frac{m^*}{2\pi \hbar^2}$, we find
\begin{equation}
g(\varepsilon) \simeq \frac{m^*}{\pi \hbar^2} \left\{1+\sum_{l=1}^{\infty}\left(\cos[2\pi  l F_+(\varepsilon)]+ \cos[2\pi  l F_-(\varepsilon)] \right)\right\}. \label{tot-dos0}
\end{equation}
Using the identity, 
\begin{equation}
\cos a + \cos b = 2 \cos[(a+b)/2]\cos[(a-b)/2],
\end{equation}
we can rewrite Eq.~(\ref{tot-dos0}) as 
\begin{equation}
g(\varepsilon) \simeq \frac{m^*}{\pi \hbar^2}  \left\{1+\sum_{l=1}^{\infty}\denis{2}\cos[2\pi  l \mathcal{F}_+(\varepsilon)]
\cos[2\pi  l \mathcal{F}_-(\varepsilon)]\right\},  \label{tot-dos1}
\end{equation}
where 
\begin{equation}
\mathcal{F}_\pm(\varepsilon) = \frac{1}{2}[F_+(\varepsilon) \pm F_-(\varepsilon)].
\label{cal-F}
\end{equation}
To regain the DOS notation in the main text, we now make $g(\varepsilon) \rightarrow \mathcal{D}(\varepsilon,B)$ and use $\mathcal{D}_0 = \frac{m^*}{2\pi \hbar^2}$. Hence, Eq.~(\ref{tot-dos1}) becomes
\begin{equation}
\mathcal{D}(\varepsilon,B) \simeq 2\mathcal{D}_0  \left\{1+\denis{2}\sum_{l=1}^{\infty}\cos[2\pi  l \mathcal{F}_+(\varepsilon)]
\cos[2\pi  l \mathcal{F}_-(\varepsilon)]\right\}, \label{tot-dos2}
\end{equation}
or
\begin{equation}
\frac{\mathcal{D}(\varepsilon,B)  - \denis{2}\mathcal{D}_0}{\denis{2}\mathcal{D}_0} \simeq 2 \sum_{l=1}^{\infty}\cos[2\pi  l \mathcal{F}_+(\varepsilon)]
\cos[2\pi  l \mathcal{F}_-(\varepsilon)], 
\label{dos-main-1}
\end{equation}
which is Eq.~(\ref{eq8}) in the main text.} 

\subsection{Density of states including Landau level broadening}

\carlos{We can account for LL broadening in the DOS calculation by considering Lorentzian or Gaussian functions 
as particular representations of the ideal $\delta$ functions describing the discrete levels. We consider a simple phenomenological 
description which assumes that all LLs have the same spin-independent broadening $\Gamma$. }

\subsubsection{Lorentzian DOS case}

\carlos{Here we take the delta function representing the DOS of a single LL as, 
\begin{equation}
\delta(\varepsilon - \varepsilon_n) = \lim_{\Gamma\rightarrow 0} \frac{1}{\pi}\frac{\Gamma/2}{(\varepsilon-\varepsilon_n)^2 \denis{+} (\Gamma/2)^2}=\lim_{\Gamma\rightarrow 0} L_\Gamma(\varepsilon - \varepsilon_n), \label{d-l}
\end{equation}
where
\begin{equation}
 L_\Gamma(\varepsilon)=\frac{1}{\pi}\frac{\Gamma/2}{\varepsilon^2\denis{+} (\Gamma/2)^2}, \label{lorentzian}
\end{equation}
with 
\begin{equation}
\int_{-\infty}^\infty L_\Gamma(\varepsilon) d\varepsilon = 1.
\end{equation}
Note that 
\begin{equation}
\int_{-\infty}^\infty L_\Gamma(\varepsilon) e^{-2\pi i l \varepsilon} d\varepsilon = e^{-\Gamma \pi |l|}=\tilde L_\Gamma(k),
\label{L-shift}
\end{equation}
where $\tilde L_\Gamma(k)$ is the Fourier transform (FT) of $L_\Gamma(\varepsilon)$ and $l \in \mathbb{Z}$. Using the shifting property of FTs, it follows that the FT of $L_\Gamma(\varepsilon-x)$ is $e^{-2\pi i k x} \tilde L_\Gamma(k)$.  Generalizing Eq.~(\ref{dos}) for Lorentzian-broadened levels we have (we will add a subindex $s$ later on)
\begin{equation}
g(\varepsilon) =  \lim_{\Gamma\rightarrow 0}\sum_{n=0}^{\infty} D(n) L_\Gamma(\varepsilon-\varepsilon_n), \label{dos-L}
\end{equation}
which we can rewrite as,
\begin{equation}
g(\varepsilon) = \lim_{\Gamma\rightarrow 0} \sum_{n=0}^{\infty} \int_{-\infty}^\infty{D(n)}L_\Gamma(\varepsilon - x) \delta(x-\varepsilon_n)dx. 
\end{equation}
Considering that $D(n)$ is independent of $n$ and using Eq.~(\ref{delta3}) with the replacement $\varepsilon \rightarrow F(\varepsilon)$, 
we obtain
\begin{equation}
g(\varepsilon) =  \lim_{\Gamma\rightarrow 0} \int_{-\infty}^\infty  D(F(x)) |F^\prime(x)| \sum_{l=-\infty}^{\infty} e^{2\pi i l F(x)} L_\Gamma(\varepsilon-x) dx.
\end{equation}
Since $L_\Gamma(\varepsilon - x)=L_\Gamma(x-\varepsilon)$ is peaked at $x=\varepsilon$, it is convenient to expand $F(x)$ around this point. Then $g(\varepsilon)$ becomes
\begin{align}
g(\varepsilon) =  \lim_{\Gamma\rightarrow 0} \int_{-\infty}^\infty  D[F(\varepsilon)+|F^\prime(\varepsilon)|(x-\varepsilon)] 
[|F^\prime(\varepsilon)|+\nonumber \\ |F^{\prime\prime}(\varepsilon)|(x-\varepsilon) ]
\sum_{l=-\infty}^{\infty} e^{2\pi i l \left[ F(\varepsilon)+|F^\prime(\varepsilon)|(x-\varepsilon)\right]} L_\Gamma(\varepsilon-x) dx.
\end{align}
Neglecting the contribution $|F^{\prime\prime}(\varepsilon)|(x-\varepsilon)$ [as a matter of fact, this contribution vanishes identically in the limit $L_\Gamma(x-\varepsilon) \rightarrow \delta(x-\varepsilon)$, because $\int_{-\infty}^{\infty} f(x) \delta(x-x_0)dx=f(x_0)$], we have
\begin{align}
g(\varepsilon) = D(\varepsilon) \lim_{\Gamma\rightarrow 0}
|F^\prime(\varepsilon)| \sum_{l=-\infty}^{\infty} e^{2\pi i l F(\varepsilon)} \times  \nonumber \\
 \int_{-\infty}^\infty   e^{2\pi i l |F^\prime(\varepsilon)|(x-\varepsilon)} L_\Gamma(x-\varepsilon) d(x-\varepsilon).
 \label{eq:dg_Integral}
\end{align}
Using Eq.~(\ref{L-shift}), we can write
\begin{equation}
g(\varepsilon) =  D(\varepsilon) \lim_{\Gamma\rightarrow 0} 
|F^\prime(\varepsilon)| \sum_{l=-\infty}^{\infty} e^{2\pi i l F(\varepsilon)} \tilde L_\Gamma(l |F^\prime(\varepsilon)|).
\end{equation}
or
\begin{equation}
g(\varepsilon) =  D(\varepsilon) \lim_{\Gamma\rightarrow 0} 
|F^\prime(\varepsilon)| \sum_{l=-\infty}^{\infty} e^{2\pi i l F(\varepsilon)}
e^{-\Gamma \pi |lF^\prime(\varepsilon)|},
\label{dos-b}
\end{equation}
where have used,
\begin{equation}
\tilde L_\Gamma(l |F^\prime(\varepsilon)|) = e^{-\Gamma \pi |lF^\prime(\varepsilon)|}
\label{eq:LGa}
\end{equation}
As before [Eq.~(\ref{dos22})], we can rewrite Eq.~(\ref{dos-b}) by adding a subindex $s$ to obtain the LL-broadened DOS per spin 
\begin{equation}
g_s(\varepsilon) =  D(\varepsilon)|F_s^\prime(\varepsilon)| 
\left\{1+2 \sum_{l=1}^\infty \cos[2 \pi l F_s(\varepsilon)]
e^{-\Gamma \pi l |F_s^\prime(\varepsilon)|}\right\}. \label{dos3-gen}
\end{equation}
In the above we have dropped the $\lim_{\Gamma \rightarrow 0}$, since a real system has a finite $\Gamma$. Interestingly, the broadened DOS  in Eq.~(\ref{dos3-gen}) can be obtained directly from the case without broadening [Eq.~(\ref{dos3})] by simply multiplying the oscillating components (harmonics) in the latter by the exponential (``Dingle") factor $e^{-\Gamma \pi l |F_s^\prime(\varepsilon)|}$.} 

\carlos{Here again, for the systems of interest here $F_s^\prime(\varepsilon)\simeq 1/\hbar\omega_c$ and the exponential factor in Eq.~(\ref{dos3-gen}) becomes
\begin{equation}
e^{-\Gamma \pi l |F_s^\prime(\varepsilon)|}= e^{-\pi l \Gamma/\hbar \omega_c},
\end{equation}
where $\Gamma \equiv \hbar/\tau_q$, $\tau_q$ is the quantum lifetime of the LL. Summing over the (pseudo) spin index $s$ 
Eq.~(\ref{dos3-gen}) becomes
\begin{equation}
g(\varepsilon) =  \frac{m^*}{\pi \hbar^2}
\left\{1+ \denis{2}\sum_{l=1}^{\infty}\cos[2\pi  l \mathcal{F}_+(\varepsilon)]
\cos[2\pi  l \mathcal{F}_-(\varepsilon)]
e^{- \denis{\frac{\pi l \Gamma}{\hbar \omega_c}} }\right\}. \label{dos-tot-L}
\end{equation}
In the notation of the main text we have 
\begin{equation}
\frac{\mathcal{D}(\varepsilon,B)  - \denis{2}\mathcal{D}_0}{\denis{2}\mathcal{D}_0} \simeq   2
 \sum_{l=1}^{\infty}\cos[2\pi  l \mathcal{F}_+(\varepsilon)]
\cos[2\pi  l \mathcal{F}_-(\varepsilon)]
e^{-\denis{\frac{\pi l \Gamma}{\hbar\omega_c}}}, \label{dos-tot-LP}
\end{equation}
which is the Eq.~(\ref{sdh}) of the main text, but written for the Lorentzian broadening case. }

\subsubsection{Gaussian DOS case}

\carlos{The Gaussian-broadened case can be treated similarly by considering the delta function representation
\begin{equation}
\delta(\varepsilon - \varepsilon_n) = \lim_{\Gamma\rightarrow 0} \frac{1}{\sqrt{2\pi}\Gamma}e^{-\frac{(\varepsilon-\varepsilon_n)^2}{2\Gamma^2}}.
\label{g-l}
\end{equation}
From this we can evaluate the integral in Eq.\ (\ref{L-shift}) which results in the Gaussian version of Eq.\ (\ref{eq:LGa}):
\begin{equation}
\tilde{L}_\Gamma \left ( l|F' (\epsilon)|  \right )=
e^{-2 \pi^2 (l|F' (\epsilon)| )^2 \Gamma^2}.
\end{equation}
This reduces to Eq.\ (\ref{eq:GaussBq}) for $l=1$ (fundamental frequency) and $|F' (\epsilon)|=1/\hbar \omega_c$.}

\subsubsection{Calculating the F-function and its derivative $F^\prime(\varepsilon)$}

\carlos{Here we illustrate the calculation of $F_s(\varepsilon)$ and its derivative with respect to $\varepsilon$, $F^\prime(\varepsilon)$, in the presence of SO interaction. For simplicity, we consider the pure Rashba case (no Zeeman).} \carlos{To determine the F-functions we need to invert $\denis{\varepsilon_{n,s}} =\varepsilon$, where 
\denis{\begin{align}
    \frac{\varepsilon_{n,s}}{\hbar\omega_{c}}=&\left(n+\frac{1}{2}+\frac{s}{2}\right)\label{eq:EnsR-Ap}\\
    &-\frac{s}{2}\frac{1-\tilde{\Delta}}{|1-\tilde{\Delta}|}\sqrt{\left(1-\tilde{\Delta}\right)^{2}+16 \alpha_{B}^{2}\left(n+\frac{1}{2}+\frac{s}{2}\right)},\nonumber 
\end{align}}
is the pure Rashba energy, Eq.~(\denis{\ref{pureR}}) in the main text. Squaring $\varepsilon - \tilde n \hbar \omega_c$, with $\tilde n=n+(1+s)/2$, we find
\begin{eqnarray}
[\varepsilon - \tilde n\hbar \omega_c]^2 &=& \frac{1}{4}(\hbar \omega_c - \Delta)^2 + 4 \varepsilon_R\hbar \omega_c \tilde n \nonumber \\
\varepsilon^2 - 2\varepsilon \hbar \omega_c \tilde n &+& \tilde n^2 \hbar^2 \omega_c^2 = \frac{1}{4}(\hbar \omega_c - \Delta)^2 + 4 \varepsilon_R\hbar \omega_c \tilde n \nonumber \\
 \tilde n^2 \hbar^2 \omega_c^2 - (2\varepsilon \hbar \omega_c &+&4 \varepsilon_R\hbar \omega_c) \tilde n - \frac{1}{4}(\hbar \omega_c - \Delta)^2 + \varepsilon^2=0 \nonumber \\
 \tilde n^2 - \left(\frac{2\varepsilon} {\hbar \omega_c} + \frac{4\varepsilon_R}{\hbar \omega_c} \right) \tilde n &-& \left(\frac{1}{2} - \frac{\Delta}{2\hbar \omega_c}\right)^2 + \left(\frac{\varepsilon}{\hbar \omega_c}\right)^2=0 \label{eq-2nd}
\end{eqnarray}
We can easily solve (\ref{eq-2nd}) for $\tilde n_s(\varepsilon) \Rightarrow n_s (\varepsilon) =- (1+s)/2 + \tilde n_s (\varepsilon) = f_s^{-1} = F_s(\varepsilon)$ 
\begin{eqnarray}
F_s(\varepsilon) &=& -\frac{1+s}{2} + \frac{\varepsilon} {\hbar \omega_c} + \frac{2\varepsilon_R}{\hbar \omega_c}  \\ &+&  s  \sqrt{\left(\frac{\varepsilon} {\hbar \omega_c} + \frac{2\varepsilon_R}{\hbar \omega_c} \right)^2 + \left(\frac{1}{2} - \frac{\Delta}{2\hbar \omega_c}\right)^2 - \left(\frac{\varepsilon}{\hbar \omega_c}\right)^2 }. \nonumber\label{F_pm-R}
\end{eqnarray}
We obtain   $F_\pm^\prime(\varepsilon)$ by   differentiating (\ref{F_pm-R}),
\begin{eqnarray}
F_s^\prime(\varepsilon) &=& \frac{1} {\hbar \omega_c} +s \frac{1}{2} \frac{2\left(\frac{\varepsilon} {\hbar \omega_c} + \frac{2\varepsilon_R}{\hbar \omega_c} \right)\frac{1}{\hbar \omega_c}-\frac{2\varepsilon}{\hbar^2 \omega_c^2}}{\sqrt{\left(\frac{\varepsilon} {\hbar \omega_c} + \frac{2\varepsilon_R}{\hbar \omega_c}\right)^2 + \left(\frac{1}{2} - \frac{\Delta}{2\hbar \omega_c}\right)^2 -\left(\frac{\varepsilon}{\hbar \omega_c}\right)^2 }}, \nonumber \\
\hbox{\noindent or} \\
F_s^\prime(\varepsilon) &=& \frac{1}{\hbar \omega_c} +s \frac{\frac{2\varepsilon_R}{\hbar \omega_c}}{\sqrt{\frac{4\varepsilon\varepsilon_R}{\hbar^2 \omega_c^2} + \left(\frac{2\varepsilon_R}{\hbar \omega_c} \right)^2 + \left(\frac{1}{2} - \frac{\Delta}{2\hbar \omega_c}\right)^2 }}. \label{F-deriv}
\end{eqnarray}
}
\carlos{As mentioned earlier, the leading term in $F_s^\prime(\varepsilon)$ is $1/\hbar\omega_c$. By expanding the above expression, we can easily find $\mathcal{O} \left( \varepsilon_R/\varepsilon_F\right) = \mathcal{O}[(\alpha m^*\ell_c)^{2}/\hbar]$ corrections. The above calculation also holds for the Dresselhaus case. The general case with simultaneous and arbitrary Rashba and Dresselhaus couplings lead to the corrections $\mathcal{O}[(\alpha m^*\ell_c)^{2}/\hbar]+\mathcal{O}[(\beta m^*\ell_c)^2/\hbar)]$ mentioned following Eq.~(\ref{eq12-dos-main-text}).} 

\section{Orthogonal subspaces $\mathcal{P}$ \label{sec:Orthogonal-subspaces-}}
\label{app:chi}
When both Rashba and Dresselhaus are present neither ${\cal N}_{+}$ nor ${\cal N}_{-}$
are conserved, i.e.\ $[{\cal N}_{\pm},{\cal \tilde{H}}]\neq0$. This will result in mixing of states, e.g.\ the pure Rashba states will get couple to each other when a finite $\beta$ is introduced, and vice versa.
However, there is a conserved quantity that can be constructed
from ${\cal N}_{\pm}$ by defining \cite{casanova2010deep,braak11:100401}
\begin{gather}
\mathcal{P}_{\pm}=\exp(i\pi (\mathcal{N}_{\pm}+\sfrac{1}{2})).
\end{gather}
Using the definition of ${\cal N}_{+}=a^{\dagger}a+\frac{1}{2}\sigma_{z}$
we can show that 
\begin{eqnarray}
\mathcal{P}_{+} & = & \exp \left (i\pi(a^{\dagger}a+\frac{1}{2}\sigma_{z}) \right )=\exp \left (i\pi(a^{\dagger}a-\frac{1}{2}\sigma_{z} +\sigma_z) \right )\nonumber \\
 & = & \mathcal{P}_{-}\exp(i\pi\sigma_{z})=-\mathcal{P}_{-},
\end{eqnarray}
where we used $\exp(i\pi\sigma_{z})=-1$. Since $\mathcal{P}_{\pm}$ have eigenvalue $\pm 1$, we only need to
consider $\mathcal{P}=\mathcal{P}_{+}=-\mathcal{P}_{-}$.
First, we look at how the operator $\mathcal{P}$ affects the operators $a$, and $\sigma_+$:
\begin{eqnarray}
\mathcal{P} \sigma_{+} \mathcal{P}^\dagger &=&e^{i \frac{\pi}{2} \sigma_z }\sigma_{+} 
e^{-i \frac{\pi}{2} \sigma_z }=e^{i\pi}\sigma_+ =-\sigma_+ \\
\mathcal{P} a \mathcal{P}^\dagger &=&e^{i\pi a^\dagger a }a 
e^{-i \pi a^\dagger a }=e^{i \pi}a =-a
\end{eqnarray} 
The Hamiltonians in both Eqs. (\ref{eq:LL-RD_real}) and 
Eq.\ (\ref{eq:D3}) contain diagonal terms ($a^\dagger a$ and $\sigma_z$) that commute with $\mathcal{P}$, and non-diagonal terms that involve {\em odd} power $a, a^\dagger$ multiplying $\sigma_+, \sigma_-$, so then its straightforward to show that $[\mathcal{H},{\cal P}]=0$.  Note that $\mathcal{P}$ is unitary so the condition $[\mathcal{H},{\cal P}]=0$,  can be rewritten as $\mathcal{P} \mathcal{H}\mathcal{P}^\dagger=\mathcal{H}$.  Focusing on the spin-orbit part of Eq.\ (\ref{eq:LL-RD_real})  one obtains
\begin{eqnarray}
& &\mathcal{P} \left (\alpha_{B} a^{\dagger}\sigma_{-}+ \beta_{B} a^{\dagger}\sigma_{+} \right ) \mathcal{P}^\dagger  + \mathrm{h.c.} \nonumber \\
&=&\left (\alpha_{B}  {\cal P}a^{\dagger}\mathcal{P}^\dagger \mathcal{P} \sigma_{-} \mathcal{P}^\dagger+ \beta_{B} \mathcal{P} a^{\dagger}\mathcal{P}^\dagger \mathcal{P} \sigma_{+}\mathcal{P}^\dagger  \right )  + \mathrm{h.c.} \nonumber \\
&=&  \left (\alpha_{B} (-a^{\dagger})(-\sigma_{-})+ \beta_{B} (-a^{\dagger})(-\sigma_{+}) \right )   + \mathrm{h.c.} \nonumber \\
&=&\left (\alpha_{B} a^{\dagger}\sigma_{-}+ \beta_{B} a^{\dagger}\sigma_{+} \right )   + \mathrm{h.c.} ,
\label{eq:Hcommchi}
\end{eqnarray}
which shows that $\mathcal{P}  \mathcal{H}\mathcal{P} ^\dagger=\mathcal{H}$, since the diagonal terms in $\mathcal{H}$ trivially commute with $\mathcal{P}$.  

The practical results of having a diagonal operator $\mathcal{P} $ that commutes with $\mathcal{H}$ is that the Hamiltonian can be diagonalized using two {\em separate} sets of basis states:
\begin{eqnarray}
\mathcal{P} =+1&:& \quad \{ |0, \uparrow \rangle, |1, \downarrow \rangle,  |2, \uparrow \rangle, |3, \downarrow \rangle, |4, \uparrow \rangle, \dots \} \nonumber  \\
\mathcal{P} =-1&:& \quad \{ |0, \downarrow \rangle, |1, \uparrow \rangle,  |2, \downarrow \rangle, |3, \uparrow \rangle, |4, \downarrow \rangle, \dots \} \nonumber
\end{eqnarray}
Diagonalizing $\mathcal{H}$ in either of the $\mathcal{P} = +1$, or $-1$, subspaces will result in a set of states that all anticross.
We can connect these sets of states to $\mathcal{N}_+$ eigenstates
\begin{eqnarray}
\mathcal{P} =+1&:& \quad \{ \overbrace{|0, \uparrow \rangle, |1, \downarrow \rangle}^{|0,+\rangle,|1,-\rangle}, \overbrace{  |2, \uparrow \rangle, |3, \downarrow \rangle}^{|2,+\rangle,|3,-\rangle}, |4, \uparrow \rangle, \dots \} \nonumber  \\
\mathcal{P} =-1&:& \quad \{ |0, \downarrow \rangle, \underbrace{|1, \uparrow \rangle,  |2, \downarrow \rangle}_{|1,+\rangle,|2,-\rangle}, \underbrace{|3, \uparrow \rangle, |4, \downarrow \rangle }_{|3,+\rangle,|4,-\rangle} , \dots \} \nonumber,
\end{eqnarray}
and similarly for the $\mathcal{N}_-$ eigenstates
\begin{eqnarray}
\mathcal{P} =+1&:& \quad \{ |0, \uparrow \rangle,  \overbrace{|1, \downarrow \rangle,  |2, \uparrow \rangle}^{|1,-\rangle,|2,+\rangle}, \overbrace{|3, \downarrow \rangle, |4, \uparrow \rangle }^{|3,-\rangle,|4,+\rangle}, \dots \} \nonumber  \\
\mathcal{P} =-1&:& \quad \{
\underbrace{|0, \downarrow \rangle, |1, \uparrow \rangle}_{|0,-\rangle,|1,+\rangle}, 
\underbrace{|2, \downarrow \rangle,|3, \uparrow \rangle}_{|2,-\rangle,|3,+\rangle}, |4, \downarrow \rangle  , \dots \} \nonumber
\end{eqnarray}
Note that $\mathcal{P} $ also commutes with the cubic Dresselhaus terms as is discussed in App.\ \ref{app:D3}.

\section{Cubic Dresselhaus}
\label{app:D3}
The Hamiltonian in Eq.\ (\ref{eq:LL-RD_real}) describes a 2DEG with both Rashba and {\em linear} Dresselhaus.  For the numerical part we also include the {\em full} cubic Dresselhaus contribution.  Starting from Eq.\ (6.1) in Ref.\ \onlinecite{winkler03:1}, and projecting down to the lowest transverse level results in 
\begin{eqnarray}
\mathcal{H}_{D3}&=&\frac{(-\gamma  \langle k_{z}^2 \rangle)}{\hbar} \left (  \left [ \frac{1}{2}
\Pi_+ \sigma_+ \right . \right . \nonumber \\
& & \left . \left .
- \frac{1}{8 \hbar^2\langle k_{z}^2 \rangle }
\{
\Pi_+^2-\Pi_-^2,\Pi_-
\}  
\right ] + \mathrm{h.c.}
\right ) ,
\end{eqnarray}
where $\Pi_\pm=\Pi_x \pm i \Pi_y$, and  $\langle \Pi_{z}^2 \rangle = \hbar^2 \langle k_{z}^2 \rangle$.  
Note that now the Dresselhaus spin-orbit coupling is parametrized by two parameters $\gamma$ and $\langle k_z^2 \rangle$, while for the linear approximation, only the single parameter $\beta=(-\gamma) \langle k_z^2 \rangle$ is required.
Using the definition in Eqs.\ (\ref{a}) and (\ref{adagger}) the full Dresselhaus Hamiltonian becomes
\begin{eqnarray}
\mathcal{H}_{D3}&=&\frac{(-\gamma  \langle k_{z}^2 \rangle)}{\hbar} \left\{ 
\left [ \left ( 1- \frac{1}{2 \langle k_{z}^2 \rangle \ell_c^2 }a^\dagger a \right )
a^\dagger \sigma_+ \right . \right . \nonumber \\
& & \left . \left .
+ \frac{1}{2\langle k_{z}^2 \rangle \ell_c^2} a^3 \sigma_+
\right ] + \mathrm{h.c.}
\right\} .
\label{eq:D3asigma}
\end{eqnarray}
In the absence of spin-orbit interaction $a^\dagger a$ can be replaced by its eigenvalue $n$, which in turn is related to the ratio of the Fermi energy and $\hbar  \omega_c$ (valid for $\varepsilon_F \ll \hbar \omega_c$)
\begin{eqnarray}
\frac{1}{\ell_c^2}a^\dagger a \rightarrow \frac{1}{\ell_c^2}n &\approx &\frac{1}{\ell_c^2}\frac{\varepsilon_F}{\hbar \omega_c} =\frac{ k_F^2}{2}=\pi n_{2D}.
\end{eqnarray}
In the presence of spin-orbit we can still formally rewrite Eq.\ (\ref{eq:D3asigma}) as
\begin{eqnarray}
\mathcal{H}_{D3}
&=&\frac{(-\gamma ) \left (\langle k_{z}^2 \rangle - \frac{\pi}{2  }  n_{2D} \right )}{\hbar} \left\{ 
\left [ \frac{\langle k_{z}^2 \rangle - \frac{1}{2  \ell_c^2 }a^\dagger a }{ \langle k_{z}^2 \rangle - \frac{\pi }{2 } n_{2D}}
a^\dagger \sigma_+ \right . \right . \nonumber \\
& & \left . \left .
+ \frac{1}{2 \ell_c^2} \frac{1}{ \langle k_{z}^2 \rangle - \frac{\pi }{2 } n_{2D}} a^3 \sigma_+
\right ] + \mathrm{h.c.}
\right\} .
\label{eq:D3asigmaRenorm}
\end{eqnarray}
The prefactor $-\gamma \left (\langle k_{z}^2 \rangle - \frac{\pi}{2  }  n_{2D} \right )$ is {\em defined} as
\begin{eqnarray}
\beta&=&\beta_1-\beta_3 \nonumber \\
&=& \left [(-\gamma) \langle k_{z}^2 \rangle \right ] -  \left [ (-\gamma) \frac{\pi}{2  }  n_{2D} \right ],
\end{eqnarray}
which reduces to the traditional definition of $\beta$ for low density samples as considered in Sec.\ \ref{sec:Analysis}.

The parity operator $\mathcal{P} $ introduced in App.\ \ref{app:chi} also commutes with the Hamiltonian in Eq.\  (\ref{eq:D3asigma}), since the spin-orbit terms involve odd powers of $a,a^\dagger$ multiplied by either $\sigma_+$ or $\sigma_-$, and the sign introduced the unitary transformation gets cancelled.

\section{The numerical procedure for finding the $F$-function}
\label{app:numerical}
For fixed parameter values, the eigenenergies of the Hamiltonian Eq.\ (\ref{eq:LL-RD_real}) take discrete values.  They are obtained numerically by diagonalizing the Hamiltonian matrix using a large enough set of basis states.
Finding the $F$-function as described in Eq.\ (\ref{eq:defFfun}) is equivalent to a root finding problem for the function 
\begin{equation}
g_s(n)=\varepsilon_{n,s}(B)-\varepsilon_F =0.
\end{equation}
This requires the quantum number $n$ to be a continuous variable.  
which leads to a minor modification of the Hamiltonian diagonlization procedure. 
The standard diagonalization proceedure is to construct a $2N_L$ matrix from $N_L$ harmonic oscillator eigenstates, in addition to the spin degree of freedom.
The Pauli matrices form $2\times 2$ blocks that are coupled by the ladder operators $a$ and $a^\dagger$, leading to block tri-diagonal matrix with $2 \times 2$ block matrices
\begin{eqnarray}
h_{l,l}&=&
(l-1)
\left [
\begin{array}{cc}
   1  & 0 \\
   0  & 1
\end{array}  
\right ] +
\left [
\begin{array}{cc}
  \frac{1-\tilde{\Delta}}{2}  & 0 \\
  0  & \frac{1+\tilde{\Delta}}{2}
\end{array} 
\right ] \label{eq:hll} \\
h_{l,l+1}&=& \sqrt{l+1}\left [
\begin{array}{cc}
   0  & 2 \alpha_\beta \\
   2 \beta_B   & 0
\end{array}  
\right ],\label{eq:hllp1}
\end{eqnarray}
where $l$ runs from 1 to $N_L$ (number of Landau levels used in the calculations).  To obtain a continuous version of Eqs. (\ref{eq:hll}) and (\ref{eq:hllp1}) a variable $x$ is added to the index $l$, resulting in
\begin{eqnarray}
h_{l,l}(x)&=&
(l+x-1)
\left [
\begin{array}{cc}
   1  & 0 \\
   0  & 1
\end{array}  
\right ] +
\left [
\begin{array}{cc}
  \frac{1-\tilde{\Delta}}{2}  & 0 \\
  0  & \frac{1+\tilde{\Delta}}{2}
\end{array} 
\right ] \label{eq:hxii} \\
h_{l,l+1}(x)&=& \sqrt{l+x+1}\left [
\begin{array}{cc}
   0  & 2 \alpha_\beta \\
   2 \beta_B   & 0
\end{array}  
\right ].\label{eq:hxiip1}
\end{eqnarray}
The full block-tridiagonal matrix based on the submatrices in Eqs.\ (\ref{eq:hxii}) and (\ref{eq:hxiip1}) will then yield a spectrum $\varepsilon_{n+x,s}$, for $x \in [-1,1]$. To further simply the calculations the basis states can be split into $\mathcal{P}=\pm 1$ subspaces.  Each $\mathcal{P}$-subspace contains ordered states $\{ \epsilon_{0+x}, \epsilon_{1+x}, \dots \}$.  For each subspace, one chooses the two adjecent eigenenergies determined by the condition $\epsilon_{n+x} < \frac{\varepsilon_F}{\hbar \omega_c} <\epsilon_{n+x+1}$. Subsequently the value of $x$ is found by solving $g_s(n+x)=0$.

\section{Perturbation theory and ``Bogoliubov-de Gennes Hamiltonian''}
\label{perturbation}

Here we solve the Hamiltonian Eq.~(\ref{eq:LL-ga_de}) through a perturbative approach. As the Hamiltonian due to the spin-orbit terms are generally much smaller than the Hamiltonian corresponding to free electron gas, we rewrite Eq.~(\ref{eq:LL-ga_de}) as
\begin{align*}
\frac{{\cal \tilde{H}}}{\hbar\omega_{c}} & =\overset{{\cal H}_{0}/\hbar\omega_{c}}{\overbrace{a^{\dagger}a+\frac{1}{2}+\frac{\tilde{\Delta}}{2}\sigma_{z}}}+\overset{{\cal V}/\hbar\omega_{c}}{\overbrace{\gamma\left(a^{\dagger}+a\right)\sigma_{x}+i\delta\left(a-a^{\dagger}\right)\sigma_{y}}}\\
 & ={\cal H}_{0}/\hbar\omega_{c}+{\cal V}/\hbar\omega_{c}.
\end{align*}
with corresponding unperturbed Hamiltonian and perturbation, ${\cal H}_0$ and ${\cal V}$, respectively. Using now the Schrieffer--Wolff transformation~\cite{PhysRev.149.491,bravyi11:2793}, defined by $e^{S}$, with the constraint ${\cal V}+\left[{\cal S},{\cal H}_{0}\right]=0$, we obtain an effective Hamiltonian given by ${\cal H}_{eff}={\cal H}_{0}+\frac{1}{2}\left[{\cal S},{\cal V}\right]+{\cal O}\left({\cal V}^{3}\right)$. For our system we find ${\cal S}={\cal S}_\gamma+{\cal S}_\delta$, with
\begin{align}
{\cal S}_{\gamma} & =-\frac{\gamma}{1-\tilde{\Delta}^{2}}\left\{ a\left(\sigma_{x}+i\tilde{\Delta}\sigma_{y}\right)-a^{\dagger}\left(\sigma_{x}-i\tilde{\Delta}\sigma_{y}\right)\right\} ,\\
{\cal S}_{\delta} & =-\frac{i\delta}{1-\tilde{\Delta}^{2}}\left\{ a\left(\sigma_{y}-i\tilde{\Delta}\sigma_{x}\right)+a^{\dagger}\left(\sigma_{y}+i\tilde{\Delta}\sigma_{x}\right)\right\} ,
\end{align}
yielding 
\begin{align}
    \frac{\tilde{{\cal H}}_{eff}}{\hbar\omega_{c}}&=\frac{1}{2}\left(1+\tilde{\Delta}\sigma_{z}\right)-\Omega-\Lambda\sigma_{z}+\left(1-2\Lambda\sigma_{z}\right)a^{\dagger}a\nonumber \\
    &+\Gamma\left(aa+a^{\dagger}a^{\dagger}\right)\sigma_{z}, \label{HBdG}
\end{align}
{with}
\begin{align}
    \Omega&=\frac{\left(\gamma^{2}+\delta^{2}\right)+2\delta\gamma\tilde{\Delta}}{1-\tilde{\Delta}^{2}},\\
    \Lambda&=\frac{\left(\gamma^{2}+\delta^{2}\right)\tilde{\Delta}+2\delta\gamma}{1-\tilde{\Delta}^{2}},\\
    \varGamma&={\frac{\delta^{2}-\gamma^{2}}{1-\tilde{\Delta}^{2}}\tilde{\Delta}}.
    \end{align}
    
The Hamiltonian Eq.~(\ref{HBdG}) can be rewritten in the Bogoliubov-de Gennes form as
\begin{align}
    \frac{\tilde{{\cal H}}_{eff}}{\hbar\omega_{c}}&=\frac{1}{2}(1+\tilde{\Delta}\sigma_{z})-\Omega-\Lambda\sigma_{z}-\frac{1}{2}\left(1-2\Lambda\sigma_{z}\right)\nonumber\\
    &+\frac{1}{2}\left(a^{\dagger}\quad a\right)\left[\begin{array}{cc}
1-2\Lambda\sigma_z & {2\Gamma\sigma_z}\\
{2\Gamma\sigma_z} & 1-2\Lambda\sigma_z
\end{array}\right]\left(\begin{array}{c}
a\\
a^{\dagger}
\end{array}\right),
\end{align}
which can be diagonalized by a $2\times2$ Bogoliubov-de Gennes transformation, and reads
\begin{widetext}
\begin{equation}
    \frac{\tilde{{\cal H}}_{eff}}{\hbar\omega_{c}}=\frac{1}{2}(1+\tilde{\Delta}\sigma_{z})-\Omega-\Lambda\sigma_{z}-\frac{1}{2}\left(1-2\Lambda\sigma_{z}\right) +\frac{1}{2}\left(\tilde{a}^{\dagger}\quad \tilde{a}\right)\left[\begin{array}{cc}
\sqrt{(1-2\Lambda\sigma_z)^2-4\Gamma^2} & 0  \\
0 & \sqrt{(1-2\Lambda\sigma_z)^2-4\Gamma^2}
\end{array}\right]\left(\begin{array}{c}
\tilde{a}\\
\tilde{a}^{\dagger}
\end{array}\right),
\end{equation}
\end{widetext}
with the diagonal operators $\tilde{a}$ and $\tilde{a}^\dagger$. For most semiconductors, we have $\Omega,\Lambda,\Gamma\ll 1$. By neglecting the fourth order or higher spin-orbit terms, i.e., $\delta^i \gamma^j$ with $i+j\geq 4$, we obtain  
\begin{align}
     \frac{\tilde{{\cal H}}_{eff}}{\hbar\omega_{c}}=\frac{\tilde{\Delta}}{2}\sigma_{z}+\left|1-2\Lambda\sigma_{z}\right|\left(\tilde{a}^{\dagger}\tilde{a}+\frac{1}{2}\right)-\Omega
\end{align}
with energies 
\begin{equation}
    \frac{\varepsilon_{l,s}}{\hbar\omega_{c}}=\frac{s}{2}\tilde{\Delta}-\Omega+\left|1-2\Lambda s\right|\left(l+\frac{1}{2}\right).
\end{equation}
For $1-2\Lambda>0$ we obtain 
\begin{equation}
    \frac{\varepsilon_{l,s}}{\hbar\omega_{c}}=\left(l+\frac{1}{2}+\tilde{\Delta}\frac{s}{2}\right)-2s\Lambda\left(l+\frac{1}{2}\right)-\Omega,
\end{equation}
which is Eq.~(\ref{eq:pertEn}) in the main text.

\section{Approximations leading to Eqs.~(\ref{Fp-app-RD}) and (\ref{Fm-app-RD})}
\label{appox-ffunction}
 Starting from Eq.\ (\ref{eq:pertEnSqrt}) one can obtain the the $F$-function by inverting the energy levels \carlos{to obtain $l$}, for each value of $s$.  The resulting equations are
{\small \begin{widetext}
\begin{align}
{\cal F}_{+} & =\frac{\varepsilon}{\hbar\omega_{c}}-\frac{1}{2}+\left|\Lambda\right|\left(1-\frac{\Lambda}{\left|\Lambda\right|}\tilde{\Delta}\right)+\frac{1}{4}\frac{\Lambda}{\left|\Lambda\right|}\frac{1-\frac{\Lambda}{\left|\Lambda\right|}\tilde{\Delta}}{\left|1-\frac{\Lambda}{\left|\Lambda\right|}\tilde{\Delta}\right|}\sqrt{\left(1-\frac{\Lambda}{\left|\Lambda\right|}\tilde{\Delta}\right)^{2}+8\left(1-\frac{\Lambda}{\left|\Lambda\right|}\tilde{\Delta}\right)\left[\frac{\varepsilon}{\hbar\omega_{c}}\left|\Lambda\right|+\frac{\left|\Lambda\right|^{2}}{2}\left(1-\frac{\Lambda}{\left|\Lambda\right|}\tilde{\Delta}\right)+\frac{1}{2}\left(\,\Omega\frac{\Lambda}{\left|\Lambda\right|}-\Lambda\right)\right]}\nonumber \\
 & -\frac{1}{4}\frac{\Lambda}{\left|\Lambda\right|}\frac{1-\frac{\Lambda}{\left|\Lambda\right|}\tilde{\Delta}}{\left|1-\frac{\Lambda}{\left|\Lambda\right|}\tilde{\Delta}\right|}\sqrt{\left(1-\frac{\Lambda}{\left|\Lambda\right|}\tilde{\Delta}\right)^{2}+8\left(1-\frac{\Lambda}{\left|\Lambda\right|}\tilde{\Delta}\right)\left[\frac{\varepsilon}{\hbar\omega_{c}}\left|\Lambda\right|+\frac{\left|\Lambda\right|^{2}}{2}\left(1-\frac{\Lambda}{\left|\Lambda\right|}\tilde{\Delta}\right)-\frac{1}{2}\left(\,\Omega\frac{\Lambda}{\left|\Lambda\right|}-\Lambda\right)\right]}\label{Fp-RD}\\
\nonumber \\
{\cal F}_{-} & =-\frac{1}{2}\frac{\Lambda}{\left|\Lambda\right|}+\frac{1}{4}\frac{\Lambda}{\left|\Lambda\right|}\frac{1-\frac{\Lambda}{\left|\Lambda\right|}\tilde{\Delta}}{\left|1-\frac{\Lambda}{\left|\Lambda\right|}\tilde{\Delta}\right|}\sqrt{\left(1-\frac{\Lambda}{\left|\Lambda\right|}\tilde{\Delta}\right)^{2}+8\left(1-\frac{\Lambda}{\left|\Lambda\right|}\tilde{\Delta}\right)\left[\frac{\varepsilon}{\hbar\omega_{c}}\left|\Lambda\right|+\frac{\left|\Lambda\right|^{2}}{2}\left(1-\frac{\Lambda}{\left|\Lambda\right|}\tilde{\Delta}\right)+\frac{1}{2}\left(\,\Omega\frac{\Lambda}{\left|\Lambda\right|}-\Lambda\right)\right]}\nonumber \\
 & +\frac{1}{4}\frac{\Lambda}{\left|\Lambda\right|}\frac{1-\frac{\Lambda}{\left|\Lambda\right|}\tilde{\Delta}}{\left|1-\frac{\Lambda}{\left|\Lambda\right|}\tilde{\Delta}\right|}\sqrt{\left(1-\frac{\Lambda}{\left|\Lambda\right|}\tilde{\Delta}\right)^{2}+8\left(1-\frac{\Lambda}{\left|\Lambda\right|}\tilde{\Delta}\right)\left[\frac{\varepsilon}{\hbar\omega_{c}}\left|\Lambda\right|+\frac{\left|\Lambda\right|^{2}}{2}\left(1-\frac{\Lambda}{\left|\Lambda\right|}\tilde{\Delta}\right)-\frac{1}{2}\left(\,\Omega\frac{\Lambda}{\left|\Lambda\right|}-\Lambda\right)\right]}\label{Fm-RD}
\end{align}
\end{widetext}}

We will further simplify these equations by approximating Eqs.~(\ref{Fp-RD}) and (\ref{Fm-RD}) \carlos{up to second order in the spin-orbit parameters $\Lambda$ and $\Omega$ (or fourth order in $\gamma$ and $\delta$).} 
Accordingly, we rewrite these equations as
\begin{align}
{\cal F}_{+} &=\frac{\varepsilon}{\hbar\omega_{c}}-\frac{1}{2}+|\Lambda|\left(1-\frac{\Lambda}{|\Lambda|}\tilde{\Delta}\right)\nonumber\\
&+\frac{1}{4}\frac{\Lambda}{|\Lambda|}\frac{1-\frac{\Lambda}{\left|\Lambda\right|}\tilde{\Delta}}{\left|1-\frac{\Lambda}{\left|\Lambda\right|}\tilde{\Delta}\right|}\left(\sqrt{A+B}-\sqrt{A-B}\right),\label{Fp-square} \\
{\cal F}_{-} & =-\frac{1}{2}\frac{\Lambda}{|\Lambda|}+\frac{1}{4}\frac{\Lambda}{|\Lambda|}\frac{1-\frac{\Lambda}{\left|\Lambda\right|}\tilde{\Delta}}{\left|1-\frac{\Lambda}{\left|\Lambda\right|}\tilde{\Delta}\right|}\left(\sqrt{A+B}+\sqrt{A-B}\right), \label{Fm-square}
\end{align}
where $A=A_{0}+A_{1}+A_{2}$ and $B=B_{1}$, with 
\begin{align}
A_{0} & =\left(1-\frac{\Lambda}{\left|\Lambda\right|}\tilde{\Delta}\right)^{2},\\
A_{1} & =8\frac{\varepsilon}{\hbar\omega_{c}}\left|\Lambda\right|\left(1-\frac{\Lambda}{\left|\Lambda\right|}\tilde{\Delta}\right),\\
A_{2} & =4\left|\Lambda\right|^{2}\left(1-\frac{\Lambda}{\left|\Lambda\right|}\tilde{\Delta}\right)^{2},\\
B_{1} & =4\left(1-\frac{\Lambda}{\left|\Lambda\right|}\tilde{\Delta}\right)\left(\,\Omega\frac{\Lambda}{\left|\Lambda\right|}-\Lambda\right).
\end{align}
Here, the nominal values of the \carlos{subindices} of $A_i$ and $B_j$ indicate their order on the spin-orbit terms $\Lambda$ and $\Omega$. Accordingly, we expand the square roots of Eqs.~(\ref{Fp-square}) and (\ref{Fm-square}) and keep only terms up to second order in either $\Lambda$ or $\Omega$, yielding 
\begin{align}
\sqrt{A+B}+\sqrt{A-B} & \approx 2\sqrt{A_{0}+A_{1}}\left( 1+\frac{1}{2}\frac{A_{2}}{A_{0}+A_{1}}\right) ,\nonumber \\
& = 2\sqrt{A_0+A_1+A_2} \\
\sqrt{A+B}-\sqrt{A-B} &\approx\thinspace \frac{B_{1}}{\sqrt{A_{0}}}.
\end{align}
As a consequence, we can finally write
\begin{widetext}
\begin{align}
{\cal F}_{+}  &=\frac{\varepsilon}{\hbar\omega_{c}}-\frac{1}{2}+\Omega-\Lambda\tilde{\Delta}  \\
{\cal F}_{-} & =-\frac{1}{2}\frac{\Lambda}{|\Lambda|}+\frac{1}{2}\frac{\Lambda}{|\Lambda|}\frac{1-\frac{\Lambda}{\left|\Lambda\right|}\tilde{\Delta}}{\left|1-\frac{\Lambda}{\left|\Lambda\right|}\tilde{\Delta}\right|}\times \sqrt{\left(1-\frac{\Lambda}{\left|\Lambda\right|}\tilde{\Delta}\right)^{2}+8\left|\Lambda\right|\left(1-\frac{\Lambda}{\left|\Lambda\right|}\tilde{\Delta}\right)\left[\frac{\varepsilon}{\hbar\omega_{c}}+\frac{1}{2}\left|\Lambda\right|\left(1-\frac{\Lambda}{\left|\Lambda\right|}\tilde{\Delta}\right)\right]}, 
\end{align}
\end{widetext}
which are Eqs.\ (\ref{Fp-app-RD}) and (\ref{Fm-app-RD})\carlos{, respectively}.

\section{Temperature dependence of the normalized differential resistivity}
\label{app-temp}

In this section we derive the general temperature dependence of the normalized differential magnetoresistivity in Eq.~(\ref{normdiffmag}) for the systems studied in this work.
\begin{align}
\delta \rho_{xx}(B)&=2\sum_{l=1}^{\infty} \int d\varepsilon 
\tilde{L}_\Gamma \left ( l
\frac{\Gamma}{\hbar \omega_c}\right ) \left(-\frac{df_{0}(\varepsilon)}{d\varepsilon}\right)\nonumber \\ 
&\times\cos(2\pi l\mathcal{F}_{-})\cos(2\pi l\mathcal{F}_{+}). \label{norm}
\end{align}
At $T=0$~K, we have $-df_0(\varepsilon)/d\varepsilon \rightarrow \delta(\varepsilon-\varepsilon_F)$, which simplifies Eq.~(\ref{norm}) to
\begin{equation}
\delta \rho_{xx}(B)=2\sum_{l=1}^{\infty}
\tilde{L}_\Gamma \left ( l
\frac{\Gamma}{\hbar \omega_c}\right ) \left. \cos(2\pi l\mathcal{F}_{-})\cos(2\pi l\mathcal{F}_{+}) \right|_{\varepsilon=\varepsilon_F},
\end{equation}
\carlos{being} obviously temperature independent. When the temperature is finite but small, i.e., \carlos{$ k_B T\ll\mu\sim \varepsilon_F$}, we have a temperature \carlos{dependent} $\delta \rho_{xx}(B)$. We now analyze the relevant case for low-density semiconductors, but with $\varepsilon_F\gg\varepsilon_{R},\varepsilon_{D}$, and high number of populated Landau levels, i.e., $\varepsilon_F/\hbar\omega_c\gg1$. With these conditions, all the different cases analyzed in this manuscript present ${\cal F}_{\pm}$-functions constant or linearly dependent on the energy, so we write here, ${\cal F}_{\pm}\propto \varepsilon+cte$, see, for example Eqs.~(\ref{FpmZ}), (\ref{eq:frpSdH}), (\ref{frm2}), (\ref{Fp-app-RD}), and (\ref{Fm-app-RD}).
Using $2\pi l{\cal F}_+=2\Lambda_{+}^{l}\varepsilon+\phi_{+}^{l}$, ${2\pi l{\cal F}_-=2\Lambda_{-}^{l}\varepsilon+\phi_{-}^{l}}$, \carlos{with $\phi_{\pm}^l$ properly defined by comparison with these equations,} and assuming an energy-independent Dingle factor (\carlos{only true for Lorentzian broadening.}), we need to calculate integrals of the following form,
\begin{widetext}
\begin{equation}
\int_{0}^{\infty}d\varepsilon\left(-\frac{\partial f^{0}}{\partial\varepsilon}\right)\cos\left(2\Lambda_{+}^{l}\varepsilon+\phi_{+}^{l}\right)\cos\left(2\Lambda_{-}^{l}\varepsilon+\phi_{-}^{l}\right)=\int_{-\frac{\mu}{2k_{B}T}}^{\infty}dx\frac{\cos\left(4\Lambda_{+}^{l}k_{B}Tx+2\Lambda_{+}^{l}\mu+\phi_{+}^{l}\right)\cos\left(4\Lambda_{-}^{l}k_{B}Tx+2\Lambda_{-}^{l}\mu+\phi_{-}^{l}\right)}{2\sinh^{2}x},
\label{}
\end{equation}
\end{widetext}
where we have introduced the dimensioness quantity $x=\frac{\varepsilon-\mu}{2k_{B}T}$. For $\mu\gg k_B T$, we obtain
\begin{widetext}
\begin{align}
\int_{0}^{\infty}d\varepsilon\left(-\frac{\partial f^{0}}{\partial\varepsilon}\right)\cos\left(2\Lambda_{+}^{l}\varepsilon+\phi_{+}^{l}\right)\cos\left(2\Lambda_{-}^{l}\varepsilon+\phi_{-}^{l}\right)=\pi k_{B}T\left\{ \frac{\left(\Lambda_{+}^{l}-\Lambda_{-}^{l}\right)\cos\left[2\mu\left(\Lambda_{+}^{l}-\Lambda_{-}^{l}\right)+\phi_{+}^{l}-\phi_{-}^{l}\right]}{\sinh\left[2\pi k_{B}T\left(\Lambda_{+}^{l}-\Lambda_{-}^{l}\right)\right]}\right.\\+\left.\frac{\left(\Lambda_{+}^{l}+\Lambda_{-}^{l}\right)\cos\left[2\mu\left(\Lambda_{+}^{l}+\Lambda_{-}^{l}\right)+\phi_{+}^{l}+\phi_{-}^{l}\right]}{\sinh\left[2\pi k_{B}T\left(\Lambda_{+}^{l}+\Lambda_{-}^{l}\right)\right]}\right\}.
\label{}
\end{align}
\end{widetext}
using 
\begin{align*}
\int_{-\infty}^{\infty}dx\frac{\cos(2\lambda_{1}+a_{1})\cos(2\lambda_{2}+a_{2})}{\cosh{x}^{2}}\\
=\frac{\pi(\lambda_{1}-\lambda_{2})\cos(a_{1}-a_{2})}{\sinh\pi(\lambda_{1}-\lambda_{2})}+\frac{\pi(\lambda_{1}+\lambda_{2})\cos(a_{1}+a_{2})}{\sinh\pi(\lambda_{1}+\lambda_{2})}
\end{align*}
For the cases treated in this work, $\Lambda_{+}^{l}\gg \Lambda_{-}^{l}$ holds, and we obtain 
\begin{align*}
\int_{0}^{\infty}d\varepsilon\left(-\frac{\partial f^{0}}{\partial\varepsilon}\right)\cos\left(2\Lambda_{+}^{l}\varepsilon+\phi_{+}^{l}\right)\cos\left(2\Lambda_{-}^{l}\varepsilon+\phi_{-}^{l}\right)\\
\approx{\cal A}_{l}(T)\cos\left(2\mu\Lambda_{+}^{l}+\phi_{+}^{l}\right)\cos\left(2\mu\Lambda_{-}^{l}+\phi_{-}^{l}\right),
\end{align*}
with
\begin{equation}
    {\cal A}_l(T)=\frac{2\pi k_{B}T\Lambda_{+}^{l}}{\sinh\left(2\pi k_{B}T\Lambda_{+}^{l}\right)},
\end{equation}
for the temperature dependent coefficient for the SdH oscillation. For all the cases \carlos{investigated in this work}, we have $\Lambda_{+}=\pi l/\hbar\omega_c$, yielding \carlos{Eq.~(\ref{T-coef}) in the main text,}
\begin{equation}
    {\cal A}_l\left(T\right)=\frac{2\pi^{2}lk_{B}T/\hbar\omega_{c}}{\sinh\left(2\pi^{2}lk_{B}T/\hbar\omega_{c}\right)}.
\end{equation}

\bibliographystyle{apsrev4-1}
\bibliography{general,books,SdH_Landau,spinOrbit,spintronicsTheo,spintronicsExp,spindynamics,spintronics}

\begin{thebibliography}{67}%
\makeatletter
\providecommand \@ifxundefined [1]{%
 \@ifx{#1\undefined}
}%
\providecommand \@ifnum [1]{%
 \ifnum #1\expandafter \@firstoftwo
 \else \expandafter \@secondoftwo
 \fi
}%
\providecommand \@ifx [1]{%
 \ifx #1\expandafter \@firstoftwo
 \else \expandafter \@secondoftwo
 \fi
}%
\providecommand \natexlab [1]{#1}%
\providecommand \enquote  [1]{``#1''}%
\providecommand \bibnamefont  [1]{#1}%
\providecommand \bibfnamefont [1]{#1}%
\providecommand \citenamefont [1]{#1}%
\providecommand \href@noop [0]{\@secondoftwo}%
\providecommand \href [0]{\begingroup \@sanitize@url \@href}%
\providecommand \@href[1]{\@@startlink{#1}\@@href}%
\providecommand \@@href[1]{\endgroup#1\@@endlink}%
\providecommand \@sanitize@url [0]{\catcode `\\12\catcode `\$12\catcode
  `\&12\catcode `\#12\catcode `\^12\catcode `\_12\catcode `\%12\relax}%
\providecommand \@@startlink[1]{}%
\providecommand \@@endlink[0]{}%
\providecommand \url  [0]{\begingroup\@sanitize@url \@url }%
\providecommand \@url [1]{\endgroup\@href {#1}{\urlprefix }}%
\providecommand \urlprefix  [0]{URL }%
\providecommand \Eprint [0]{\href }%
\providecommand \doibase [0]{http://dx.doi.org/}%
\providecommand \selectlanguage [0]{\@gobble}%
\providecommand \bibinfo  [0]{\@secondoftwo}%
\providecommand \bibfield  [0]{\@secondoftwo}%
\providecommand \translation [1]{[#1]}%
\providecommand \BibitemOpen [0]{}%
\providecommand \bibitemStop [0]{}%
\providecommand \bibitemNoStop [0]{.\EOS\space}%
\providecommand \EOS [0]{\spacefactor3000\relax}%
\providecommand \BibitemShut  [1]{\csname bibitem#1\endcsname}%
\let\auto@bib@innerbib\@empty
\bibitem [{\citenamefont {D'Yakonov}\ and\ \citenamefont
  {Perel}(1971)}]{d1971possibility}%
  \BibitemOpen
  \bibfield  {author} {\bibinfo {author} {\bibfnamefont {M.~I.}\ \bibnamefont
  {D'Yakonov}}\ and\ \bibinfo {author} {\bibfnamefont {V.}~\bibnamefont
  {Perel}},\ }\href@noop {} {\bibfield  {journal} {\bibinfo  {journal} {Soviet
  Journal of Experimental and Theoretical Physics Letters}\ }\textbf {\bibinfo
  {volume} {13}},\ \bibinfo {pages} {467} (\bibinfo {year} {1971})}\BibitemShut
  {NoStop}%
\bibitem [{\citenamefont {Dyakonov}\ and\ \citenamefont
  {Perel}(1971)}]{dyakonov1971current}%
  \BibitemOpen
  \bibfield  {author} {\bibinfo {author} {\bibfnamefont {M.~I.}\ \bibnamefont
  {Dyakonov}}\ and\ \bibinfo {author} {\bibfnamefont {V.}~\bibnamefont
  {Perel}},\ }\href@noop {} {\bibfield  {journal} {\bibinfo  {journal} {Physics
  Letters A}\ }\textbf {\bibinfo {volume} {35}},\ \bibinfo {pages} {459}
  (\bibinfo {year} {1971})}\BibitemShut {NoStop}%
\bibitem [{\citenamefont {Hirsch}(1999)}]{PhysRevLett.83.1834}%
  \BibitemOpen
  \bibfield  {author} {\bibinfo {author} {\bibfnamefont {J.~E.}\ \bibnamefont
  {Hirsch}},\ }\href {\doibase 10.1103/PhysRevLett.83.1834} {\bibfield
  {journal} {\bibinfo  {journal} {Phys. Rev. Lett.}\ }\textbf {\bibinfo
  {volume} {83}},\ \bibinfo {pages} {1834} (\bibinfo {year}
  {1999})}\BibitemShut {NoStop}%
\bibitem [{\citenamefont {Landisman}\ and\ \citenamefont
  {Connors}(2005)}]{doi:10.1126/science.1114655}%
  \BibitemOpen
  \bibfield  {author} {\bibinfo {author} {\bibfnamefont {C.~E.}\ \bibnamefont
  {Landisman}}\ and\ \bibinfo {author} {\bibfnamefont {B.~W.}\ \bibnamefont
  {Connors}},\ }\href {\doibase 10.1126/science.1114655} {\bibfield  {journal}
  {\bibinfo  {journal} {Science}\ }\textbf {\bibinfo {volume} {310}},\ \bibinfo
  {pages} {1809} (\bibinfo {year} {2005})},\ \Eprint
  {http://arxiv.org/abs/https://www.science.org/doi/pdf/10.1126/science.1114655}
  {https://www.science.org/doi/pdf/10.1126/science.1114655} \BibitemShut
  {NoStop}%
\bibitem [{\citenamefont {Schliemann}\ \emph
  {et~al.}(2003{\natexlab{a}})\citenamefont {Schliemann}, \citenamefont
  {Egues},\ and\ \citenamefont {Loss}}]{PhysRevLett.90.146801}%
  \BibitemOpen
  \bibfield  {author} {\bibinfo {author} {\bibfnamefont {J.}~\bibnamefont
  {Schliemann}}, \bibinfo {author} {\bibfnamefont {J.~C.}\ \bibnamefont
  {Egues}}, \ and\ \bibinfo {author} {\bibfnamefont {D.}~\bibnamefont {Loss}},\
  }\href {\doibase 10.1103/PhysRevLett.90.146801} {\bibfield  {journal}
  {\bibinfo  {journal} {Phys. Rev. Lett.}\ }\textbf {\bibinfo {volume} {90}},\
  \bibinfo {pages} {146801} (\bibinfo {year} {2003}{\natexlab{a}})}\BibitemShut
  {NoStop}%
\bibitem [{\citenamefont {Bernevig}\ \emph
  {et~al.}(2006{\natexlab{a}})\citenamefont {Bernevig}, \citenamefont
  {Orenstein},\ and\ \citenamefont {Zhang}}]{PhysRevLett.97.236601}%
  \BibitemOpen
  \bibfield  {author} {\bibinfo {author} {\bibfnamefont {B.~A.}\ \bibnamefont
  {Bernevig}}, \bibinfo {author} {\bibfnamefont {J.}~\bibnamefont {Orenstein}},
  \ and\ \bibinfo {author} {\bibfnamefont {S.-C.}\ \bibnamefont {Zhang}},\
  }\href {\doibase 10.1103/PhysRevLett.97.236601} {\bibfield  {journal}
  {\bibinfo  {journal} {Phys. Rev. Lett.}\ }\textbf {\bibinfo {volume} {97}},\
  \bibinfo {pages} {236601} (\bibinfo {year} {2006}{\natexlab{a}})}\BibitemShut
  {NoStop}%
\bibitem [{\citenamefont {Fu}\ \emph {et~al.}(2016)\citenamefont {Fu},
  \citenamefont {Penteado}, \citenamefont {Hachiya}, \citenamefont {Loss},\
  and\ \citenamefont {Egues}}]{PhysRevLett.117.226401}%
  \BibitemOpen
  \bibfield  {author} {\bibinfo {author} {\bibfnamefont {J.}~\bibnamefont
  {Fu}}, \bibinfo {author} {\bibfnamefont {P.~H.}\ \bibnamefont {Penteado}},
  \bibinfo {author} {\bibfnamefont {M.~O.}\ \bibnamefont {Hachiya}}, \bibinfo
  {author} {\bibfnamefont {D.}~\bibnamefont {Loss}}, \ and\ \bibinfo {author}
  {\bibfnamefont {J.~C.}\ \bibnamefont {Egues}},\ }\href {\doibase
  10.1103/PhysRevLett.117.226401} {\bibfield  {journal} {\bibinfo  {journal}
  {Phys. Rev. Lett.}\ }\textbf {\bibinfo {volume} {117}},\ \bibinfo {pages}
  {226401} (\bibinfo {year} {2016})}\BibitemShut {NoStop}%
\bibitem [{\citenamefont {Kane}\ and\ \citenamefont
  {Mele}(2005)}]{PhysRevLett.95.226801}%
  \BibitemOpen
  \bibfield  {author} {\bibinfo {author} {\bibfnamefont {C.~L.}\ \bibnamefont
  {Kane}}\ and\ \bibinfo {author} {\bibfnamefont {E.~J.}\ \bibnamefont
  {Mele}},\ }\href {\doibase 10.1103/PhysRevLett.95.226801} {\bibfield
  {journal} {\bibinfo  {journal} {Phys. Rev. Lett.}\ }\textbf {\bibinfo
  {volume} {95}},\ \bibinfo {pages} {226801} (\bibinfo {year}
  {2005})}\BibitemShut {NoStop}%
\bibitem [{\citenamefont {Bernevig}\ and\ \citenamefont
  {Zhang}(2006)}]{PhysRevLett.96.106802}%
  \BibitemOpen
  \bibfield  {author} {\bibinfo {author} {\bibfnamefont {B.~A.}\ \bibnamefont
  {Bernevig}}\ and\ \bibinfo {author} {\bibfnamefont {S.-C.}\ \bibnamefont
  {Zhang}},\ }\href {\doibase 10.1103/PhysRevLett.96.106802} {\bibfield
  {journal} {\bibinfo  {journal} {Phys. Rev. Lett.}\ }\textbf {\bibinfo
  {volume} {96}},\ \bibinfo {pages} {106802} (\bibinfo {year}
  {2006})}\BibitemShut {NoStop}%
\bibitem [{\citenamefont {Bernevig}\ \emph
  {et~al.}(2006{\natexlab{b}})\citenamefont {Bernevig}, \citenamefont
  {Hughes},\ and\ \citenamefont {Zhang}}]{doi:10.1126/science.1133734}%
  \BibitemOpen
  \bibfield  {author} {\bibinfo {author} {\bibfnamefont {B.~A.}\ \bibnamefont
  {Bernevig}}, \bibinfo {author} {\bibfnamefont {T.~L.}\ \bibnamefont
  {Hughes}}, \ and\ \bibinfo {author} {\bibfnamefont {S.-C.}\ \bibnamefont
  {Zhang}},\ }\href {\doibase 10.1126/science.1133734} {\bibfield  {journal}
  {\bibinfo  {journal} {Science}\ }\textbf {\bibinfo {volume} {314}},\ \bibinfo
  {pages} {1757} (\bibinfo {year} {2006}{\natexlab{b}})},\ \Eprint
  {http://arxiv.org/abs/https://www.science.org/doi/pdf/10.1126/science.1133734}
  {https://www.science.org/doi/pdf/10.1126/science.1133734} \BibitemShut
  {NoStop}%
\bibitem [{\citenamefont {Kitaev}(2001)}]{Kitaev_2001}%
  \BibitemOpen
  \bibfield  {author} {\bibinfo {author} {\bibfnamefont {A.~Y.}\ \bibnamefont
  {Kitaev}},\ }\href {\doibase 10.1070/1063-7869/44/10S/S29} {\bibfield
  {journal} {\bibinfo  {journal} {Physics-Uspekhi}\ }\textbf {\bibinfo {volume}
  {44}},\ \bibinfo {pages} {131} (\bibinfo {year} {2001})}\BibitemShut
  {NoStop}%
\bibitem [{\citenamefont {Fu}\ and\ \citenamefont
  {Kane}(2009)}]{PhysRevB.79.161408}%
  \BibitemOpen
  \bibfield  {author} {\bibinfo {author} {\bibfnamefont {L.}~\bibnamefont
  {Fu}}\ and\ \bibinfo {author} {\bibfnamefont {C.~L.}\ \bibnamefont {Kane}},\
  }\href {\doibase 10.1103/PhysRevB.79.161408} {\bibfield  {journal} {\bibinfo
  {journal} {Phys. Rev. B}\ }\textbf {\bibinfo {volume} {79}},\ \bibinfo
  {pages} {161408} (\bibinfo {year} {2009})}\BibitemShut {NoStop}%
\bibitem [{\citenamefont {Candido}\ \emph {et~al.}(2018)\citenamefont
  {Candido}, \citenamefont {Flatt\'e},\ and\ \citenamefont
  {Egues}}]{candido18:256804}%
  \BibitemOpen
  \bibfield  {author} {\bibinfo {author} {\bibfnamefont {D.~R.}\ \bibnamefont
  {Candido}}, \bibinfo {author} {\bibfnamefont {M.~E.}\ \bibnamefont
  {Flatt\'e}}, \ and\ \bibinfo {author} {\bibfnamefont {J.~C.}\ \bibnamefont
  {Egues}},\ }\href {\doibase 10.1103/PhysRevLett.121.256804} {\bibfield
  {journal} {\bibinfo  {journal} {Phys. Rev. Lett.}\ }\textbf {\bibinfo
  {volume} {121}},\ \bibinfo {pages} {256804} (\bibinfo {year}
  {2018})}\BibitemShut {NoStop}%
\bibitem [{\citenamefont {Pal}(2011)}]{doi:10.1119/1.3549729}%
  \BibitemOpen
  \bibfield  {author} {\bibinfo {author} {\bibfnamefont {P.~B.}\ \bibnamefont
  {Pal}},\ }\href {\doibase 10.1119/1.3549729} {\bibfield  {journal} {\bibinfo
  {journal} {American Journal of Physics}\ }\textbf {\bibinfo {volume} {79}},\
  \bibinfo {pages} {485} (\bibinfo {year} {2011})},\ \Eprint
  {http://arxiv.org/abs/https://doi.org/10.1119/1.3549729}
  {https://doi.org/10.1119/1.3549729} \BibitemShut {NoStop}%
\bibitem [{\citenamefont {Gilbertsson}\ \emph {et~al.}(2008)\citenamefont
  {Gilbertsson}, \citenamefont {Fearn}, \citenamefont {Jefferson},
  \citenamefont {Murdin}, \citenamefont {Buckle},\ and\ \citenamefont
  {Cohen}}]{gilbertsson08:165335}%
  \BibitemOpen
  \bibfield  {author} {\bibinfo {author} {\bibfnamefont {A.}~\bibnamefont
  {Gilbertsson}}, \bibinfo {author} {\bibfnamefont {M.}~\bibnamefont {Fearn}},
  \bibinfo {author} {\bibfnamefont {J.}~\bibnamefont {Jefferson}}, \bibinfo
  {author} {\bibfnamefont {B.}~\bibnamefont {Murdin}}, \bibinfo {author}
  {\bibfnamefont {P.}~\bibnamefont {Buckle}}, \ and\ \bibinfo {author}
  {\bibfnamefont {L.}~\bibnamefont {Cohen}},\ }\href@noop {} {\bibfield
  {journal} {\bibinfo  {journal} {Phys.\ Rev.\ B}\ }\textbf {\bibinfo {volume}
  {77}},\ \bibinfo {pages} {165335} (\bibinfo {year} {2008})}\BibitemShut
  {NoStop}%
\bibitem [{\citenamefont {Shubnikov}\ and\ \citenamefont
  {De~Haas}(1930)}]{sdh-original}%
  \BibitemOpen
  \bibfield  {author} {\bibinfo {author} {\bibfnamefont {L.}~\bibnamefont
  {Shubnikov}}\ and\ \bibinfo {author} {\bibfnamefont {W.}~\bibnamefont
  {De~Haas}},\ }\href@noop {} {\bibfield  {journal} {\bibinfo  {journal} {J. de
  Haas, Leiden Commun. 207a, 207c}\ }\textbf {\bibinfo {volume} {207}},\
  \bibinfo {pages} {210a} (\bibinfo {year} {1930})}\BibitemShut {NoStop}%
\bibitem [{\citenamefont {Shubnikov}\ and\ \citenamefont
  {de~Haas}(1930)}]{sdh-original2}%
  \BibitemOpen
  \bibfield  {author} {\bibinfo {author} {\bibfnamefont {L.}~\bibnamefont
  {Shubnikov}}\ and\ \bibinfo {author} {\bibfnamefont {W.}~\bibnamefont
  {de~Haas}},\ }in\ \href@noop {} {\emph {\bibinfo {booktitle} {Proc.
  Netherlands Roy. Acad. Sci}}},\ Vol.~\bibinfo {volume} {33}\ (\bibinfo {year}
  {1930})\ p.\ \bibinfo {pages} {363}\BibitemShut {NoStop}%
\bibitem [{\citenamefont {Smole\ifmmode~\acute{n}\else \'{n}\fi{}ski}\ \emph
  {et~al.}(2019)\citenamefont {Smole\ifmmode~\acute{n}\else \'{n}\fi{}ski},
  \citenamefont {Cotlet}, \citenamefont {Popert}, \citenamefont {Back},
  \citenamefont {Shimazaki}, \citenamefont {Kn\"uppel}, \citenamefont
  {Dietler}, \citenamefont {Taniguchi}, \citenamefont {Watanabe}, \citenamefont
  {Kroner},\ and\ \citenamefont {Imamoglu}}]{PhysRevLett.123.097403}%
  \BibitemOpen
  \bibfield  {author} {\bibinfo {author} {\bibfnamefont {T.}~\bibnamefont
  {Smole\ifmmode~\acute{n}\else \'{n}\fi{}ski}}, \bibinfo {author}
  {\bibfnamefont {O.}~\bibnamefont {Cotlet}}, \bibinfo {author} {\bibfnamefont
  {A.}~\bibnamefont {Popert}}, \bibinfo {author} {\bibfnamefont
  {P.}~\bibnamefont {Back}}, \bibinfo {author} {\bibfnamefont {Y.}~\bibnamefont
  {Shimazaki}}, \bibinfo {author} {\bibfnamefont {P.}~\bibnamefont
  {Kn\"uppel}}, \bibinfo {author} {\bibfnamefont {N.}~\bibnamefont {Dietler}},
  \bibinfo {author} {\bibfnamefont {T.}~\bibnamefont {Taniguchi}}, \bibinfo
  {author} {\bibfnamefont {K.}~\bibnamefont {Watanabe}}, \bibinfo {author}
  {\bibfnamefont {M.}~\bibnamefont {Kroner}}, \ and\ \bibinfo {author}
  {\bibfnamefont {A.}~\bibnamefont {Imamoglu}},\ }\href {\doibase
  10.1103/PhysRevLett.123.097403} {\bibfield  {journal} {\bibinfo  {journal}
  {Phys. Rev. Lett.}\ }\textbf {\bibinfo {volume} {123}},\ \bibinfo {pages}
  {097403} (\bibinfo {year} {2019})}\BibitemShut {NoStop}%
\bibitem [{\citenamefont {Korm\'anyos}\ \emph {et~al.}(2015)\citenamefont
  {Korm\'anyos}, \citenamefont {Rakyta},\ and\ \citenamefont
  {Burkard}}]{Kormanyos_2015}%
  \BibitemOpen
  \bibfield  {author} {\bibinfo {author} {\bibfnamefont {A.}~\bibnamefont
  {Korm\'anyos}}, \bibinfo {author} {\bibfnamefont {P.}~\bibnamefont {Rakyta}},
  \ and\ \bibinfo {author} {\bibfnamefont {G.}~\bibnamefont {Burkard}},\ }\href
  {\doibase 10.1088/1367-2630/17/10/103006} {\bibfield  {journal} {\bibinfo
  {journal} {New Journal of Physics}\ }\textbf {\bibinfo {volume} {17}},\
  \bibinfo {pages} {103006} (\bibinfo {year} {2015})}\BibitemShut {NoStop}%
\bibitem [{\citenamefont {Cui}\ \emph {et~al.}(2015)\citenamefont {Cui},
  \citenamefont {Lee}, \citenamefont {Kim}, \citenamefont {Arefe},
  \citenamefont {Huang}, \citenamefont {Lee}, \citenamefont {Chenet},
  \citenamefont {Zhang}, \citenamefont {Wang}, \citenamefont {Ye} \emph
  {et~al.}}]{cui2015multi}%
  \BibitemOpen
  \bibfield  {author} {\bibinfo {author} {\bibfnamefont {X.}~\bibnamefont
  {Cui}}, \bibinfo {author} {\bibfnamefont {G.-H.}\ \bibnamefont {Lee}},
  \bibinfo {author} {\bibfnamefont {Y.~D.}\ \bibnamefont {Kim}}, \bibinfo
  {author} {\bibfnamefont {G.}~\bibnamefont {Arefe}}, \bibinfo {author}
  {\bibfnamefont {P.~Y.}\ \bibnamefont {Huang}}, \bibinfo {author}
  {\bibfnamefont {C.-H.}\ \bibnamefont {Lee}}, \bibinfo {author} {\bibfnamefont
  {D.~A.}\ \bibnamefont {Chenet}}, \bibinfo {author} {\bibfnamefont
  {X.}~\bibnamefont {Zhang}}, \bibinfo {author} {\bibfnamefont
  {L.}~\bibnamefont {Wang}}, \bibinfo {author} {\bibfnamefont {F.}~\bibnamefont
  {Ye}},  \emph {et~al.},\ }\href@noop {} {\bibfield  {journal} {\bibinfo
  {journal} {Nature nanotechnology}\ }\textbf {\bibinfo {volume} {10}},\
  \bibinfo {pages} {534} (\bibinfo {year} {2015})}\BibitemShut {NoStop}%
\bibitem [{\citenamefont {Xu}\ \emph {et~al.}(2016)\citenamefont {Xu},
  \citenamefont {Wu}, \citenamefont {Lu}, \citenamefont {Han}, \citenamefont
  {Long}, \citenamefont {Chen}, \citenamefont {Han}, \citenamefont {Ye},
  \citenamefont {Wu}, \citenamefont {Lin} \emph {et~al.}}]{xu2016universal}%
  \BibitemOpen
  \bibfield  {author} {\bibinfo {author} {\bibfnamefont {S.}~\bibnamefont
  {Xu}}, \bibinfo {author} {\bibfnamefont {Z.}~\bibnamefont {Wu}}, \bibinfo
  {author} {\bibfnamefont {H.}~\bibnamefont {Lu}}, \bibinfo {author}
  {\bibfnamefont {Y.}~\bibnamefont {Han}}, \bibinfo {author} {\bibfnamefont
  {G.}~\bibnamefont {Long}}, \bibinfo {author} {\bibfnamefont {X.}~\bibnamefont
  {Chen}}, \bibinfo {author} {\bibfnamefont {T.}~\bibnamefont {Han}}, \bibinfo
  {author} {\bibfnamefont {W.}~\bibnamefont {Ye}}, \bibinfo {author}
  {\bibfnamefont {Y.}~\bibnamefont {Wu}}, \bibinfo {author} {\bibfnamefont
  {J.}~\bibnamefont {Lin}},  \emph {et~al.},\ }\href@noop {} {\bibfield
  {journal} {\bibinfo  {journal} {2D Materials}\ }\textbf {\bibinfo {volume}
  {3}},\ \bibinfo {pages} {021007} (\bibinfo {year} {2016})}\BibitemShut
  {NoStop}%
\bibitem [{\citenamefont {Cui}\ \emph {et~al.}(2017)\citenamefont {Cui},
  \citenamefont {Shih}, \citenamefont {Jauregui}, \citenamefont {Chae},
  \citenamefont {Kim}, \citenamefont {Li}, \citenamefont {Seo}, \citenamefont
  {Pistunova}, \citenamefont {Yin}, \citenamefont {Park} \emph
  {et~al.}}]{cui2017low}%
  \BibitemOpen
  \bibfield  {author} {\bibinfo {author} {\bibfnamefont {X.}~\bibnamefont
  {Cui}}, \bibinfo {author} {\bibfnamefont {E.-M.}\ \bibnamefont {Shih}},
  \bibinfo {author} {\bibfnamefont {L.~A.}\ \bibnamefont {Jauregui}}, \bibinfo
  {author} {\bibfnamefont {S.~H.}\ \bibnamefont {Chae}}, \bibinfo {author}
  {\bibfnamefont {Y.~D.}\ \bibnamefont {Kim}}, \bibinfo {author} {\bibfnamefont
  {B.}~\bibnamefont {Li}}, \bibinfo {author} {\bibfnamefont {D.}~\bibnamefont
  {Seo}}, \bibinfo {author} {\bibfnamefont {K.}~\bibnamefont {Pistunova}},
  \bibinfo {author} {\bibfnamefont {J.}~\bibnamefont {Yin}}, \bibinfo {author}
  {\bibfnamefont {J.-H.}\ \bibnamefont {Park}},  \emph {et~al.},\ }\href@noop
  {} {\bibfield  {journal} {\bibinfo  {journal} {Nano letters}\ }\textbf
  {\bibinfo {volume} {17}},\ \bibinfo {pages} {4781} (\bibinfo {year}
  {2017})}\BibitemShut {NoStop}%
\bibitem [{\citenamefont {Rhodes}\ \emph {et~al.}(2019)\citenamefont {Rhodes},
  \citenamefont {Chae}, \citenamefont {Ribeiro-Palau},\ and\ \citenamefont
  {Hone}}]{rhodes2019disorder}%
  \BibitemOpen
  \bibfield  {author} {\bibinfo {author} {\bibfnamefont {D.}~\bibnamefont
  {Rhodes}}, \bibinfo {author} {\bibfnamefont {S.~H.}\ \bibnamefont {Chae}},
  \bibinfo {author} {\bibfnamefont {R.}~\bibnamefont {Ribeiro-Palau}}, \ and\
  \bibinfo {author} {\bibfnamefont {J.}~\bibnamefont {Hone}},\ }\href@noop {}
  {\bibfield  {journal} {\bibinfo  {journal} {Nature materials}\ }\textbf
  {\bibinfo {volume} {18}},\ \bibinfo {pages} {541} (\bibinfo {year}
  {2019})}\BibitemShut {NoStop}%
\bibitem [{\citenamefont {Masseroni}\ \emph {et~al.}(2023)\citenamefont
  {Masseroni}, \citenamefont {Qu}, \citenamefont {Taniguchi}, \citenamefont
  {Watanabe}, \citenamefont {Ihn},\ and\ \citenamefont
  {Ensslin}}]{PhysRevResearch.5.013113}%
  \BibitemOpen
  \bibfield  {author} {\bibinfo {author} {\bibfnamefont {M.}~\bibnamefont
  {Masseroni}}, \bibinfo {author} {\bibfnamefont {T.}~\bibnamefont {Qu}},
  \bibinfo {author} {\bibfnamefont {T.}~\bibnamefont {Taniguchi}}, \bibinfo
  {author} {\bibfnamefont {K.}~\bibnamefont {Watanabe}}, \bibinfo {author}
  {\bibfnamefont {T.}~\bibnamefont {Ihn}}, \ and\ \bibinfo {author}
  {\bibfnamefont {K.}~\bibnamefont {Ensslin}},\ }\href {\doibase
  10.1103/PhysRevResearch.5.013113} {\bibfield  {journal} {\bibinfo  {journal}
  {Phys. Rev. Res.}\ }\textbf {\bibinfo {volume} {5}},\ \bibinfo {pages}
  {013113} (\bibinfo {year} {2023})}\BibitemShut {NoStop}%
\bibitem [{\citenamefont {Slizovskiy}\ \emph {et~al.}(2023)\citenamefont
  {Slizovskiy}, \citenamefont {Tomi{\'c}}, \citenamefont {Kumar}, \citenamefont
  {Garcia-Ruiz}, \citenamefont {Zheng}, \citenamefont {Portol{\'e}s},
  \citenamefont {Ponomarenko}, \citenamefont {Geim}, \citenamefont {Watanabe},
  \citenamefont {Taniguchi} \emph {et~al.}}]{slizovskiy2023kagom}%
  \BibitemOpen
  \bibfield  {author} {\bibinfo {author} {\bibfnamefont {S.}~\bibnamefont
  {Slizovskiy}}, \bibinfo {author} {\bibfnamefont {P.}~\bibnamefont
  {Tomi{\'c}}}, \bibinfo {author} {\bibfnamefont {R.~K.}\ \bibnamefont
  {Kumar}}, \bibinfo {author} {\bibfnamefont {A.}~\bibnamefont {Garcia-Ruiz}},
  \bibinfo {author} {\bibfnamefont {G.}~\bibnamefont {Zheng}}, \bibinfo
  {author} {\bibfnamefont {E.}~\bibnamefont {Portol{\'e}s}}, \bibinfo {author}
  {\bibfnamefont {L.~A.}\ \bibnamefont {Ponomarenko}}, \bibinfo {author}
  {\bibfnamefont {A.~K.}\ \bibnamefont {Geim}}, \bibinfo {author}
  {\bibfnamefont {K.}~\bibnamefont {Watanabe}}, \bibinfo {author}
  {\bibfnamefont {T.}~\bibnamefont {Taniguchi}},  \emph {et~al.},\ }\href@noop
  {} {\bibfield  {journal} {\bibinfo  {journal} {arXiv preprint
  arXiv:2303.06403}\ } (\bibinfo {year} {2023})}\BibitemShut {NoStop}%
\bibitem [{\citenamefont {Ferreira}\ \emph {et~al.}(2022)\citenamefont
  {Ferreira}, \citenamefont {Candido}, \citenamefont {Hernandez}, \citenamefont
  {Gusev}, \citenamefont {Olshanetsky}, \citenamefont {Mikhailov},\ and\
  \citenamefont {Dvoretsky}}]{Ferreira2022}%
  \BibitemOpen
  \bibfield  {author} {\bibinfo {author} {\bibfnamefont {G.~J.}\ \bibnamefont
  {Ferreira}}, \bibinfo {author} {\bibfnamefont {D.~R.}\ \bibnamefont
  {Candido}}, \bibinfo {author} {\bibfnamefont {F.~G.~G.}\ \bibnamefont
  {Hernandez}}, \bibinfo {author} {\bibfnamefont {G.~M.}\ \bibnamefont
  {Gusev}}, \bibinfo {author} {\bibfnamefont {E.~B.}\ \bibnamefont
  {Olshanetsky}}, \bibinfo {author} {\bibfnamefont {N.~N.}\ \bibnamefont
  {Mikhailov}}, \ and\ \bibinfo {author} {\bibfnamefont {S.~A.}\ \bibnamefont
  {Dvoretsky}},\ }\href {\doibase 10.1038/s41598-022-06431-0} {\bibfield
  {journal} {\bibinfo  {journal} {Scientific Reports}\ }\textbf {\bibinfo
  {volume} {12}} (\bibinfo {year} {2022}),\
  10.1038/s41598-022-06431-0}\BibitemShut {NoStop}%
\bibitem [{\citenamefont {Cao}\ \emph {et~al.}(2018{\natexlab{a}})\citenamefont
  {Cao}, \citenamefont {Fatemi}, \citenamefont {Fang}, \citenamefont
  {Watanabe}, \citenamefont {Taniguchi}, \citenamefont {Kaxiras},\ and\
  \citenamefont {Jarillo-Herrero}}]{cao2018unconventional}%
  \BibitemOpen
  \bibfield  {author} {\bibinfo {author} {\bibfnamefont {Y.}~\bibnamefont
  {Cao}}, \bibinfo {author} {\bibfnamefont {V.}~\bibnamefont {Fatemi}},
  \bibinfo {author} {\bibfnamefont {S.}~\bibnamefont {Fang}}, \bibinfo {author}
  {\bibfnamefont {K.}~\bibnamefont {Watanabe}}, \bibinfo {author}
  {\bibfnamefont {T.}~\bibnamefont {Taniguchi}}, \bibinfo {author}
  {\bibfnamefont {E.}~\bibnamefont {Kaxiras}}, \ and\ \bibinfo {author}
  {\bibfnamefont {P.}~\bibnamefont {Jarillo-Herrero}},\ }\href@noop {}
  {\bibfield  {journal} {\bibinfo  {journal} {Nature}\ }\textbf {\bibinfo
  {volume} {556}},\ \bibinfo {pages} {43} (\bibinfo {year}
  {2018}{\natexlab{a}})}\BibitemShut {NoStop}%
\bibitem [{\citenamefont {Cao}\ \emph {et~al.}(2018{\natexlab{b}})\citenamefont
  {Cao}, \citenamefont {Fatemi}, \citenamefont {Demir}, \citenamefont {Fang},
  \citenamefont {Tomarken}, \citenamefont {Luo}, \citenamefont
  {Sanchez-Yamagishi}, \citenamefont {Watanabe}, \citenamefont {Taniguchi},
  \citenamefont {Kaxiras} \emph {et~al.}}]{cao2018correlated}%
  \BibitemOpen
  \bibfield  {author} {\bibinfo {author} {\bibfnamefont {Y.}~\bibnamefont
  {Cao}}, \bibinfo {author} {\bibfnamefont {V.}~\bibnamefont {Fatemi}},
  \bibinfo {author} {\bibfnamefont {A.}~\bibnamefont {Demir}}, \bibinfo
  {author} {\bibfnamefont {S.}~\bibnamefont {Fang}}, \bibinfo {author}
  {\bibfnamefont {S.~L.}\ \bibnamefont {Tomarken}}, \bibinfo {author}
  {\bibfnamefont {J.~Y.}\ \bibnamefont {Luo}}, \bibinfo {author} {\bibfnamefont
  {J.~D.}\ \bibnamefont {Sanchez-Yamagishi}}, \bibinfo {author} {\bibfnamefont
  {K.}~\bibnamefont {Watanabe}}, \bibinfo {author} {\bibfnamefont
  {T.}~\bibnamefont {Taniguchi}}, \bibinfo {author} {\bibfnamefont
  {E.}~\bibnamefont {Kaxiras}},  \emph {et~al.},\ }\href@noop {} {\bibfield
  {journal} {\bibinfo  {journal} {Nature}\ }\textbf {\bibinfo {volume} {556}},\
  \bibinfo {pages} {80} (\bibinfo {year} {2018}{\natexlab{b}})}\BibitemShut
  {NoStop}%
\bibitem [{\citenamefont {Hu}\ \emph {et~al.}(2016)\citenamefont {Hu},
  \citenamefont {Tang}, \citenamefont {Liu}, \citenamefont {Liu}, \citenamefont
  {Zhu}, \citenamefont {Graf}, \citenamefont {Myhro}, \citenamefont {Tran},
  \citenamefont {Lau}, \citenamefont {Wei} \emph {et~al.}}]{hu2016evidence}%
  \BibitemOpen
  \bibfield  {author} {\bibinfo {author} {\bibfnamefont {J.}~\bibnamefont
  {Hu}}, \bibinfo {author} {\bibfnamefont {Z.}~\bibnamefont {Tang}}, \bibinfo
  {author} {\bibfnamefont {J.}~\bibnamefont {Liu}}, \bibinfo {author}
  {\bibfnamefont {X.}~\bibnamefont {Liu}}, \bibinfo {author} {\bibfnamefont
  {Y.}~\bibnamefont {Zhu}}, \bibinfo {author} {\bibfnamefont {D.}~\bibnamefont
  {Graf}}, \bibinfo {author} {\bibfnamefont {K.}~\bibnamefont {Myhro}},
  \bibinfo {author} {\bibfnamefont {S.}~\bibnamefont {Tran}}, \bibinfo {author}
  {\bibfnamefont {C.~N.}\ \bibnamefont {Lau}}, \bibinfo {author} {\bibfnamefont
  {J.}~\bibnamefont {Wei}},  \emph {et~al.},\ }\href@noop {} {\bibfield
  {journal} {\bibinfo  {journal} {Physical Review Letters}\ }\textbf {\bibinfo
  {volume} {117}},\ \bibinfo {pages} {016602} (\bibinfo {year}
  {2016})}\BibitemShut {NoStop}%
\bibitem [{\citenamefont {Murakawa}\ \emph {et~al.}(2013)\citenamefont
  {Murakawa}, \citenamefont {Bahramy}, \citenamefont {Tokunaga}, \citenamefont
  {Kohama}, \citenamefont {Bell}, \citenamefont {Kaneko}, \citenamefont
  {Nagaosa}, \citenamefont {Hwang},\ and\ \citenamefont
  {Tokura}}]{doi:10.1126/science.1242247}%
  \BibitemOpen
  \bibfield  {author} {\bibinfo {author} {\bibfnamefont {H.}~\bibnamefont
  {Murakawa}}, \bibinfo {author} {\bibfnamefont {M.~S.}\ \bibnamefont
  {Bahramy}}, \bibinfo {author} {\bibfnamefont {M.}~\bibnamefont {Tokunaga}},
  \bibinfo {author} {\bibfnamefont {Y.}~\bibnamefont {Kohama}}, \bibinfo
  {author} {\bibfnamefont {C.}~\bibnamefont {Bell}}, \bibinfo {author}
  {\bibfnamefont {Y.}~\bibnamefont {Kaneko}}, \bibinfo {author} {\bibfnamefont
  {N.}~\bibnamefont {Nagaosa}}, \bibinfo {author} {\bibfnamefont {H.~Y.}\
  \bibnamefont {Hwang}}, \ and\ \bibinfo {author} {\bibfnamefont
  {Y.}~\bibnamefont {Tokura}},\ }\href {\doibase 10.1126/science.1242247}
  {\bibfield  {journal} {\bibinfo  {journal} {Science}\ }\textbf {\bibinfo
  {volume} {342}},\ \bibinfo {pages} {1490} (\bibinfo {year} {2013})},\ \Eprint
  {http://arxiv.org/abs/https://www.science.org/doi/pdf/10.1126/science.1242247}
  {https://www.science.org/doi/pdf/10.1126/science.1242247} \BibitemShut
  {NoStop}%
\bibitem [{\citenamefont {Datta}\ \emph {et~al.}(2019)\citenamefont {Datta},
  \citenamefont {Adak}, \citenamefont {kun Shi}, \citenamefont {Watanabe},
  \citenamefont {Taniguchi}, \citenamefont {Song},\ and\ \citenamefont
  {Deshmukh}}]{doi:10.1126/sciadv.aax6550}%
  \BibitemOpen
  \bibfield  {author} {\bibinfo {author} {\bibfnamefont {B.}~\bibnamefont
  {Datta}}, \bibinfo {author} {\bibfnamefont {P.~C.}\ \bibnamefont {Adak}},
  \bibinfo {author} {\bibfnamefont {L.}~\bibnamefont {kun Shi}}, \bibinfo
  {author} {\bibfnamefont {K.}~\bibnamefont {Watanabe}}, \bibinfo {author}
  {\bibfnamefont {T.}~\bibnamefont {Taniguchi}}, \bibinfo {author}
  {\bibfnamefont {J.~C.~W.}\ \bibnamefont {Song}}, \ and\ \bibinfo {author}
  {\bibfnamefont {M.~M.}\ \bibnamefont {Deshmukh}},\ }\href {\doibase
  10.1126/sciadv.aax6550} {\bibfield  {journal} {\bibinfo  {journal} {Science
  Advances}\ }\textbf {\bibinfo {volume} {5}},\ \bibinfo {pages} {eaax6550}
  (\bibinfo {year} {2019})},\ \Eprint
  {http://arxiv.org/abs/https://www.science.org/doi/pdf/10.1126/sciadv.aax6550}
  {https://www.science.org/doi/pdf/10.1126/sciadv.aax6550} \BibitemShut
  {NoStop}%
\bibitem [{\citenamefont {Alexandradinata}\ \emph {et~al.}(2018)\citenamefont
  {Alexandradinata}, \citenamefont {Wang}, \citenamefont {Duan},\ and\
  \citenamefont {Glazman}}]{alexandradinata2018revealing}%
  \BibitemOpen
  \bibfield  {author} {\bibinfo {author} {\bibfnamefont {A.}~\bibnamefont
  {Alexandradinata}}, \bibinfo {author} {\bibfnamefont {C.}~\bibnamefont
  {Wang}}, \bibinfo {author} {\bibfnamefont {W.}~\bibnamefont {Duan}}, \ and\
  \bibinfo {author} {\bibfnamefont {L.}~\bibnamefont {Glazman}},\ }\href@noop
  {} {\bibfield  {journal} {\bibinfo  {journal} {Physical Review X}\ }\textbf
  {\bibinfo {volume} {8}},\ \bibinfo {pages} {011027} (\bibinfo {year}
  {2018})}\BibitemShut {NoStop}%
\bibitem [{\citenamefont {Ihn}(2010)}]{ihn10:book}%
  \BibitemOpen
  \bibfield  {author} {\bibinfo {author} {\bibfnamefont {T.}~\bibnamefont
  {Ihn}},\ }\href@noop {} {\emph {\bibinfo {title} {Semiconductor
  Nanostructures}}}\ (\bibinfo  {publisher} {Oxford University Press},\
  \bibinfo {year} {2010})\BibitemShut {NoStop}%
\bibitem [{\citenamefont {Das}\ \emph {et~al.}(1989)\citenamefont {Das},
  \citenamefont {Miller}, \citenamefont {Datta}, \citenamefont {Reifenberg},
  \citenamefont {Hong}, \citenamefont {Bhattacharya}, \citenamefont {Singh},\
  and\ \citenamefont {Jaffe}}]{das89:1411}%
  \BibitemOpen
  \bibfield  {author} {\bibinfo {author} {\bibfnamefont {B.}~\bibnamefont
  {Das}}, \bibinfo {author} {\bibfnamefont {D.}~\bibnamefont {Miller}},
  \bibinfo {author} {\bibfnamefont {S.}~\bibnamefont {Datta}}, \bibinfo
  {author} {\bibfnamefont {R.}~\bibnamefont {Reifenberg}}, \bibinfo {author}
  {\bibfnamefont {W.}~\bibnamefont {Hong}}, \bibinfo {author} {\bibfnamefont
  {P.}~\bibnamefont {Bhattacharya}}, \bibinfo {author} {\bibfnamefont
  {J.}~\bibnamefont {Singh}}, \ and\ \bibinfo {author} {\bibfnamefont
  {M.}~\bibnamefont {Jaffe}},\ }\href@noop {} {\bibfield  {journal} {\bibinfo
  {journal} {Phys.\ Rev.\ B}\ }\textbf {\bibinfo {volume} {39}},\ \bibinfo
  {pages} {1411} (\bibinfo {year} {1989})}\BibitemShut {NoStop}%
\bibitem [{\citenamefont {Nitta}\ \emph {et~al.}(1997)\citenamefont {Nitta},
  \citenamefont {Akazaki}, \citenamefont {Takayanagi},\ and\ \citenamefont
  {Enoki}}]{PhysRevLett.78.1335}%
  \BibitemOpen
  \bibfield  {author} {\bibinfo {author} {\bibfnamefont {J.}~\bibnamefont
  {Nitta}}, \bibinfo {author} {\bibfnamefont {T.}~\bibnamefont {Akazaki}},
  \bibinfo {author} {\bibfnamefont {H.}~\bibnamefont {Takayanagi}}, \ and\
  \bibinfo {author} {\bibfnamefont {T.}~\bibnamefont {Enoki}},\ }\href
  {\doibase 10.1103/PhysRevLett.78.1335} {\bibfield  {journal} {\bibinfo
  {journal} {Phys. Rev. Lett.}\ }\textbf {\bibinfo {volume} {78}},\ \bibinfo
  {pages} {1335} (\bibinfo {year} {1997})}\BibitemShut {NoStop}%
\bibitem [{\citenamefont {Engels}\ \emph {et~al.}(1997)\citenamefont {Engels},
  \citenamefont {Lange}, \citenamefont {Sch\"{a}pers},\ and\ \citenamefont
  {L\"{u}th}}]{engels97:1958R}%
  \BibitemOpen
  \bibfield  {author} {\bibinfo {author} {\bibfnamefont {G.}~\bibnamefont
  {Engels}}, \bibinfo {author} {\bibfnamefont {J.}~\bibnamefont {Lange}},
  \bibinfo {author} {\bibfnamefont {T.}~\bibnamefont {Sch\"{a}pers}}, \ and\
  \bibinfo {author} {\bibfnamefont {H.}~\bibnamefont {L\"{u}th}},\ }\href@noop
  {} {\bibfield  {journal} {\bibinfo  {journal} {Phys.\ Rev.\ B}\ }\textbf
  {\bibinfo {volume} {55}},\ \bibinfo {pages} {R1958} (\bibinfo {year}
  {1997})}\BibitemShut {NoStop}%
\bibitem [{\citenamefont {Sch\"apers}\ \emph {et~al.}(1998)\citenamefont
  {Sch\"apers}, \citenamefont {Engels}, \citenamefont {Lange}, \citenamefont
  {Klocke}, \citenamefont {Hollfelder},\ and\ \citenamefont
  {Lüth}}]{10.1063/1.367192}%
  \BibitemOpen
  \bibfield  {author} {\bibinfo {author} {\bibfnamefont {T.}~\bibnamefont
  {Sch\"apers}}, \bibinfo {author} {\bibfnamefont {G.}~\bibnamefont {Engels}},
  \bibinfo {author} {\bibfnamefont {J.}~\bibnamefont {Lange}}, \bibinfo
  {author} {\bibfnamefont {T.}~\bibnamefont {Klocke}}, \bibinfo {author}
  {\bibfnamefont {M.}~\bibnamefont {Hollfelder}}, \ and\ \bibinfo {author}
  {\bibfnamefont {H.}~\bibnamefont {Lüth}},\ }\href {\doibase
  10.1063/1.367192} {\bibfield  {journal} {\bibinfo  {journal} {Journal of
  Applied Physics}\ }\textbf {\bibinfo {volume} {83}},\ \bibinfo {pages} {4324}
  (\bibinfo {year} {1998})},\ \Eprint
  {http://arxiv.org/abs/https://pubs.aip.org/aip/jap/article-pdf/83/8/4324/10592945/4324\_1\_online.pdf}
  {https://pubs.aip.org/aip/jap/article-pdf/83/8/4324/10592945/4324\_1\_online.pdf}
  \BibitemShut {NoStop}%
\bibitem [{\citenamefont {Akabori}\ \emph {et~al.}(2006)\citenamefont
  {Akabori}, \citenamefont {Sunouchi}, \citenamefont {Kakegawa}, \citenamefont
  {Sato}, \citenamefont {Suzuki},\ and\ \citenamefont
  {Yamada}}]{akabori06:413}%
  \BibitemOpen
  \bibfield  {author} {\bibinfo {author} {\bibfnamefont {M.}~\bibnamefont
  {Akabori}}, \bibinfo {author} {\bibfnamefont {T.}~\bibnamefont {Sunouchi}},
  \bibinfo {author} {\bibfnamefont {T.}~\bibnamefont {Kakegawa}}, \bibinfo
  {author} {\bibfnamefont {T.}~\bibnamefont {Sato}}, \bibinfo {author}
  {\bibfnamefont {T.}~\bibnamefont {Suzuki}}, \ and\ \bibinfo {author}
  {\bibfnamefont {Y.}~\bibnamefont {Yamada}},\ }\href@noop {} {\bibfield
  {journal} {\bibinfo  {journal} {Physica E}\ }\textbf {\bibinfo {volume}
  {34}},\ \bibinfo {pages} {413} (\bibinfo {year} {2006})}\BibitemShut
  {NoStop}%
\bibitem [{\citenamefont {Akabori}\ \emph {et~al.}(2008)\citenamefont
  {Akabori}, \citenamefont {Guzenko}, \citenamefont {Sato}, \citenamefont
  {Sch\"{a}pers}, \citenamefont {Suzuki},\ and\ \citenamefont
  {Yamada}}]{akabori08:205320}%
  \BibitemOpen
  \bibfield  {author} {\bibinfo {author} {\bibfnamefont {M.}~\bibnamefont
  {Akabori}}, \bibinfo {author} {\bibfnamefont {V.}~\bibnamefont {Guzenko}},
  \bibinfo {author} {\bibfnamefont {T.}~\bibnamefont {Sato}}, \bibinfo {author}
  {\bibfnamefont {T.}~\bibnamefont {Sch\"{a}pers}}, \bibinfo {author}
  {\bibfnamefont {T.}~\bibnamefont {Suzuki}}, \ and\ \bibinfo {author}
  {\bibfnamefont {S.}~\bibnamefont {Yamada}},\ }\href@noop {} {\bibfield
  {journal} {\bibinfo  {journal} {Phys.\ Rev.\ B}\ }\textbf {\bibinfo {volume}
  {77}},\ \bibinfo {pages} {205320} (\bibinfo {year} {2008})}\BibitemShut
  {NoStop}%
\bibitem [{\citenamefont {Yang}\ and\ \citenamefont
  {Chang}(2006)}]{yang06:045303}%
  \BibitemOpen
  \bibfield  {author} {\bibinfo {author} {\bibfnamefont {W.}~\bibnamefont
  {Yang}}\ and\ \bibinfo {author} {\bibfnamefont {K.}~\bibnamefont {Chang}},\
  }\href@noop {} {\bibfield  {journal} {\bibinfo  {journal} {Phys. Rev. B}\
  }\textbf {\bibinfo {volume} {73}},\ \bibinfo {pages} {045303} (\bibinfo
  {year} {2006})}\BibitemShut {NoStop}%
\bibitem [{\citenamefont {Beukman}\ \emph {et~al.}(2017)\citenamefont
  {Beukman}, \citenamefont {de~Vries}, \citenamefont {van Veen}, \citenamefont
  {Skolasinski}, \citenamefont {Wimmer}, \citenamefont {Qu}, \citenamefont
  {de~Vries}, \citenamefont {Nguyen}, \citenamefont {Yi}, \citenamefont
  {Kiselev}, \citenamefont {Sokolich}, \citenamefont {Manfra}, \citenamefont
  {Nichele}, \citenamefont {Marcus},\ and\ \citenamefont
  {Kouwenhoven}}]{beukmann17:241401}%
  \BibitemOpen
  \bibfield  {author} {\bibinfo {author} {\bibfnamefont {A.~J.~A.}\
  \bibnamefont {Beukman}}, \bibinfo {author} {\bibfnamefont {F.~K.}\
  \bibnamefont {de~Vries}}, \bibinfo {author} {\bibfnamefont {J.}~\bibnamefont
  {van Veen}}, \bibinfo {author} {\bibfnamefont {R.}~\bibnamefont
  {Skolasinski}}, \bibinfo {author} {\bibfnamefont {M.}~\bibnamefont {Wimmer}},
  \bibinfo {author} {\bibfnamefont {F.}~\bibnamefont {Qu}}, \bibinfo {author}
  {\bibfnamefont {D.~T.}\ \bibnamefont {de~Vries}}, \bibinfo {author}
  {\bibfnamefont {B.-M.}\ \bibnamefont {Nguyen}}, \bibinfo {author}
  {\bibfnamefont {W.}~\bibnamefont {Yi}}, \bibinfo {author} {\bibfnamefont
  {A.~A.}\ \bibnamefont {Kiselev}}, \bibinfo {author} {\bibfnamefont
  {M.}~\bibnamefont {Sokolich}}, \bibinfo {author} {\bibfnamefont {M.~J.}\
  \bibnamefont {Manfra}}, \bibinfo {author} {\bibfnamefont {F.}~\bibnamefont
  {Nichele}}, \bibinfo {author} {\bibfnamefont {C.~M.}\ \bibnamefont {Marcus}},
  \ and\ \bibinfo {author} {\bibfnamefont {L.~P.}\ \bibnamefont
  {Kouwenhoven}},\ }\href@noop {} {\bibfield  {journal} {\bibinfo  {journal}
  {Phys. Rev. B}\ }\textbf {\bibinfo {volume} {96}},\ \bibinfo {pages} {241401}
  (\bibinfo {year} {2017})}\BibitemShut {NoStop}%
\bibitem [{\citenamefont {Winkler}\ \emph {et~al.}(2000)\citenamefont
  {Winkler}, \citenamefont {Papadakis}, \citenamefont {De~Poortere},\ and\
  \citenamefont {Shayegan}}]{winkler2000anomalous}%
  \BibitemOpen
  \bibfield  {author} {\bibinfo {author} {\bibfnamefont {R.}~\bibnamefont
  {Winkler}}, \bibinfo {author} {\bibfnamefont {S.}~\bibnamefont {Papadakis}},
  \bibinfo {author} {\bibfnamefont {E.}~\bibnamefont {De~Poortere}}, \ and\
  \bibinfo {author} {\bibfnamefont {M.}~\bibnamefont {Shayegan}},\ }\href@noop
  {} {\bibfield  {journal} {\bibinfo  {journal} {Physical review letters}\
  }\textbf {\bibinfo {volume} {84}},\ \bibinfo {pages} {713} (\bibinfo {year}
  {2000})}\BibitemShut {NoStop}%
\bibitem [{\citenamefont {Raikh}\ and\ \citenamefont
  {Shahbazyan}(1994)}]{raikh94:5531}%
  \BibitemOpen
  \bibfield  {author} {\bibinfo {author} {\bibfnamefont {M.}~\bibnamefont
  {Raikh}}\ and\ \bibinfo {author} {\bibfnamefont {T.}~\bibnamefont
  {Shahbazyan}},\ }\href@noop {} {\bibfield  {journal} {\bibinfo  {journal}
  {Phys.\ Rev.\ B}\ }\textbf {\bibinfo {volume} {49}},\ \bibinfo {pages} {5531}
  (\bibinfo {year} {1994})}\BibitemShut {NoStop}%
\bibitem [{\citenamefont {Averkiev}\ \emph {et~al.}(2005)\citenamefont
  {Averkiev}, \citenamefont {Glazov},\ and\ \citenamefont
  {Tarasenko}}]{averkiev05:543}%
  \BibitemOpen
  \bibfield  {author} {\bibinfo {author} {\bibfnamefont {N.}~\bibnamefont
  {Averkiev}}, \bibinfo {author} {\bibfnamefont {M.}~\bibnamefont {Glazov}}, \
  and\ \bibinfo {author} {\bibfnamefont {S.}~\bibnamefont {Tarasenko}},\
  }\href@noop {} {\bibfield  {journal} {\bibinfo  {journal} {Solid State
  Commun.}\ }\textbf {\bibinfo {volume} {133}},\ \bibinfo {pages} {543}
  (\bibinfo {year} {2005})}\BibitemShut {NoStop}%
\bibitem [{\citenamefont {Cimpoiasu}\ \emph {et~al.}(2019)\citenamefont
  {Cimpoiasu}, \citenamefont {Dunphy}, \citenamefont {Mack}, \citenamefont
  {Christodoulides}, \citenamefont {Lunsford-Poe},\ and\ \citenamefont
  {Bennett}}]{cimpoiasu19:075704}%
  \BibitemOpen
  \bibfield  {author} {\bibinfo {author} {\bibfnamefont {E.}~\bibnamefont
  {Cimpoiasu}}, \bibinfo {author} {\bibfnamefont {B.}~\bibnamefont {Dunphy}},
  \bibinfo {author} {\bibfnamefont {S.}~\bibnamefont {Mack}}, \bibinfo {author}
  {\bibfnamefont {J.}~\bibnamefont {Christodoulides}}, \bibinfo {author}
  {\bibfnamefont {B.}~\bibnamefont {Lunsford-Poe}}, \ and\ \bibinfo {author}
  {\bibfnamefont {B.}~\bibnamefont {Bennett}},\ }\href@noop {} {\bibfield
  {journal} {\bibinfo  {journal} {Journal of App.\ Phys.}\ }\textbf {\bibinfo
  {volume} {126}},\ \bibinfo {pages} {075704} (\bibinfo {year}
  {2019})}\BibitemShut {NoStop}%
\bibitem [{\citenamefont {Dettwiler}\ \emph {et~al.}(2017)\citenamefont
  {Dettwiler}, \citenamefont {Fu}, \citenamefont {Mack}, \citenamefont
  {Weigele}, \citenamefont {Egues}, \citenamefont {Awschalom},\ and\
  \citenamefont {Zumb\"uhl}}]{psh-prx}%
  \BibitemOpen
  \bibfield  {author} {\bibinfo {author} {\bibfnamefont {F.}~\bibnamefont
  {Dettwiler}}, \bibinfo {author} {\bibfnamefont {J.}~\bibnamefont {Fu}},
  \bibinfo {author} {\bibfnamefont {S.}~\bibnamefont {Mack}}, \bibinfo {author}
  {\bibfnamefont {P.~J.}\ \bibnamefont {Weigele}}, \bibinfo {author}
  {\bibfnamefont {J.~C.}\ \bibnamefont {Egues}}, \bibinfo {author}
  {\bibfnamefont {D.~D.}\ \bibnamefont {Awschalom}}, \ and\ \bibinfo {author}
  {\bibfnamefont {D.~M.}\ \bibnamefont {Zumb\"uhl}},\ }\href {\doibase
  10.1103/PhysRevX.7.031010} {\bibfield  {journal} {\bibinfo  {journal} {Phys.
  Rev. X}\ }\textbf {\bibinfo {volume} {7}},\ \bibinfo {pages} {031010}
  (\bibinfo {year} {2017})}\BibitemShut {NoStop}%
\bibitem [{\citenamefont {Brack}\ and\ \citenamefont
  {Bhaduri}(1997)}]{brack97:book}%
  \BibitemOpen
  \bibfield  {author} {\bibinfo {author} {\bibfnamefont {M.}~\bibnamefont
  {Brack}}\ and\ \bibinfo {author} {\bibfnamefont {R.}~\bibnamefont
  {Bhaduri}},\ }\href@noop {} {\emph {\bibinfo {title} {Semiclassical
  physics}}}\ (\bibinfo  {publisher} {Addison-Wesley Publishing},\ \bibinfo
  {year} {1997})\BibitemShut {NoStop}%
\bibitem [{\citenamefont {Bychkov}\ and\ \citenamefont
  {Rashba}(1984)}]{bychkov84:6039}%
  \BibitemOpen
  \bibfield  {author} {\bibinfo {author} {\bibfnamefont {Y.}~\bibnamefont
  {Bychkov}}\ and\ \bibinfo {author} {\bibfnamefont {E.}~\bibnamefont
  {Rashba}},\ }\href@noop {} {\bibfield  {journal} {\bibinfo  {journal} {J.\
  Phys.\ C}\ }\textbf {\bibinfo {volume} {17}},\ \bibinfo {pages} {6039}
  (\bibinfo {year} {1984})}\BibitemShut {NoStop}%
\bibitem [{\citenamefont {Dresselhaus}(1955)}]{dresselhaus55:580}%
  \BibitemOpen
  \bibfield  {author} {\bibinfo {author} {\bibfnamefont {G.}~\bibnamefont
  {Dresselhaus}},\ }\href@noop {} {\bibfield  {journal} {\bibinfo  {journal}
  {Phys.\ Rev.}\ }\textbf {\bibinfo {volume} {100}},\ \bibinfo {pages} {580}
  (\bibinfo {year} {1955})}\BibitemShut {NoStop}%
\bibitem [{\citenamefont {Winkler}(2003)}]{winkler03:1}%
  \BibitemOpen
  \bibfield  {author} {\bibinfo {author} {\bibfnamefont {R.}~\bibnamefont
  {Winkler}},\ }\href@noop {} {\emph {\bibinfo {title} {Spin-Orbit Coupling
  Effects in Two-Dimensional Electron and Hole Systems}}}\ (\bibinfo
  {publisher} {Springer Verlag},\ \bibinfo {year} {2003})\BibitemShut {NoStop}%
\bibitem [{\citenamefont {Braak}(2011)}]{braak11:100401}%
  \BibitemOpen
  \bibfield  {author} {\bibinfo {author} {\bibfnamefont {D.}~\bibnamefont
  {Braak}},\ }\href@noop {} {\bibfield  {journal} {\bibinfo  {journal} {Phys.\
  Rev.\ Lett.}\ }\textbf {\bibinfo {volume} {107}},\ \bibinfo {pages} {100401}
  (\bibinfo {year} {2011})}\BibitemShut {NoStop}%
\bibitem [{\citenamefont {Casanova}\ \emph {et~al.}(2010)\citenamefont
  {Casanova}, \citenamefont {Romero}, \citenamefont {Lizuain}, \citenamefont
  {Garc{\'\i}a-Ripoll},\ and\ \citenamefont {Solano}}]{casanova2010deep}%
  \BibitemOpen
  \bibfield  {author} {\bibinfo {author} {\bibfnamefont {J.}~\bibnamefont
  {Casanova}}, \bibinfo {author} {\bibfnamefont {G.}~\bibnamefont {Romero}},
  \bibinfo {author} {\bibfnamefont {I.}~\bibnamefont {Lizuain}}, \bibinfo
  {author} {\bibfnamefont {J.~J.}\ \bibnamefont {Garc{\'\i}a-Ripoll}}, \ and\
  \bibinfo {author} {\bibfnamefont {E.}~\bibnamefont {Solano}},\ }\href@noop {}
  {\bibfield  {journal} {\bibinfo  {journal} {Physical Review Letters}\
  }\textbf {\bibinfo {volume} {105}},\ \bibinfo {pages} {263603} (\bibinfo
  {year} {2010})}\BibitemShut {NoStop}%
\bibitem [{\citenamefont {Schliemann}\ \emph
  {et~al.}(2003{\natexlab{b}})\citenamefont {Schliemann}, \citenamefont
  {Egues},\ and\ \citenamefont {Loss}}]{schliemann03:085302}%
  \BibitemOpen
  \bibfield  {author} {\bibinfo {author} {\bibfnamefont {J.}~\bibnamefont
  {Schliemann}}, \bibinfo {author} {\bibfnamefont {J.~C.}\ \bibnamefont
  {Egues}}, \ and\ \bibinfo {author} {\bibfnamefont {D.}~\bibnamefont {Loss}},\
  }\href@noop {} {\bibfield  {journal} {\bibinfo  {journal} {Phys.\ Rev.\ B}\
  }\textbf {\bibinfo {volume} {67}},\ \bibinfo {pages} {085302} (\bibinfo
  {year} {2003}{\natexlab{b}})}\BibitemShut {NoStop}%
\bibitem [{\citenamefont {Zarea}\ and\ \citenamefont
  {Ulloa}(2005)}]{zarea05:085342}%
  \BibitemOpen
  \bibfield  {author} {\bibinfo {author} {\bibfnamefont {M.}~\bibnamefont
  {Zarea}}\ and\ \bibinfo {author} {\bibfnamefont {S.}~\bibnamefont {Ulloa}},\
  }\href@noop {} {\bibfield  {journal} {\bibinfo  {journal} {Phys.\ Rev.\ B}\
  }\textbf {\bibinfo {volume} {72}},\ \bibinfo {pages} {085342} (\bibinfo
  {year} {2005})}\BibitemShut {NoStop}%
\bibitem [{\citenamefont {Wang}\ and\ \citenamefont
  {Vasilopoulos}(2005)}]{wang05:085344}%
  \BibitemOpen
  \bibfield  {author} {\bibinfo {author} {\bibfnamefont {X.}~\bibnamefont
  {Wang}}\ and\ \bibinfo {author} {\bibfnamefont {P.}~\bibnamefont
  {Vasilopoulos}},\ }\href@noop {} {\bibfield  {journal} {\bibinfo  {journal}
  {Phys.\ Rev.\ B}\ }\textbf {\bibinfo {volume} {72}},\ \bibinfo {pages}
  {085344} (\bibinfo {year} {2005})}\BibitemShut {NoStop}%
\bibitem [{\citenamefont {Zhang}(2006)}]{zhang06:L477}%
  \BibitemOpen
  \bibfield  {author} {\bibinfo {author} {\bibfnamefont {D.}~\bibnamefont
  {Zhang}},\ }\href@noop {} {\bibfield  {journal} {\bibinfo  {journal} {J.\
  Phys.\ A: Math.\ Gen.}\ }\textbf {\bibinfo {volume} {39}},\ \bibinfo {pages}
  {L477} (\bibinfo {year} {2006})}\BibitemShut {NoStop}%
\bibitem [{\citenamefont {Erlingsson}\ \emph {et~al.}(2010)\citenamefont
  {Erlingsson}, \citenamefont {Egues},\ and\ \citenamefont
  {Loss}}]{erlingsson10:155456}%
  \BibitemOpen
  \bibfield  {author} {\bibinfo {author} {\bibfnamefont {S.}~\bibnamefont
  {Erlingsson}}, \bibinfo {author} {\bibfnamefont {J.}~\bibnamefont {Egues}}, \
  and\ \bibinfo {author} {\bibfnamefont {D.}~\bibnamefont {Loss}},\ }\href@noop
  {} {\bibfield  {journal} {\bibinfo  {journal} {Phys.\ Rev.\ B}\ }\textbf
  {\bibinfo {volume} {82}},\ \bibinfo {pages} {155456} (\bibinfo {year}
  {2010})}\BibitemShut {NoStop}%
\bibitem [{\citenamefont {Tarasenko}(2002)}]{tarasenko02:1769}%
  \BibitemOpen
  \bibfield  {author} {\bibinfo {author} {\bibfnamefont {S.}~\bibnamefont
  {Tarasenko}},\ }\href@noop {} {\bibfield  {journal} {\bibinfo  {journal}
  {Physics of Solid State}\ }\textbf {\bibinfo {volume} {44}},\ \bibinfo
  {pages} {1769} (\bibinfo {year} {2002})}\BibitemShut {NoStop}%
\bibitem [{\citenamefont {Laikhtman}\ and\ \citenamefont
  {Altshuler}(1994)}]{laikhtman94:1994332}%
  \BibitemOpen
  \bibfield  {author} {\bibinfo {author} {\bibfnamefont {B.}~\bibnamefont
  {Laikhtman}}\ and\ \bibinfo {author} {\bibfnamefont {B.}~\bibnamefont
  {Altshuler}},\ }\href {\doibase https://doi.org/10.1006/aphy.1994.1056}
  {\bibfield  {journal} {\bibinfo  {journal} {Annals of Physics}\ }\textbf
  {\bibinfo {volume} {232}},\ \bibinfo {pages} {332} (\bibinfo {year}
  {1994})}\BibitemShut {NoStop}%
\bibitem [{\citenamefont {Lei}\ \emph {et~al.}(2020)\citenamefont {Lei},
  \citenamefont {Lehner}, \citenamefont {Rubi}, \citenamefont {Cheah},
  \citenamefont {Karalic}, \citenamefont {Mittag}, \citenamefont {Alt},
  \citenamefont {Scharnetzky}, \citenamefont {M\"arki}, \citenamefont
  {Zeitler}, \citenamefont {Wegscheider}, \citenamefont {Ihn},\ and\
  \citenamefont {Ensslin}}]{lei20:033213}%
  \BibitemOpen
  \bibfield  {author} {\bibinfo {author} {\bibfnamefont {Z.}~\bibnamefont
  {Lei}}, \bibinfo {author} {\bibfnamefont {C.~A.}\ \bibnamefont {Lehner}},
  \bibinfo {author} {\bibfnamefont {K.}~\bibnamefont {Rubi}}, \bibinfo {author}
  {\bibfnamefont {E.}~\bibnamefont {Cheah}}, \bibinfo {author} {\bibfnamefont
  {M.}~\bibnamefont {Karalic}}, \bibinfo {author} {\bibfnamefont
  {C.}~\bibnamefont {Mittag}}, \bibinfo {author} {\bibfnamefont
  {L.}~\bibnamefont {Alt}}, \bibinfo {author} {\bibfnamefont {J.}~\bibnamefont
  {Scharnetzky}}, \bibinfo {author} {\bibfnamefont {P.}~\bibnamefont
  {M\"arki}}, \bibinfo {author} {\bibfnamefont {U.}~\bibnamefont {Zeitler}},
  \bibinfo {author} {\bibfnamefont {W.}~\bibnamefont {Wegscheider}}, \bibinfo
  {author} {\bibfnamefont {T.}~\bibnamefont {Ihn}}, \ and\ \bibinfo {author}
  {\bibfnamefont {K.}~\bibnamefont {Ensslin}},\ }\href {\doibase
  10.1103/PhysRevResearch.2.033213} {\bibfield  {journal} {\bibinfo  {journal}
  {Phys. Rev. Research}\ }\textbf {\bibinfo {volume} {2}},\ \bibinfo {pages}
  {033213} (\bibinfo {year} {2020})}\BibitemShut {NoStop}%
\bibitem [{Note1()}]{Note1}%
  \BibitemOpen
  \bibinfo {note} {InAs has the value of $\protect \mathaccentV
  {tilde}07E{\Delta } \approx 0.24$, but one would have a symmetric structure
  to minimize the spin-orbit contribution}\BibitemShut {NoStop}%
\bibitem [{\citenamefont {Tarasenko}\ and\ \citenamefont
  {Averkiev}(2002)}]{tarasenko02:552}%
  \BibitemOpen
  \bibfield  {author} {\bibinfo {author} {\bibfnamefont {S.}~\bibnamefont
  {Tarasenko}}\ and\ \bibinfo {author} {\bibfnamefont {N.}~\bibnamefont
  {Averkiev}},\ }\href@noop {} {\bibfield  {journal} {\bibinfo  {journal} {JETP
  Lett.}\ }\textbf {\bibinfo {volume} {75}},\ \bibinfo {pages} {552} (\bibinfo
  {year} {2002})}\BibitemShut {NoStop}%
\bibitem [{Note2()}]{Note2}%
  \BibitemOpen
  \bibinfo {note} {Here we emphasize that depending on the values of $\Lambda $
  and $\Omega $, the argument of square root of Eq.~(\ref {eq:pertEnSqrt})
  becomes negative, thus yielding imaginary energies. This happens when $x\gg
  1$, which violates the assumption of writing Eq.~(\ref
  {eq:pertEnSqrt})}\BibitemShut {NoStop}%
\bibitem [{\citenamefont {Gilbertsson}\ \emph {et~al.}(2009)\citenamefont
  {Gilbertsson}, \citenamefont {Branford}, \citenamefont {M.\~Fearn},
  \citenamefont {Ashley},\ and\ \citenamefont {Cohen}}]{gilbertsson09:235333}%
  \BibitemOpen
  \bibfield  {author} {\bibinfo {author} {\bibfnamefont {A.}~\bibnamefont
  {Gilbertsson}}, \bibinfo {author} {\bibfnamefont {W.}~\bibnamefont
  {Branford}}, \bibinfo {author} {\bibfnamefont {L.~B.}\ \bibnamefont
  {M.\~Fearn}}, \bibinfo {author} {\bibfnamefont {T.}~\bibnamefont {Ashley}}, \
  and\ \bibinfo {author} {\bibfnamefont {L.}~\bibnamefont {Cohen}},\
  }\href@noop {} {\bibfield  {journal} {\bibinfo  {journal} {Phys.\ Rev.\ B}\
  }\textbf {\bibinfo {volume} {79}},\ \bibinfo {pages} {235333} (\bibinfo
  {year} {2009})}\BibitemShut {NoStop}%
\bibitem [{\citenamefont {Fu}\ \emph {et~al.}(2020)\citenamefont {Fu},
  \citenamefont {Penteado}, \citenamefont {Candido}, \citenamefont {Ferreira},
  \citenamefont {Pires}, \citenamefont {Bernardes},\ and\ \citenamefont
  {Egues}}]{fu20:134416}%
  \BibitemOpen
  \bibfield  {author} {\bibinfo {author} {\bibfnamefont {J.}~\bibnamefont
  {Fu}}, \bibinfo {author} {\bibfnamefont {P.~H.}\ \bibnamefont {Penteado}},
  \bibinfo {author} {\bibfnamefont {D.~R.}\ \bibnamefont {Candido}}, \bibinfo
  {author} {\bibfnamefont {G.~J.}\ \bibnamefont {Ferreira}}, \bibinfo {author}
  {\bibfnamefont {D.~P.}\ \bibnamefont {Pires}}, \bibinfo {author}
  {\bibfnamefont {E.}~\bibnamefont {Bernardes}}, \ and\ \bibinfo {author}
  {\bibfnamefont {J.~C.}\ \bibnamefont {Egues}},\ }\href {\doibase
  10.1103/PhysRevB.101.134416} {\bibfield  {journal} {\bibinfo  {journal}
  {Phys. Rev. B}\ }\textbf {\bibinfo {volume} {101}},\ \bibinfo {pages}
  {134416} (\bibinfo {year} {2020})}\BibitemShut {NoStop}%
\bibitem [{\citenamefont {Schrieffer}\ and\ \citenamefont
  {Wolff}(1966)}]{PhysRev.149.491}%
  \BibitemOpen
  \bibfield  {author} {\bibinfo {author} {\bibfnamefont {J.~R.}\ \bibnamefont
  {Schrieffer}}\ and\ \bibinfo {author} {\bibfnamefont {P.~A.}\ \bibnamefont
  {Wolff}},\ }\href {\doibase 10.1103/PhysRev.149.491} {\bibfield  {journal}
  {\bibinfo  {journal} {Phys. Rev.}\ }\textbf {\bibinfo {volume} {149}},\
  \bibinfo {pages} {491} (\bibinfo {year} {1966})}\BibitemShut {NoStop}%
\bibitem [{\citenamefont {Sergey~Bravyi}(2011)}]{bravyi11:2793}%
  \BibitemOpen
  \bibfield  {author} {\bibinfo {author} {\bibfnamefont {D.~L.}\ \bibnamefont
  {Sergey~Bravyi}, \bibfnamefont {David~DiVincenzo}},\ }\href@noop {}
  {\bibfield  {journal} {\bibinfo  {journal} {Annals of Physics}\ }\textbf
  {\bibinfo {volume} {326}},\ \bibinfo {pages} {2793} (\bibinfo {year}
  {2011})}\BibitemShut {NoStop}%
\end{thebibliography}%

\end{document}